%% file: main.tex
\documentclass[useAMS,usenatbib]{mn2e}

\usepackage{graphicx}
\usepackage{times}
\usepackage{tabularx}
\usepackage{natbib}
\usepackage{url} 
\usepackage{xcolor}
\usepackage{caption}
\usepackage{amsmath}
\usepackage{amssymb}
\usepackage{float}

\DeclareUnicodeCharacter{2212}{\textendash}

\input{commands.tex}

\begin{document}

\title[Substructures of radio relics]{A morphological analysis of the substructures in radio relics}
\author[D. Wittor et al.]
{D. Wittor$^{1,2}$\thanks{E-mail: dwittor@hs.uni-hamburg.de}, M. Brüggen$^{1}$, P. Grete$^{1}$, K. Rajpurohit$^{3,2,4}$ \\
$^{1}$ Hamburger Sternwarte, University of Hamburg, Gojenbergsweg 112, 21029 Hamburg, Germany \\
$^{2}$ Dipartimento di Fisica e Astronomia, Universita di Bologna, Via Gobetti 93/2, 40122, Bologna, Italy  \\
$^{3}$ Center for Astrophysics-Harvard \& Smithsonian, 60 Garden Street, Cambridge, MA 02138, USA \\
$^{4}$ Th\"uringer Landessternwarte, Sternwarte 5, 07778 Tautenburg, Germany}
\date{Accepted ???. Received ???; in original form ???}
\maketitle

\begin{abstract}
 Recent observations of radio relics - diffuse radio emission in galaxy clusters - have revealed that these sources are not smooth but consist of structures in the form of threads and filaments. We investigate the origin of these filamentary structures and the role of projection effects. To this end, we have developed a tool that extracts the filamentary structures from background emission. Moreover, it is capable of studying both two-dimensional and three-dimensional objects. We apply our structure extractor to, both, observations and  cosmological simulations of radio relics. Using Minkowski functionals, we determine the shape of the  identified structures. In our 2D analysis, we find that the brightest structures in the observed and simulated maps are filaments. Our analysis of the 3D simulation data shows that radio relics do not consist of sheets but only of filaments and ribbons. Furthermore, we did not find any measurable projection effects that could hide any sheet-like structures in projection. We find that, both, the magnetic field and the shock front consist of filaments and ribbons that cause filamentary radio emission.
\end{abstract}
\label{firstpage}

 
\section{Introduction}

Observations of diffuse radio emission reveal the presence of highly energetic cosmic-ray electrons, with energies from $\MeV$ to $\GeV$, as well as magnetic fields of a few $\mu \mathrm{G}$ in the intracluster medium (ICM) \citep[see][and references therein for a recent review]{vanweeren2019review}. Owing to their improving resolution and sensitivities, new radio observations reveal that the surface brightness in diffuse sources is not smooth. In most cases, the radio emission is filamentary, suggesting a much more complex picture of the sources' morphologies \citep[][]{Brienza_2021_Nest,Knowles_2022_MeerKAT,2022ApJ...934...49G,digennaro2018saus}. This, in turn, has consequences for our understanding of cosmic-ray acceleration processes and magnetic field evolution in the ICM. In this study, we focus on the filamentary structures in radio relics.
 
Radio relics are elongated objects that are normally observed at the periphery of galaxy clusters. As most relics are co-located with shock waves in the ICM, it is widely accepted that relics are produced by the shock (re-)acceleration of cosmic-ray electrons \citep[][]{ensslin1998}.  Relics are predicted to have a bright radio edge at the shock wave's location with a decreasing radio surface brightness into the downstream region \citep{2007MNRAS.375...77H}. However, recent observations have shown that the leading edges of radio relics, i.e. where particles are accelerated by a shock wave, are not uniform but consist of filaments and threads \citep[e.g][]{rajpurohit2020toothbrush,digennaro2018saus,deGasperin_2022_A3667}. Moreover, the relics' downstream regions are filled with filamentary structures - i.e. radio filaments that are brighter than their surroundings \citep[e.g][]{Rajpurohit_2022_A2256,deGasperin_2022_A3667}. The origin of these filamentary structures is still unclear. In particular, it is not understood whether they trace the complex morphology of the shock front or whether they are determined by the morphology of the underlying magnetic field \citep[e.g.][and references therein]{Wittor_2021_Spec}. Moreover, it is unknown if these filaments are also filaments in 3D or if they are actually sheets that appear as filaments in projection.

In this work, we focus on the geometrical characterization of the filament morphologies. Specifically, we have developed a tool that extracts structures from background emission. We apply this structure extractor to, both, observation and cosmological simulations. Using Minkowski functionals, we analyse the geometrical shape of the extracted structures. This analysis allows us to draw first conclusions about the origin of the filamentary structures in radio relics.

This paper is structured as follows: in Sec.~\ref{sec::filament_finder}, we describe the structure extractor that we have developed for this work. In Sec.~\ref{sec::shape_parameters}, we present the shape parameters that we use to characterize the geometrical shape of the extracted structures. In Sec.~\ref{sec::observations}, we apply the developed tools to radio data. In Sec.~\ref{sec::simulations} and in Sec.~\ref{sec::results}, we summarize the simulations that we have analysed and we present the corresponding results, respectively. We summarize and conclude our work in Sec.~\ref{sec::conclusion}.

\section{\texttt{Sub-X}: A structure extractor}\label{sec::filament_finder}
 
The detection of structures in 2D radio maps and 3D data cubes, e.g. from simulations, is a non-trivial task. Hence, the first goal of this work is the development of a robust structure extractor that can be applied to any radio observation without a detailed fine-tuning of different input parameters a priori. 

Algorithms that depend on the specific values of the radio emission appear not useful for our analysis. Such algorithms could be based on the isocontours, a minimum threshold or the gradient of the radio emission. However, they require a significant amount of fine-tuning of the input parameters. For example, if the isocontour levels are set too high, faint structures are not detected. Vice versa, if the isocontour levels are set too low, two separate bright structures might be identified as a single structure. Consequently, such structure extractor require some object-specific input parameters, making a unification of the results more difficult. One possibility to remove some biases is a machine learning approach. However, machine learning also requires some input parameters. Here, we want to develop a structure extractor that does not require any input parameters or fine-tuning a priori, but only depends on the mathematical properties of the data. 

Our structure extractor, called \texttt{Sub-X (Substructure Extractor)}, is based on two key properties: First, a structure is defined as a local maximum. Second, two structures are separated by a local minimum. By definition, a turning point lies between a local maximum and a local minimum. Here, we define \textit{the turning point as the physical boundary of the structure}. 

\begin{figure*}
 \includegraphics[width = \textwidth]{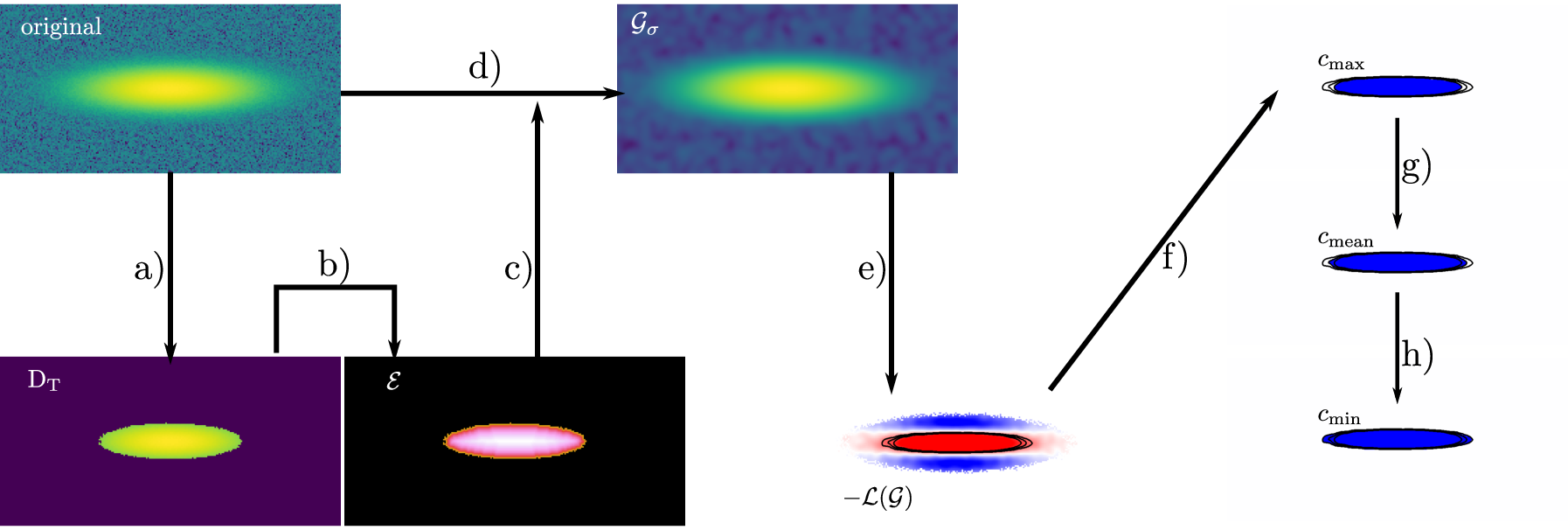}
 \caption{Sketch of \texttt{Sub-X}'s workflow. The first panel shows the original map. a) We remove emission below the threshold $S_{\mathrm{lim}}$. b) We compute the Euclidean distance transform, $\mathcal{E}$. c) and d) Using the round up of the median of the Euclidean distance transform $\tilde{\mathcal{E}}$ as filter's standard deviation, we some smooth the map using a Gaussian filter, $\mathcal{G}_{\sigma = \lceil \Tilde{\mathcal{E}} \rceil}$. e) We compute the negative of the  transform of the smoothed map, $-\mathcal{L}(\mathcal{G})$. f) We group all pixels that have $-\mathcal{L}(\mathcal{G}) > c_{\max}$. f) We group all pixels that have $-\mathcal{L}(\mathcal{G}) > c_{\mathrm{mean}}$ and $-\mathcal{L}(\mathcal{G}) < c_{\max}$, and we assign them to the $c_{\max}$ regions. h) Following the local gradient, we assign all pixels with $-\mathcal{L}(\mathcal{G}) > c_{\min}$ and $-\mathcal{L}(\mathcal{G}) < c_{\mathrm{mean}}$ to a structure.}
 \label{fig::howtotag}
\end{figure*}

In the following, we summarize the workflow of our \texttt{Sub-X}. For illustration, Fig.~\ref{fig::howtotag} displays how \texttt{Sub-X} detects a 2D Gaussian bell that has been stretched into one direction. The first panel in Fig.~\ref{fig::howtotag} shows the original image from which we want to isolate a structure. To make this example more realistic, we added some random noise to the image. \texttt{Sub-X} is split into three parts: \textit{smoothing}, \textit{Laplace transform} and \textit{grouping}, that we explain in the following.

\textit{Smoothing:} In reality, the filamentary structures in radio relics also show small fluctuations. Some of these fluctuations might be caused by physical effects in the plasma, but the majority of them could be due to low signal to noise ratios. \texttt{Sub-X} would identify the small fluctuations caused by the noise as individual structures. Hence, instead of identifying the correct structure, multiple smaller structures would be identified as individuals. Consequently, any information on the true structure would be lost. To remove these small-scale fluctuations, we smooth the map with a Gaussian filter with standard deviation $\sigma$ that is determined as follows: 

\begin{enumerate}
 \item We compute the average of the map, $S_{\mathrm{lim}}$, and we set all pixels with a value below $S_{\mathrm{lim}}$ to zero, giving the map $\mathrm{D_T}$. (Step \textit{a)} in Fig.~\ref{fig::howtotag}.)
 \item On $\mathrm{D_T}$, we perform a \textit{Euclidean distance transform}, $\mathcal{E}$ that determines the shortest distance to a cell with zero value \citep{DANIELSSON1980227,scipy}. (Step \textit{b} in Fig.~\ref{fig::howtotag}.)
 \item We compute the median of the Euclidean distance transform $\Tilde{\mathcal{E}}$, divide it by 2 to obtain the standard deviation of the Gaussian filter, $\sigma = \max( \Tilde{\mathcal{E}}/2,2)$. (Step \textit{c} in Fig.~\ref{fig::howtotag}.)
 \item After smoothing with $\sigma$, we get the smoothed mapped $\mathcal{G}_{\sigma}$. (Step \textit{d} in Fig.~\ref{fig::howtotag}.)
\end{enumerate}

\textit{Laplace transform:} Next, we compute the second derivative of the smoothed map (Step \textit{e)} in Fig.~\ref{fig::howtotag}). To this end, we use a Laplace transform $- \mathcal{L}\left( \mathcal{G}_{\sigma} \right)$. As the Laplace transform is negative for the local maxima, we multiply $\mathcal{L}\left( \mathcal{G}_{\sigma} \right)$ with $-1$. Using the Laplace transform of the smoothed map, we determine the boundaries of the structures using contours.

\textit{Grouping:} To tag the different structures in the map of the Laplace transform, we apply a method similar to the grouping algorithm described in \citet{Eisenstein_1998}. To this end, we define three different contours: 

\begin{align}
 c_{\max} &= \max [\mathrm{mean}(L^+), \mathrm{median}(L^+)] \label{eq::cmax} \\
 c_{\mathrm{mean}} &= \mathrm{mean} [\mathrm{mean}(L^+), \mathrm{median}(L^+)]  \\
 c_{\min} &= \min [\mathrm{mean}(L^+), \mathrm{median}(L^+)]. \label{eq::cmin}
\end{align}

Here,  $\mathcal{L}^+$ are all the pixels where $- \mathcal{L}\left( \mathcal{G}_{\sigma} \right) > 0$, i.e. the pixels that are positive and, hence, above the turning points of the map. Above, we defined the turning point as the physical boundary of a substructure. This definition implies that $c_{\min}$ should be equal to zero. However, only the numerical Laplace transform is available. As a consequence, the Laplace transform does not fall below zero between two close substructures, and the two would be identified as one substructure. Hence, we choose to define $c_{\min}$ as in Eq. \ref{eq::cmin}. Using, the different contours, we tag the different filamentary structures in a three-step procedure:

\begin{enumerate}
 \item First, we group all pixels with values above $c_{\max}$ into separate groups. Each group receives its specific ID. (Step \textit{f)} in Fig.~\ref{fig::howtotag})
 \item Next, we select all pixels with values above $c_{\mathrm{mean}}$ and below $c_{\max}$. If these regions touch a region that is above $c_{\max}$, they will receive the same ID. If they do not touch such a region, they will receive their own specific ID. Moreover, if two regions above $c_{\max}$ are connected by a region that is above $c_{\mathrm{mean}}$, these regions will be merged and they receive the same ID. (Step \textit{g)} in Fig.~\ref{fig::howtotag})
 \item Finally, we select all pixels with values above $c_{\min}$ and below $c_{\mathrm{mean}}$. By following the local gradients, we assign them to a region above  $c_{\mathrm{mean}}$. If this is not possible, the pixels will be discarded. (Step \textit{h)} in Fig.~\ref{fig::howtotag})
\end{enumerate}

\begin{figure}
 \centering
 \includegraphics[width = 0.5\textwidth]{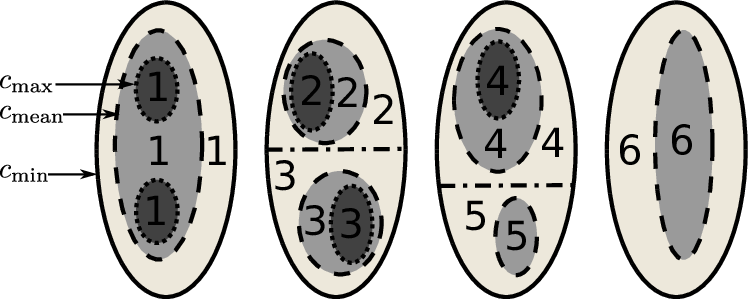}
 \caption{Sketch of how the grouping algorithm. The dotted contours give the $c_{\max}$-contours, the dashed contours give the $c_{\mathrm{mean}}$-contours and the solid contours give the $c_{\min}$-contours. Hence, everything outside the solid lines is not considered by the grouping algorithm. Structure 1 consists of two regions that are above $c_{\max}$. Yet both of them fall within the same $c_{\mathrm{mean}}$, hence, they are grouped together. Structure 2 and 3 consist of two regions that are above $c_{\max}$. Yet both of they fall within separate $c_{\mathrm{mean}}$ and are thus grouped separately. Structures 4 and 5 fall within the same $c_{\min}$. However, they fall within separate $c_{\mathrm{mean}}$. Structure 6 has a region that falls above $c_{\mathrm{mean}}$ and, hence, the surrounding $c_{\min}$ are grouped to structure 6.}
 \label{fig::how_to_contour}
\end{figure}

\begin{figure}\centering
 \includegraphics[width = 0.49\textwidth]{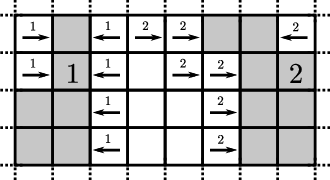}
 \caption{Sketch of how tagged cells are assigned to a structure. Gray shaded cells already belong to a structure. The shaded cells on the left are structure 1 and the shaded cells on the right are structure 2. All cells containing an arrow must still be assigned to a structure using a gradient technique. To this end, we follow the direction of the local gradients, presented by the arrows' directions. Once a shaded cell is hit, the tagged cell will be assigned to the same structure. Here all cells that contain an arrow and a small 1 will be assigned to structure 1. Similarly, all cells that contain an arrow and a number 2 will be assigned to structure 2.}
 \label{fig::howtotag4}
\end{figure}

In Fig.~\ref{fig::how_to_contour} and Fig.~\ref{fig::howtotag4}, we sketch how the grouping following the contours is done. The final product of this algorithm is a map that displays the differently tagged structures based on their ID. 

We find that \texttt{Sub-X} is very robust and only depends on the derivatives in the data. This has a number of advantages. First, \texttt{Sub-X} works on data of arbitrary dimensions. Second, \texttt{Sub-X} is not just limited to study diffuse radio emission, but it can be applied to any type of dataset. In addition, \texttt{Sub-X} only depends on the choices of $S_{\mathrm{lim}}$ and of the contours, i.e. $c_{\min}$, $c_{\mathrm{mean}}$ and $c_{\max}$. These four are inherent to each data set and, hence, they do not have to be specified in advance. In this work, we computed the four parameters as described above. Depending on the science case, one could also choose to use different values. Before continuing, we briefly describe how the four parameters determine the extraction of substructures.

The limit $S_{\mathrm{lim}}$ determines the pixels that are included in the computation of the Euclidean distance transform. Hence, $S_{\mathrm{lim}}$ determines the standard deviation of the Gaussian filter $\sigma$, that is used for the smoothing. Larger values of $S_{\mathrm{lim}}$ preserve small-scale fluctuations due to lower values of $\sigma$. Vice versa, lower values of $S_{\mathrm{lim}}$ increase $\sigma$. Consequently, small-scale substructures are smoothed out and the recovered structures become larger. However, this comes with a drawback. If the separation of two distinct substructures is below the smoothing scale, the gap in between the two is smeared out and the two substructures are identified as a single substructure.

The contour values, i.e. $c_{\min}$, $c_{\mathrm{mean}}$ and $c_{\max}$, determine how individual substructures are extracted. Specifically, $c_{\min}$ is the boundary of a substructure. Hence, lowering $c_{\min}$ results in larger substructures. $c_{\mathrm{mean}}$ is the boundary between two substructures that are not separated by a region with values below $c_{\min}$. Hence, $c_{\mathrm{mean}}$ indicates the allowed fluctuations within a substructure. Large values of $c_{\mathrm{mean}}$ do not allow a lot of fluctuations within a substructure. On the other hand, lower values of $c_{\mathrm{mean}}$ allow more fluctuations within a substructure. The latter can become problematic for substructures that are projected on top of each other.

In this work, we are interested in the small-scale filamentary structures observed in radio relics. Therefore, we have two expectations for the extracted substructures. First, two substructures that are close but distinct are still identified as two separate substructures. Second, distinct substructures that are projected on top of each other are also extracted as two substructures. Through elaborative testing of different values for $S_{\mathrm{lim}}$, $c_{\min}$, $c_{\mathrm{mean}}$ and $c_{\max}$, we have concluded that the described choices are the most optimal for our science case.

In general, maps of radio observations can contain more than one object, or the targeted object only occupies a small fraction of the entire map. In such cases, it only makes sense to reduce the area in which \texttt{Sub-X} searches for structures. Therefore, one can tell \texttt{Sub-X} to search for structures only within a certain region.

\section{Shape parameters}\label{sec::shape_parameters}
 
In this section, we present the shape parameters that we use to characterise the morphology of extracted structures. Here, we distinguish between 2D and 3D shape parameters. Independent of the dimensionality, we use Minkowski functionals to determine the geometrical shape \citep{Minkowski_1903,QuantImPy}. This approach has been successfully applied by several authors \citep[e.g][]{Sahni_1998,Schmalzing_1998,Schmalzing_1999,Sheth_2003,Bag_2019,Seta_2020}.

As we will discuss in the following, the interpretation of the different shape parameters is non-trivial. Mainly because the personal perception of various geometrical shapes differs from the mathematical definition. Therefore, we provide some examples in the appendix, App. \ref{app::fp_shapes}.
 
\subsection{2D shape parameters}
\label{ssec::2D_shape_parameters}
 
The shape of a 2D object can be characterised by its filamentarity $\fzd$ \citep[e.g.][and references therein]{Bharadwaj_2000_2D_fila}:
 
\begin{align}
 \fzd = \frac{C^2 - 4 \pi S}{C^2 + 4 \pi S}. \label{eq::filamentarity_2D}
\end{align}

Here, $C$ is the object's circumference and $S$ is its surface area. Hence, an object with $\fzd = 0$ is a perfectly filled circle, and a perfect filament has $\fzd = 1$. 

We compute, both, the circumference and the surface area, using the 2D Minkowski functionals \citep{QuantImPy}. For a 2D body $X$, with a smooth boundary $\delta X$, we use the Minkowski functionals in the form \citep{Mecke_2000_Mink_Book}
 
\begin{align}
 M_0(X) &= \int_X \dd s \\
 M_1(X) &= \frac{1}{2 \pi} \int_{\delta X} \dd c \\
 M_2(X) &= \frac{1}{2 \pi^2} \int_{\delta X} \frac{1}{R} \dd c .
\end{align}

Here, $R$ is the radius of the local curvature. The surface area $S$ and the circumference $C$ are computed from the $0^{\mathrm{th}}$ and $1^{\mathrm{st}}$ Minkowski functionals, respectively, as: 

\begin{align}
 S &= M_0(X) \\
 C &= 2 \pi M_1(X).
\end{align}

\subsection{3D shape parameters}
 
The shape of a 3D object is characterised by its  filamentarity $\fdd$ and planarity $\pdd$ \citep[e.g.][]{Sahni_1998}

\begin{align}
 \fdd &= \frac{l-w}{l+w} \label{eq::filamentarity_3D} \\
 \pdd &= \frac{w-t}{w+t} . \label{eq::planarity_3D} 
\end{align}
Here, $t$, $w$ and $l$ are the object's characteristic thickness, width and length, respectively. For these scales, following inequality must hold:

\begin{align}
 t \leq w \leq l .
\end{align}
As in the 2D case, we use Minkowski functionals to calculate $t$, $w$ and $l$ \citep{QuantImPy}. For a 3D body $X$, with smooth surface $\delta X$, the four Minkowski functionals are \citep{Mecke_2000_Mink_Book}

\begin{align}
 M_0(X) &= \int_X \dd v \\
 M_1(X) &= \frac{1}{8} \int_{\delta X} \dd s \\
 M_2(X) &= \frac{1}{2 \pi^2} \int_{\delta X} \frac{1}{2} \left( \frac{1}{R_1} + \frac{1}{R_2} \right) \dd s \\
 M_3(X) &= \frac{3}{(4 \pi)^2} \int_{\delta X} \frac{1}{R_1 R_2} \dd s  \label{eq::M3_3D} .
\end{align}

The Minkowski functionals are directly related to the object's volume $V$, its surface area $S$, its mean curvature $H$ and its Euler characteristic $\chi$ as:

\begin{align}
 V &= M_0(X) \\ 
 S &= 8 M_1(X) \\
 H &= 2 \pi^2 M_2(X) \\
 \chi &= 4 \pi / 3 M_3(X). \label{eq::chi}
\end{align}

These quantities determine an object's characteristic length scales:

\begin{align}
 l_1 &= \frac{3 V}{S} \\
 l_2 &= \frac{S}{C} \\
 l_3 = \frac{C}{4 \pi \chi} \ &\mathrm{or} \ l_3 = \frac{C}{4 \pi}.\label{eq::l3}
\end{align}
Here, we note the two different choices in the computation of $l_3$\footnote{We are also aware of the definition of $l_3$ given by \citet{Sheth_2003}, that uses the genus of an object to define $l_3$. However, this definition makes an assumption on the orientability of the object, a priori. As, we want to use the least amount of assumptions possible, we will not use this definition.}. The first choice, containing $\chi$, is taken from \citet{Schmalzing_1999} and the second choice, without $\chi$, is taken from \citet{Bag_2019}. The two expressions are equal for $\chi = 1$, i.e. a solid object that is neither hollow nor contains any holes. However, for $\chi \ne 1$, the two are different, which has implications for the object's shape. This imposes some differences, that we discuss further at the end of this section. 

The object's thickness $t$, width $w$ and length $l$ are given from the characteristic length scales as:

\begin{align}
 t &= \min([l_1,l_2,l_3]) \\ 
 w &= \mathrm{middle}([l_1,l_2,l_3]) \\
 l &= \max([l_1,l_2,l_3]). \label{eq::r3}
\end{align}

The filamentarity and planarity are computed using Eq.~\ref{eq::filamentarity_3D} and \ref{eq::planarity_3D}, respectively. From the definitions above, it directly follows that a 3D object can be classified as: 1) a filament, i.e. $ t \approx w \ll l  $, 2) a sheet, i.e. $ t \ll w \approx l  $, 3) a sphere, i.e. $ t \approx w \approx l$, or 4) a ribbon, i.e. non of the inequalities hold. These relations between $t$, $w$ and $l$ have direct implications on the planarity and the filamentarity, and an object's shape is defined as \citep[e.g.][and references therein]{Seta_2020}:

\begin{align*}
 \fdd \rightarrow 0, \ \pdd \rightarrow \infty:& \ \mathrm{sheet-like \ structure} \\
 \fdd \rightarrow \infty, \ \pdd \rightarrow 0:& \ \mathrm{filament-like \ structure} \\
 \fdd \approx 0, \ \pdd \approx 0:& \ \mathrm{sphere-like \ structure} \\
 \fdd \approx \pdd > 0:& \ \mathrm{ribbon-like \ structure}.
\end{align*}

We call the different shapes \textit{X}-like because the boundaries between the different shapes are not clear. To illustrate this classification scheme, we plot some examples in Fig.~\ref{fig::ell_shapespdf3} and we show more examples in the appendix App. \ref{app::fp_shapes}. 

In Fig.~\ref{fig::ell_shapespdf3}, we plot the contours of five different ellipsoids. Moreover in Tab. \ref{tab::shapes1}, we provide the properties of the different ellipsoid. Ellipsoid a) is a sphere, that has $(\fdd,\pdd) = (0,0)$. Ellipsoid b) is a ribbon that has $(\fdd,\pdd) = (0.11,0.12)$. Ellipsoids c) and d) are a filament with  $(\fdd,\pdd) = (0.80,0.13)$ and a sheet with $(\fdd,\pdd) = (0.10,0.82)$, respectively. While both the visual classification and the classification using $(\fdd,\pdd)$ are clear for the first four ellipsoids, they are not for ellipsoid e). Visually, one might classify object e) as a filament because it is elongated. However, it appears also compressed into one direction, which rather points to a sheet. Moreover, the filamentarity and planarity are $(\fdd,\pdd) = (0.31,0.14)$. The classification scheme based on $\fdd$ and $\pdd$ does not provide a clear classification for object e). 
 
\begin{figure}
    \centering
    \includegraphics[width = 0.5\textwidth]{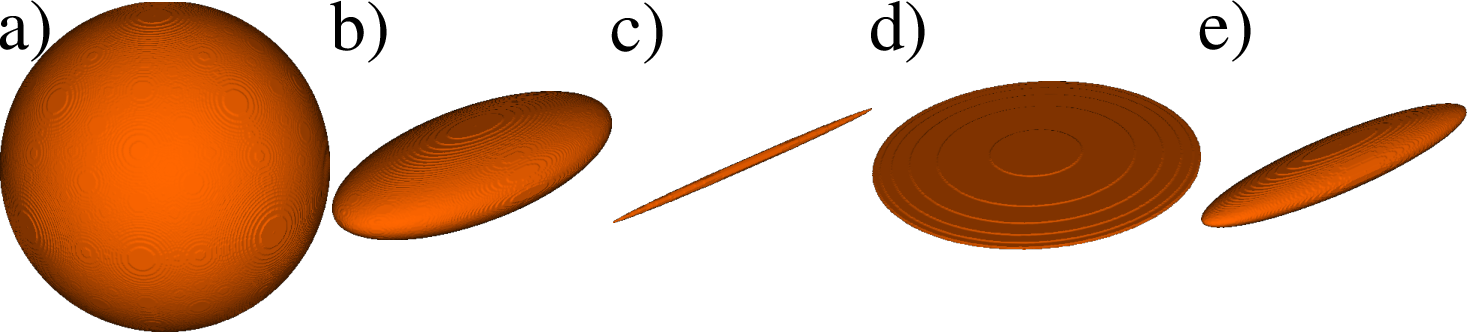}
    \caption{Examples of five different ellipsoids. The contours show the surfaces of a a) sphere, b) ribbon, c) filament and d) sheet. Object e) remains unclassified, when only using the 3D filamentarity and 3D planarity. Using the ratio of aspect ratios as well, object e) is classified as a ribbon. The properties of the different objects are summarized in Tab. \ref{tab::shapes1}}
    \label{fig::ell_shapespdf3}
\end{figure}  

\begin{table*}
 \centering
 \begin{tabular}{c|c|c|c|c|c|c|c|c|c|c}
 ID & $r_1$ & $r_2$ & $r_3$ & $\fdd$ &  $\pdd$ &  shape old & $\xi_f$ & $\xi_p$ & $a$  &  shape new \\ \hline \hline
 a  & 1.0   &  1.0       &  1.0      &  0.00   &  0.00   &  sphere    & 1.00    & 1.00    & 1.00 &  sphere  \\  
 b  & 1.0   &  0.5       &  0.25     &  0.11   &  0.12   &  ribbon    & 1.25    & 1.27    & 0.98 &  ribbon    \\ 
 c  & 1.0   &  0.03125   &  0.03125  &  0.80   &  0.13   &  filament  & 9.00    & 1.30    & 6.93 &  filament \\ 
 d  & 1.0   &  1.0       &  0.03125  &  0.10   &  0.82   &  sheet     & 1.22    & 10.11   & 0.12 &  sheet \\ 
 e  & 1.0   &  0.25      &  0.125    &  0.31   &  0.14   &  unclear   & 1.90    & 1.33    & 1.43 &  ribbon   \\ \hline
 \end{tabular}
 \caption{Properties of the five ellipsoids shown in Fig.~\ref{fig::ell_shapespdf3}. The first column gives the objects ID. The second, third and fourth column provide the length of the three principle axes of the ellipsoid. The fifth and sixth column give the 3D filamentarity, Eq.~\ref{eq::filamentarity_3D}, and planarity, Eq.~\ref{eq::planarity_3D}, respectively. The seventh column gives the classified shapes based on the "old" classification scheme. The eighth and ninth column give the two aspect ratios of the objects, Eq.~\ref{eq::xi_f} and \ref{eq::xi_p}. The tenth column gives the resulting ratio of aspect ratios, Eq.~\ref{eq::a_factor}. The last column, gives the classified shapes based on the "new" classification scheme, Tab. \ref{tab::Classification_New}. }
 \label{tab::shapes1}
\end{table*}
 
Throughout our analysis, we found that this unclear classification scheme makes a geometrical analysis based on solely the filamentarity and planarity difficult. Moreover, as shown by object e) in Fig.~\ref{fig::ell_shapespdf3} , an additional visual inspection might not always be helpful, either. The difficulty is that the two values of $\fdd$ and $\pdd$ have to be set into context to characterise a single object. Hence, we devised a new classification scheme to determine the shape of a 3D object.
 
To understand this new classification, it is useful to recall the meaning of both the planarity and the filamentarity. The filamentarity measures the aspect ratio of an objects length and width, $\xi_f$. The planarity measures the aspect ratio of an objects width and thickness, $\xi_p$:
 
\begin{align}
 \xi_f &= \frac{l}{w} = \frac{1+\fdd}{1-\fdd} \label{eq::xi_f}  \\
 \xi_p &= \frac{w}{t} = \frac{1+\pdd}{1-\pdd} . \label{eq::xi_p} 
\end{align}

Consequently, a filament has equal width and thickness, i.e. $\pdd = 0$ and $\xi_p = 1$ and $\xi_f > 1$. On the other hand, a sheet has equal width and length, i.e. $\fdd = 0$ and $\xi_f = 1$ and $\xi_p > 1$.

In order to classify an object with non-zero filamentarity and non-zero planarity, we introduce a new variable $a$, which is \textit{the ratio of the aspect ratios}:
 
\begin{align}
 a = \frac{\xi_f}{\xi_p} = \frac{1+\fdd}{1-\fdd} \frac{1-\pdd}{1+\pdd}. \label{eq::a_factor}
\end{align}

Using this ratio of the aspect ratios, we define the different objects as given in Tab.~\ref{tab::Classification_New}. This classification scheme is based on the following simple, geometrical arguments: 

\begin{itemize}
 \item A sphere has equal thickness, length and width. However, in reality equality will mostly never occur and most objects are rather sphere-like. Hence, we define an object as sphere-like, if its aspect ratios, Eq.~\ref{eq::xi_f} and \ref{eq::xi_p}, are both below $1.1$. Consequently, a sphere-like object has a filamentarity and planarity of $\fdd < 0.05$ and $\pdd < 0.05$, respectively. If one of these inequalities fails, the object is not sphere-like and one must consider the following cases.
 \item A filament-like object's length is significantly larger than, both, its width and thickness. Moreover, its width may still be larger than its thickness. However, it is required that the aspect ratio between the width and thickness is significantly smaller than the aspect ratio between the length and width. We define that the latter must be at least 1.5 times larger than the former: $\xi_f \ge 1.5 \xi_p$, yielding a ratio of the aspect ratios of $a \ge 1.5$.
 \item A sheet-like object's thickness is significantly smaller than both its length and width. Moreover, its length might still be larger than its width. Using a similar reasoning as for the filaments, we define that the aspect ratio between the length and width may be at most $2/3$ of the aspect ratio between width and thickness: $\xi_f \le 2/3 \xi_p$, yielding a ratio of the aspect ratios of $a \le 2/3$.
 \item A ribbon-like object's aspect ratio between the length and width is similar to its aspect ratio between the width and the thickness. However, this configuration will be rare. We define a ribbon as an object, where the two aspect ratios do not differ by more than $33 \ \%$, i.e. $\xi_f \ge 2/3 \xi_p$ and $\xi_f \le 1.5 \xi_p$. This yields $2/3 \le a \le 1.5$
\end{itemize}

Applying this new classification scheme to the ellipsoids plotted in Fig.~\ref{fig::ell_shapespdf3} yields the following results: The sphere is still classified as a sphere, as its planarity and filamentarity are both below 0.05. Ellipsoid b), that was previously classified as a ribbon, is still classified as a ribbon. Its two aspect ratios are rather similar and, hence, it has $a \approx 0.98$. The filament c) has a large value of $a \approx 6.93$. This is because its length is significantly larger than its width, given by its large value of $\xi_f \approx 9$. On the other hand, its width and thickness are of the same length, as shown by the small value of $\xi_p \approx 1.3$ The opposite is true for the sheet d). The sheet has the same length and width, while its thickness is significantly smaller, i.e. $\xi_f \approx 1.22$ and $\xi_p \approx 10.11$. These aspect ratios yield $a \approx 0.12$.

Using this new classification, we are finally able to classify ellipsoid e) as well. Its length is larger than its width, giving $\xi_f \approx 1.9$. Furthermore, its width is larger than its thickness $\xi_p \approx 1.33$. Moreover, the two aspect ratios are rather similar, i.e. $a \approx 1.43$. Therefore, none of the three characteristic lengths is significantly smaller or larger than the other two, and ellipsoid e) is classified as a ribbon.

We found this classification scheme to be intuitive. A structure's shape is described by a single number that already combines the 3D filamentarity and 3D planarity. Moreover, this scheme is capable of classifying objects that are difficult to be classified with solely the 3D filamentarity and 3D planarity. We have applied this classification scheme to various geometrical objects, and we found that it is very precise and robust. In App.~\ref{app::afactor}, we apply it to a larger number of ellipsoids with varying major axes.

\begin{table}
 \begin{tabular}{ccccc}
  $\fdd$ & & $\pdd$ & a & type \\ \hline \hline
  $\fdd < 0.05$ & and & $\pdd < 0.05$ & any $a$         & sphere-like \\
  $\fdd > 0.05$ & or  & $\pdd > 0.05$ & $1.5 < a$         & filament-like \\
  $\fdd > 0.05$ & or  & $\pdd > 0.05$ & $2/3 < a < 1.5$ & ribbon-like \\
  $\fdd > 0.05$ & or  & $\pdd > 0.05$ & $a < 2/3$       & sheet-like \\ \hline
 \end{tabular}
 \caption{Summary of our scheme to classify the shape of 3D objects. The first two columns give the requirements for the 3D filamentarity and 3D planarity. The third column gives the requirement for the ratio of aspect ratios. The last column give the type of morphology.}
 \label{tab::Classification_New}
\end{table}
  
Before applying our method to observed and simulated data, we want to briefly discuss the problems that occur when analysing objects with a Euler characteristic, Eq.~\ref{eq::chi}, that is different from one, i.e. $\chi \ne 1$. Characterizing the shape of such objects is difficult for several reasons.

 Essentially, the Euler characteristic is a measurement for the number of holes inside an object. For example, a solid object has $\chi = 1$, a hollow\footnote{Here, hollows means that the object as a fully connected surface, while its interior is empty. This is similar to a ping pong ball that is hollow.} object has $\chi = 2$, and a torus has $\chi = 0$. Consequently, this behaviour creates two problems for our analysis. First, $l_3$ becomes infinity for a torus-like object, and the filamentarity is not defined. Second, $l_3$ of a hollow object is a factor of 2 smaller than of the corresponding solid object\footnote{Here, corresponding solid object means that it is the same object with a filled interior.}, making the interpretation of the object's macroscopic shape difficult. Moreover, a hollow object has a larger surface area and a smaller volume than for the corresponding solid object. Also, $l_1$ and $l_2$ are different. Consequently, the filamentarity, planarity and ratio of aspect ratios of a hollow object are different than for the corresponding solid object, again making the interpretation of the object's macroscopic shape difficult.

For a better understanding, we provide a small example: A solid sphere has $(f,p) = (0,0)$, while a hollow sphere, that is empty inside half its radius, has $(f,p)_{\chi = 1} \approx (0.162, 0.557)$ or $(f,p)_{\chi \ne 1} \approx (0.470, 0.557)$, depending on the definition of $l_3$. Consequently, a hollow sphere is either classified as a sheet-like structure or a ribbon-like structure. However, macroscopically it would be still classified as a sphere.
 
As shown below, most objects that we characterise have $\chi = 1$, and they are not affected by any of this. Yet, there are some cases of $\chi \ne 1$. However, in nature, isolated radio structures should not contain holes as they are expected to be solid objects. Indeed, a visual inspection revealed that the holes or empty regions within an identified structures are normally single cells or similar. These artefacts are attributed to numerical effects. 

Such effects can occur in the last step of \texttt{Sub-X}, when grouping all pixels above $c_{\min}$. Within an identified structure the local value of the negative of the transform can fall below $c_{\min}$ in two cases. First, the smoothing step did not filter all small-scale fluctuations. Second, if the pixels/cells that contain the local maximum are surrounded by cells that contain values that are just a little bit smaller than the maximum. 

As the holes are of non-physical nature, we fill them for structures with $\chi \ne 1$. To this end, we use a combination of \textit{binary\_holes\_fill}, \textit{binary erosions} and \textit{binary dilations} \citep{scipy}. As a result, we preserve the macroscopic shape of the extracted object, while removing any microscopic irregularties. This is justified by our interest in the macroscopic shapes of the analysed objects. In App.~\ref{app::chi}, we present the filling algorithm, and verify that the filling does not affect the overall conclusions of our work.

\section{Morphology of observed relics}\label{sec::observations}

At first, we apply \texttt{Sub-X} and the morphological analysis to observed relics. To this end, we selected six widely studied filamentary relics: the relic in Abell 2256 \citep[$z \approx 0.058$,][]{2014ApJ...794...24O,Rajpurohit_2022_A2256}, the north-western relic in Abell 2744 \citep[$z \approx 0.308$,][]{2017ApJ...845...81P,rajpurohit2021A2744}, the northern and southern relics in Abell 3376 \citep[$z \approx 0.046$,][]{Knowles_2022_MeerKAT}, as well as the northern and southern relics in Abell 3667 \citep[$z \approx 0.056$,][]{Knowles_2022_MeerKAT}. We point to the given references for more details on the observations. In Tab.~\ref{tab::observations}, we summarize the key properties of the different relics.

\begin{figure*}
 \centering
 \includegraphics[width = \textwidth]{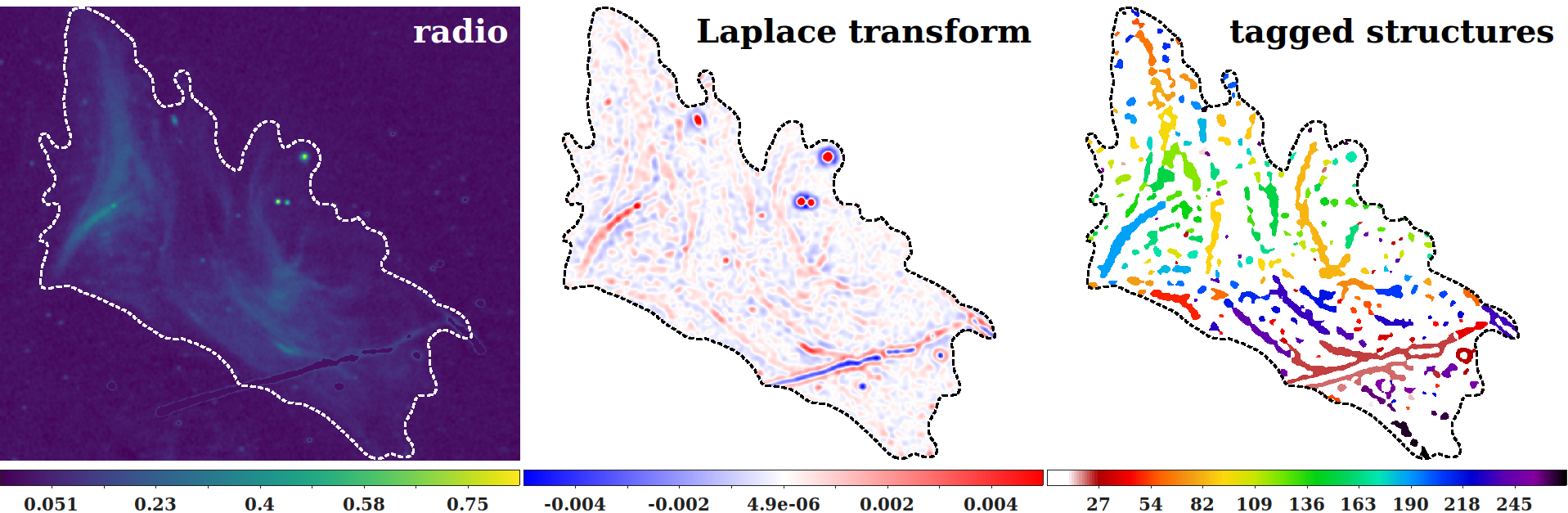}
 \caption{Observational analysis: Example of how \texttt{Sub-X} detects the filamentary structures of the radio relic in Abell 2256. The left panel shows the radio map at 1.4 GHz. The middle panel shows the negative of the  transform of the radio map. Here, the red regions have values above zero and, hence, are tagged as possible filaments. The right panel, shows the tagged structures. Here, every color is a separate filament. The dashed lines mark the region, where \texttt{Sub-X} searched for substructures. We decided these regions by hand.}
 \label{fig::A2256_paper}
\end{figure*}

We have applied \texttt{Sub-X} to the six different relics. As an example, we plot the radio map, the negative of the  transform and the corresponding tagged structures of Abell 2256 in Fig.~\ref{fig::A2256_paper}. In App. \ref{app::obs}, we plot maps of the radio emission and identified structures of the other relics. Abell 2256 hosts a radio galaxy with a tail that extends across the relic. We have masked this radio galaxy by hand from the map to emphasize the structures of the relic. A visual comparison of the three maps yields that \texttt{Sub-X} performs well to detect various structures. As expected from the mathematical definition, the negative of the  transform is positive at the location of the structures and they are encircled by zero values, i.e. the turning points. The corresponding tagged structures match the ones that one would select by eye. Moreover, \texttt{Sub-X} detects structures that are not very prominent in the radio map, due to the gentle gradients in the radio emission. 
  
\begin{figure*}
 \centering
 \includegraphics[width = 0.33\textwidth]{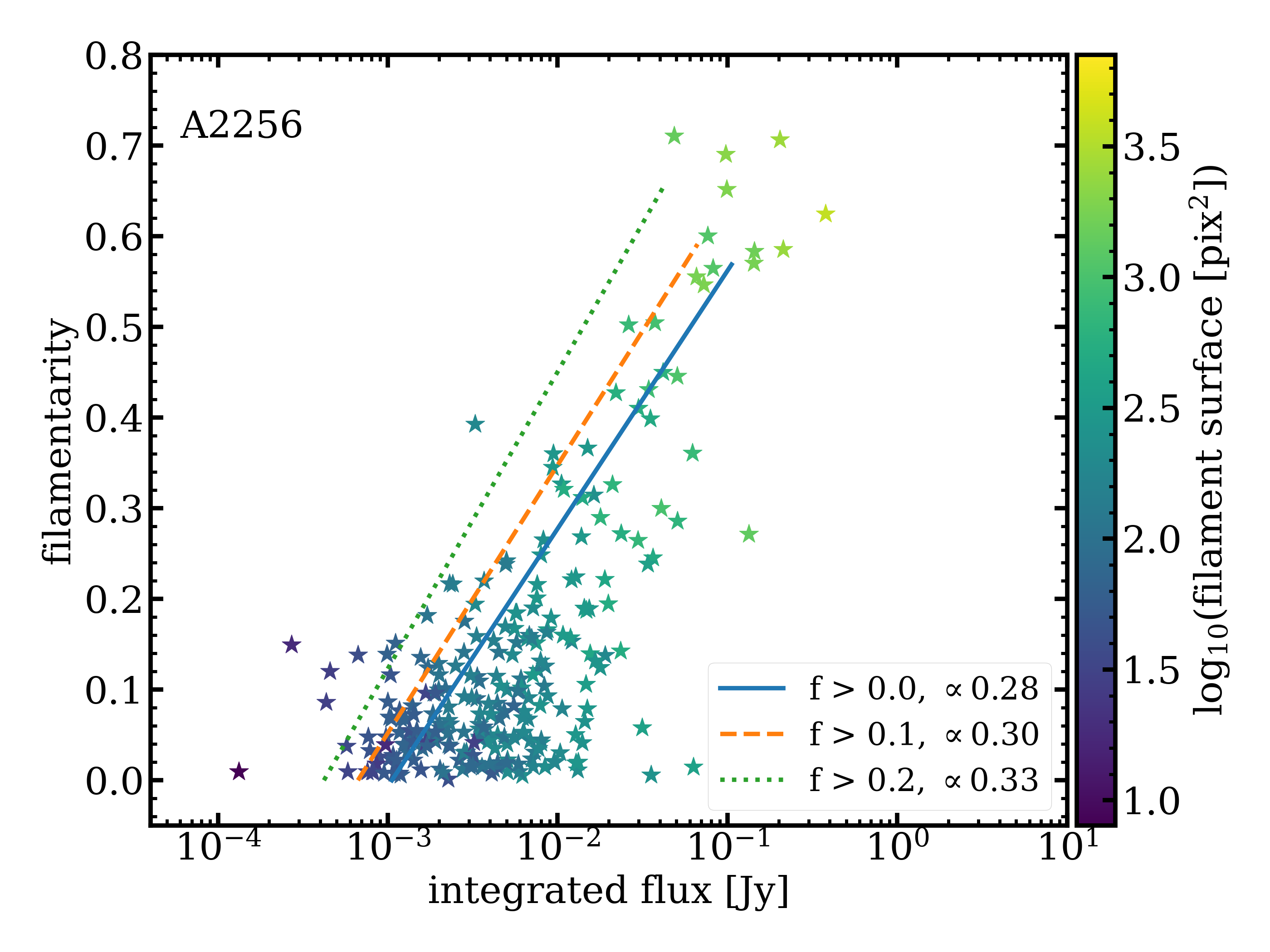} 
 \includegraphics[width = 0.33\textwidth]{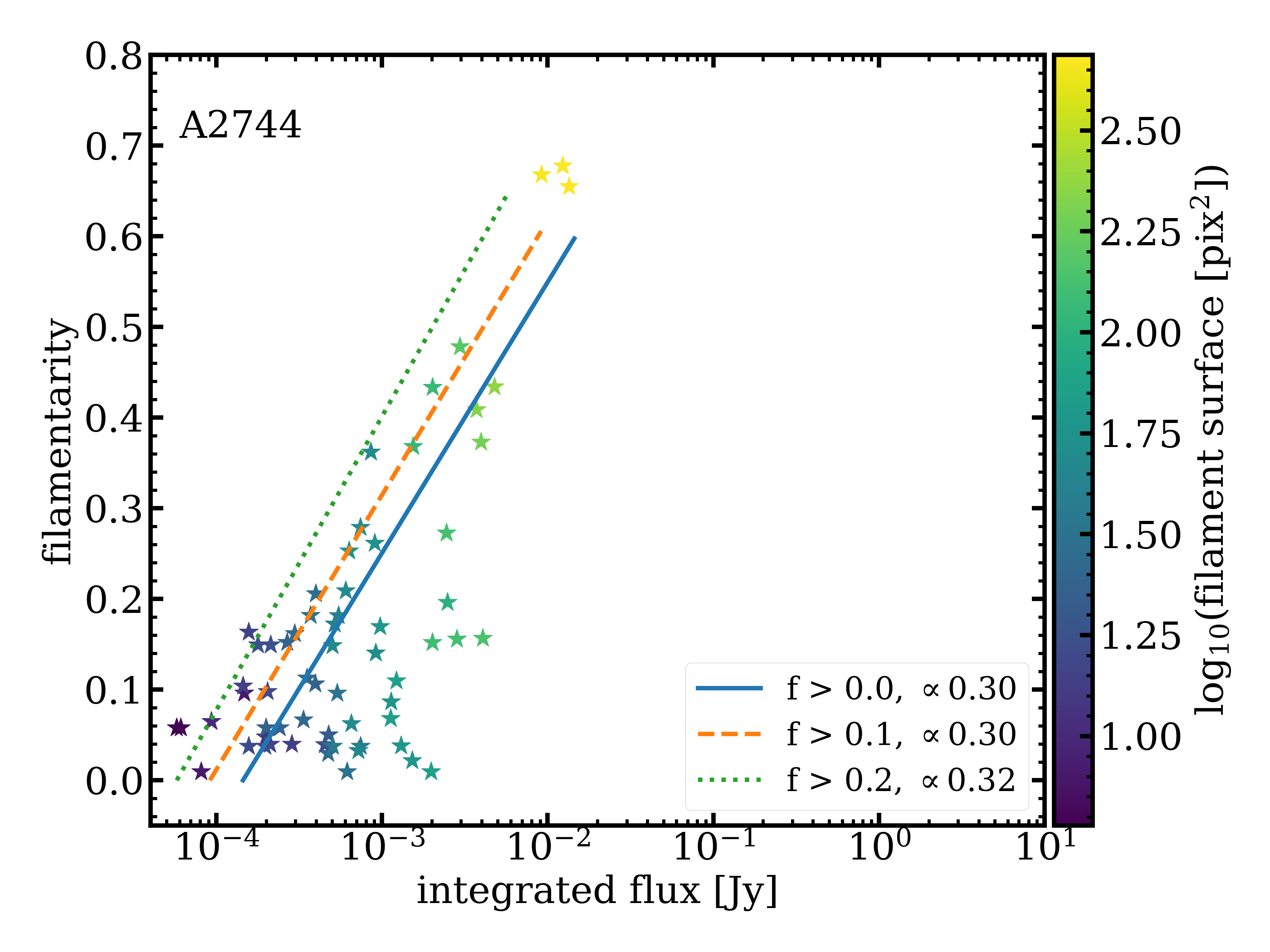}
 \includegraphics[width = 0.33\textwidth]{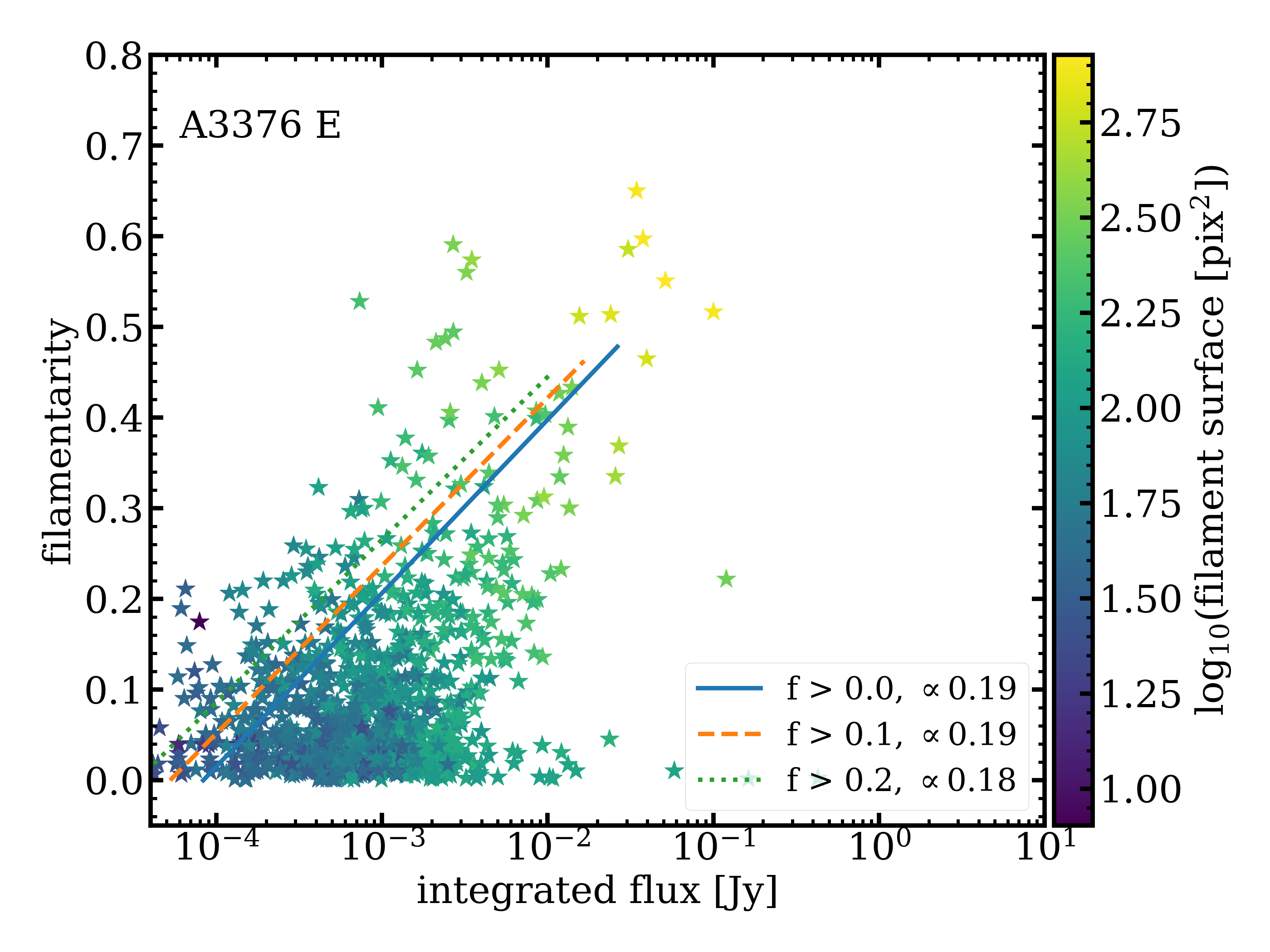} \\
 \includegraphics[width = 0.33\textwidth]{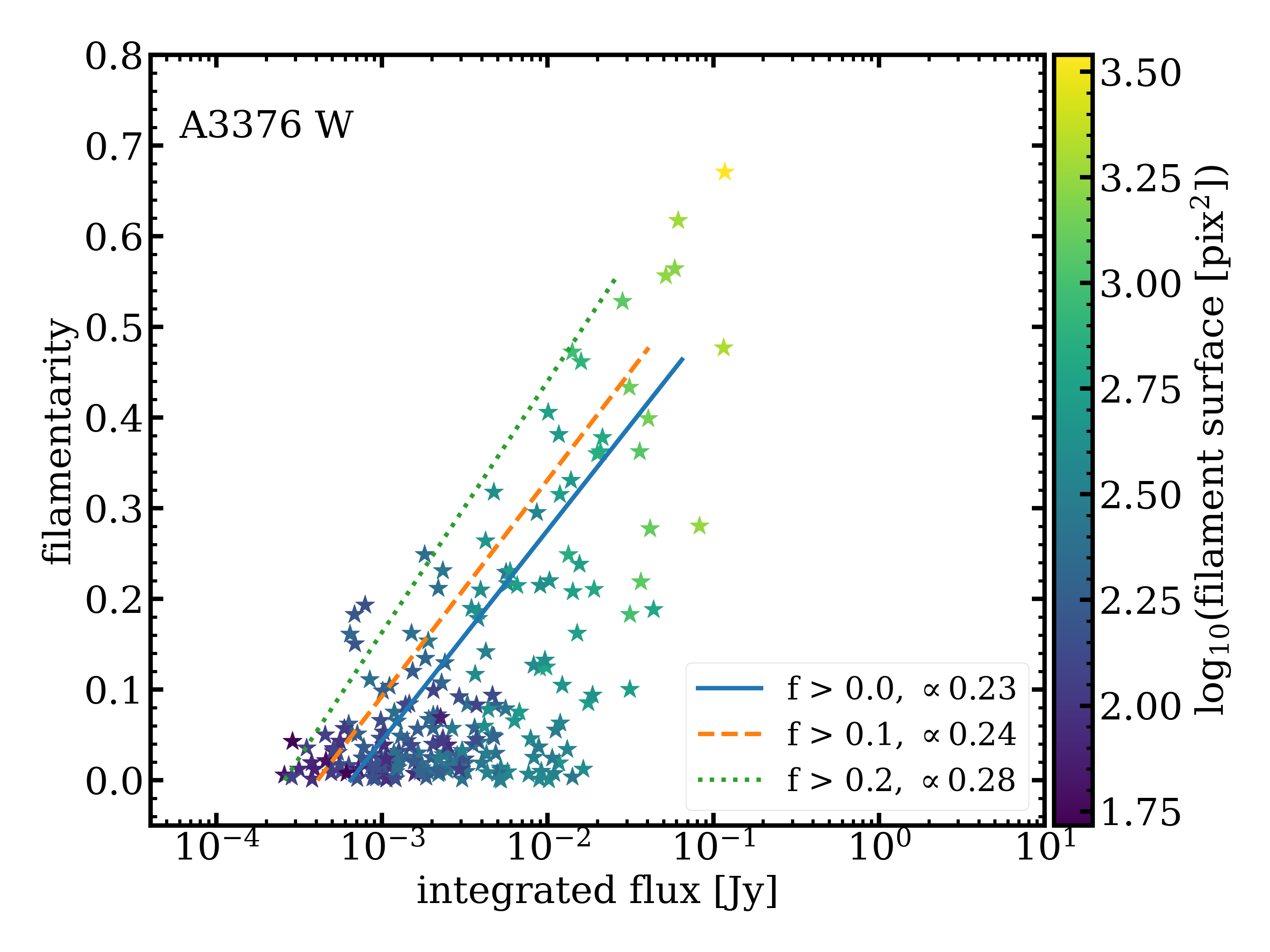}
 \includegraphics[width = 0.33\textwidth]{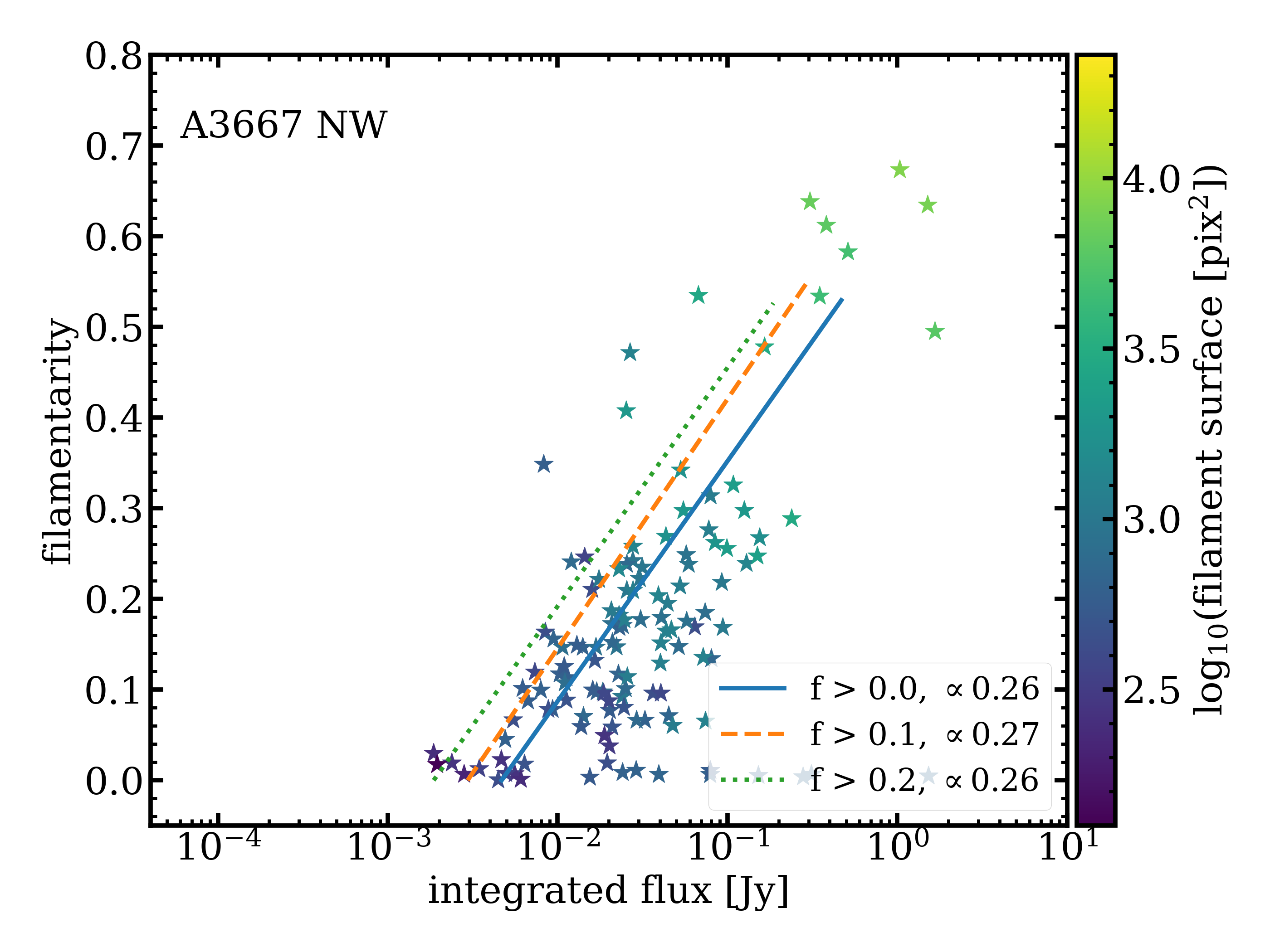}
 \includegraphics[width = 0.33\textwidth]{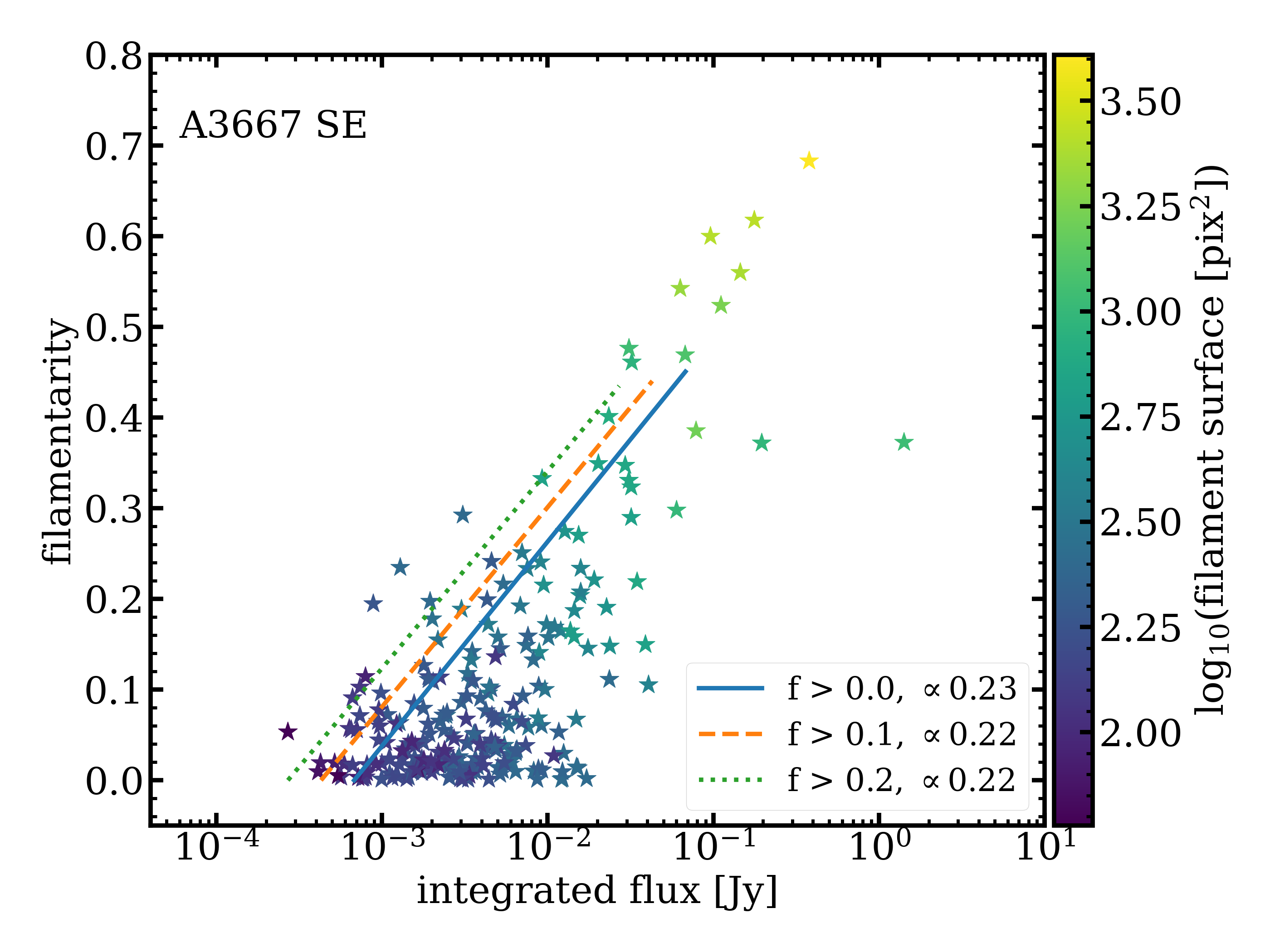}
 \caption{Observational analysis: For each observed relic, the plots show the filamentarity of each tagged structure plotted against its integrated radio flux. Furthermore, the colorcoding provides the surface area of each filament. The different lines give the slopes of each of the fits, i.e. $f_{\mathrm{2D}} \propto \kappa \log_{10}(S_{\mathrm{int}})$. }
 \label{fig::fila_vs_p14_2d_obs}
\end{figure*}

To characterise the shape of the detected structures, we computed their filamentarity using Eq.~\ref{eq::filamentarity_2D}. For each relic, we find that the detected structures have filamentarities in the range of $\sim 0.0$ to $\sim 0.8$. In Fig.~\ref{fig::fila_vs_p14_2d_obs}, we plot the filamentarity against the integrated radio flux of each structure, and colour-coded by their surface area. For the six different relics in our sample, we find similar trends. The filamentarity increases with the integrated radio flux. Moreover, the larger structures tend to have the higher filamentarity.
  
For each relic, we fit the filamentarity against the logarithm of the integrated radio flux of each structure, $\fzd \propto {\kappa} \log_{10}(S_{\mathrm{int}})$. To this end, we perform an \textit{orthogonal distance regression (ODR)}  \citep{boggs1990orthogonal}. The ODR requires an input error for the two variables. However, in our case, this error is unknown, and, hence, we take the standard deviation of each data set as the input error.

We find that the slope $\kappa$ varies between $0.19$, in case of Abell 3376 E, and $0.30$, in the case of Abell 2744. All other values are given in Tab. \ref{tab::observations} and in Fig.~\ref{fig::fila_vs_p14_2d_obs}. The average slope is $\langle \kappa \rangle \approx 0.25$. This can be interpreted as: on average, the filamentarity increases by a value of $\sim 0.25$, if the integrated radio flux increases by one order of magnitude\footnote{We note that there is a direct correlation between the surface area of a substructure and its integrated radio flux, i.e. larger substructures should have a larger integrated radio flux. According to Eq. \ref{eq::filamentarity_2D}, the filamentarity of a substructure depends not only on its surface area but also on its circumference. Hence, it is not obvious that a larger and, hence, more luminous substructure should have a higher filamentarity.}.

The scatter in the slopes might be due to the contamination by point sources. Relics such as Abell 3376 E and Abell 3376 W are significantly more contaminated by point sources than the other relics, yielding two biases. First, several of the sphere like structures might actually be point sources that reside in the background/foreground of the relics. Second, some of the relics emission is attributed to the point sources and not to the relic. Hence, the integrated flux of some detected filaments might actually be larger.

To remove any biases from small and spherical structures in the fit, we repeated the ODR only including structures that have $\fzd > 0.1$ and $\fzd > 0.2$. We find that these additional cuts only affect the slopes slightly. The average slopes are $\langle \kappa (\fzd > 0.1) \rangle \approx 0.25$ and $\langle \kappa (\fzd > 0.2) \rangle \approx 0.26$. 

In summary, we find that the substructures of the observed relics are classified as filaments. Specifically, the larger and radio brighter an identified structure is, the more likely it is to be a filament. In App. \ref{app::robustness}, we test the robustness of our results. To this end, we perform the same analysis for different choices for $c_{\max}$, $c_{\mathrm{mean}}$ and $c_{\min}$, as well as $S_{\mathrm{lim}}$ (see discussion in Sec. \ref{sec::filament_finder}). We find that our results are not affected by different choices for the parameters.

\begin{table}
 \centering
 \begin{tabular}{lccc}
  relic    & $\kappa \pm \sigma_{\kappa}$ & $\kappa \pm \sigma_{\kappa}$ & $\kappa \pm \sigma_{\kappa}$ \\
           & $\fzd > 0.0$                 & $\fzd > 0.1$                 &  $\fzd > 0.2$                \\  \hline 
  A2256    & $ 0.28 \pm 0.01 $ & $ 0.30 \pm 0.02 $ & $ 0.33 \pm 0.03 $ \\
  A2744    & $ 0.30 \pm 0.03 $ & $ 0.30 \pm 0.04 $ & $ 0.32 \pm 0.04 $ \\
  A3376 E  & $ 0.19 \pm 0.01 $ & $ 0.19 \pm 0.01 $ & $ 0.18 \pm 0.02 $ \\
  A3376 W  & $ 0.23 \pm 0.01 $ & $ 0.24 \pm 0.03 $ & $ 0.28 \pm 0.04 $ \\
  A3667 NW & $ 0.26 \pm 0.02 $ & $ 0.27 \pm 0.02 $ & $ 0.26 \pm 0.03 $ \\
  A3667 SE & $ 0.23 \pm 0.01 $ & $ 0.22 \pm 0.02 $ & $ 0.22 \pm 0.03 $ \\ \hline
  edge-on  & $ 0.26 \pm 0.02 $ & $ 0.26 \pm 0.02 $ & $ 0.25 \pm 0.04 $ \\
  side-on  & $ 0.25 \pm 0.02 $ & $ 0.22 \pm 0.02 $ & $ 0.24 \pm 0.03 $ \\
  face-on  & $ 0.24 \pm 0.01 $ & $ 0.23 \pm 0.02 $ & $ 0.23 \pm 0.03 $ \\ \hline 
 \end{tabular}
 \caption{The table summarizes the fit results for the observed radio relics that we performed in Sec.~\ref{sec::observations}. The last three columns give the slope and error of the fit $\fzd \propto {\kappa} \log_{10}(S_{\mathrm{int}})$. }
 \label{tab::observations}
\end{table}
 
\section{Simulations}\label{sec::simulations}

Observations only provide information on the 2D projections of actual 3D objects. Therefore, we perform the same analysis to a radio relic that is found in a cosmological simulation. Using a cosmological simulations has the advantage that both the 2D projection as well as the actual 3D shape of the simulated relic can be analyzed. Before discussing the morphological properties of the simulated relic, we briefly introduce the underlying cosmological simulations. 

We analysed a simulated galaxy cluster that undergoes a major merger, which gives rise a bright radio relic. We simulated the cluster with the publicly available \enzo-code \citep{ENZO_2014} and we analysed in post-processing with our Lagrangian Tracer Code \CRaTer \ \citep{2017MNRAS.464.4448W}. The simulation is part of the \textit{SanPedro}-cluster catalogue that we have already presented in detail in \citet{wittor2020gammas,Wittor_2021_Mach,Wittor_2021_Spec} and \citet{Banfi2020}. Here, we only summarize the main aspects of the simulation and point to the references for more details.

\subsection{\enzo}
 
We simulated the merging galaxy cluster with the \enzo-code \citep{ENZO_2014,2019JOSS....4.1636B}, using nested grids and adaptive mesh refinement (AMR). The root grid covers a volume of $(140 \ \Mpc / h)^3$ and it is resolved with both $256^3$ grid cells and $256^3$ dark matter particles. We further refined a $(6.56 \ \Mpc/h)^3$ centered at the galaxy cluster with 5 levels of nested grids, using \music \ \citep{music}. After redshift $z \approx 1$, we refine an additional $(3.28 \ \Mpc / h)^3$ volume centered at the cluster's region, where a giant shock wave is launched, for a total of $2^7$ refinements. On the highest AMR level, the resolution is $4.28 \ \kpc/h$. This resolution appears reasonable, when studying the small-scale structures in radio relics. As a comparison, the spectacular relic located in Abell 2256, $z \approx 0.056$, has been observed with a beam resolution of $5 \ \arcsec$ \citep{Rajpurohit_2022_A2256}. At a redshift of $0.056$, this resolution corresponds to a physical size of $3.93 \ \kpc / h$ and, hence, it is comparable to the resolution of our simulation. 
 
Here, we used \enzo's MHD-module and initialized a uniform magnetic field of $10^{-7} \ \mathrm{G}$ in each direction. To fulfill the $\nabla \cdot \mathbf{B} = 0$ condition, we applied the Dedner Cleaning \citep{2002JCoPh.175..645D}, which has produced realistic cluster magnetic fields with it in the past \citep[e.g.][]{2018MNRAS.474.1672V}. Finally, the used cosmology is in accordance with the latest results of the Planck-Collaboration \citep[][]{PlanckVI2018}: $H_0 = 67.66 \ \km \, \sek^{-1} \, \Mpc^{-1}$, $\Omega_{\Lambda} = 0.69$, $\Omega_{\mathrm{m}} = 0.31$, $\Omega_{\mathrm{b}} = 0.05$.

\subsection{\CRaTer}\label{ssec::crater}
 
The simulated cluster undergoes a major merger that produces two large-scale shock waves. Here, we analyse the radio emission associated with the larger one of the two. Using our Lagrangian Tracer code \CRaTer, we injected $N_p \approx 1.99 \cdot 10^7$ passive tracer particles into the $(1.71 \ \Mpc/h)^3$ region in front of the relic. Each tracer carries a mass of $\sim 5.7 \cdot 10^5 \ \Msun$ The tracers are then passively advected with the flow of the ICM. For details on the advection scheme see \citet{2017MNRAS.464.4448W}

The tracers detect the shock wave using a temperature-based shock finder. As this is a new \CRaTer \ module, we describe it in the appendix, App. \ref{app::shock_finder}. Once, a tracer is crossed by a shock with a Mach number above 1.3\footnote{We note that we only consider shocks with a Mach number above the threshold of 1.3, because weaker Mach numbers might be numerical artefacts.}, the tracer computes the injected cosmic-ray spectrum. Thereafter, it computes the temporal evolution of the spectrum based on the aging of the cosmic-ray electrons. The spectrum is exposed to synchrotron radiation, inverse Compton losses as well as adiabatic compression and expansion. The energy spectrum has the form:

\begin{align}
   \begin{split}
 f(E,t) &= \bar{f} E^{-\alpha} \kappa(t)^{(\alpha+2)/3} \times \\ & \times \left(1 - \left(\frac{1}{E_{\max}} + C(t)\kappa(t)^{-1/3}\right) E \right)^{\alpha-2}. \label{eq::spec}
    \end{split}
\end{align} 

Here, $\bar{f}$ is the normalization, $\kappa(t)$ accounts for adiabatic compression and expansion, $\alpha$ is the energy spectral index, $E_{\max}$ is the maximum energy to which particle are accelerated, and $C(t)$ accounts for synchrotron losses and inverse Compton losses. The corresponding radio emission is obtained by convolving the spectrum with the modified Bessel function $F(1/\tau^2)$

\begin{align}
 \frac{\dd P}{\dd V \dd \nu} = C_\mathrm{R} \int f(\tau,t) F(1/\tau^2) \dd \tau \label{eq::dPdVdv}.
\end{align}

Here, $\tau$ is a substitution for $E$, and $C_\mathrm{R}$ is a constant that depends on the magnetic field, the pitch angle and the observing frequency. For more details on the cosmic-ray model, we point to the corresponding reference \citet{Wittor_2021_Spec}.

\section{Morphology of simulated relics}\label{sec::results}

Using the models described above, we computed the radio emission at $1.4 \ \GHz$. The simulated relic has a total integrated radio power of $P \approx 2.13 \cdot 10^{30} \ \erg/\sek/\Hz$. For an easier comparison with the observations, we converted the radio power to a radio flux\footnote{To this end, we used the standard formula $S = P / (4 \pi D_{\mathrm{L}}^2)$. Here, $S$ and $P$ are the radio flux and radio power, respectively. $D_{\mathrm{L}}$ is the luminosity distance to the relic.}. To this end, we places the relic at a distance of $268.2 \ \Mpc$, which is roughly the distance to Abell 2256. The total integrated flux of the simulated radio relic is $\sim 25.72 \ \mathrm{mJy}$ at $1.4 \ \GHz$. Hence, it is significantly fainter than its observed equivalents. Though, this discrepancy between simulations and observations is common \citep[e.g.][]{2013ApJ...765...21S,Stuardi2019,wittor2019pol}. Most-likely, this discrepancy is explained by the presence of fossil electrons in the observations. Such a population of fossil electrons is commonly required to explain the brightness of most observed radio relics \citep{2020A&A...634A..64B}. Though, several studies predict that a large number of relics exists that lie below the detection limits of current radio instruments, and that will be revealed with the upcoming radio surveys \citep[e.g.][]{2012MNRAS.420.2006N,2017MNRAS.470..240N,Bruggen_2020}. In fact, new observations started to detect such low-luminous radio relics \citep{locatelli2020dsa,2020MNRAS.499..404P}.

In the following, we first present the analysis of the 2D maps and we compare them to the results from the observations. Thereafter, we analyse the properties of the relic in 3D. 
 
\subsection{Projected 2D Morphology}

\begin{figure*}
 \centering
 \includegraphics[width = \textwidth]{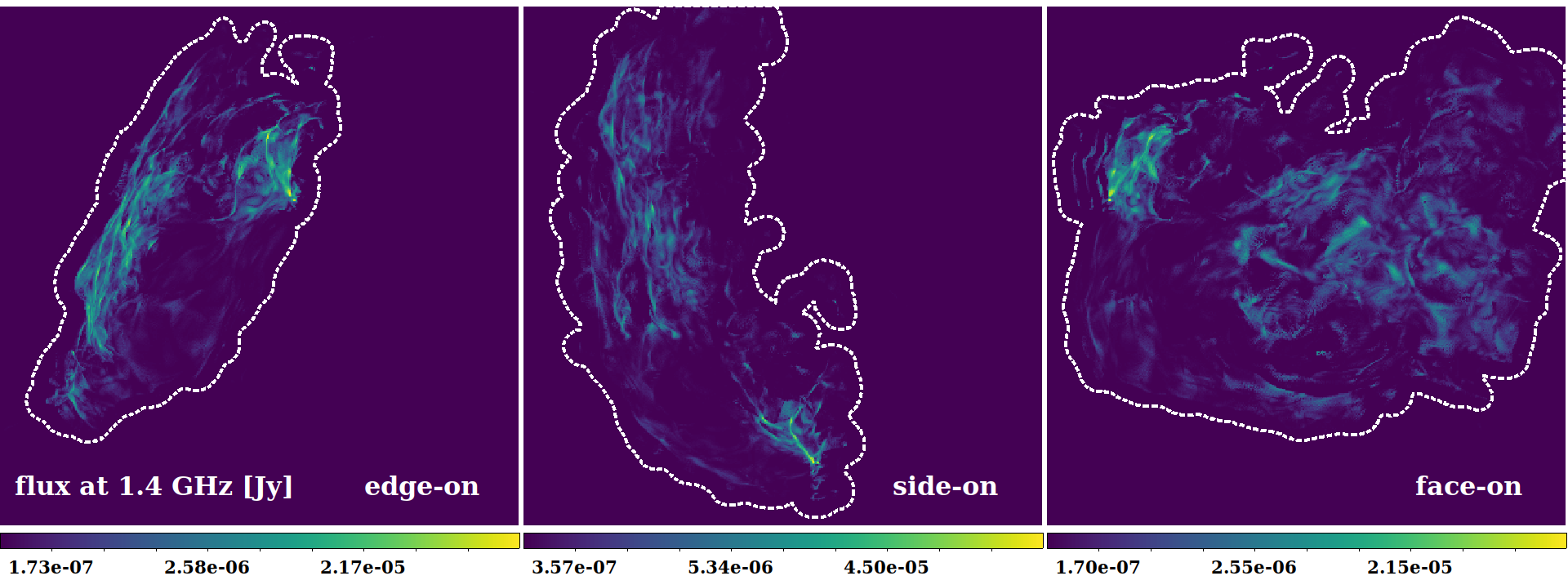} \\
 \includegraphics[width = \textwidth]{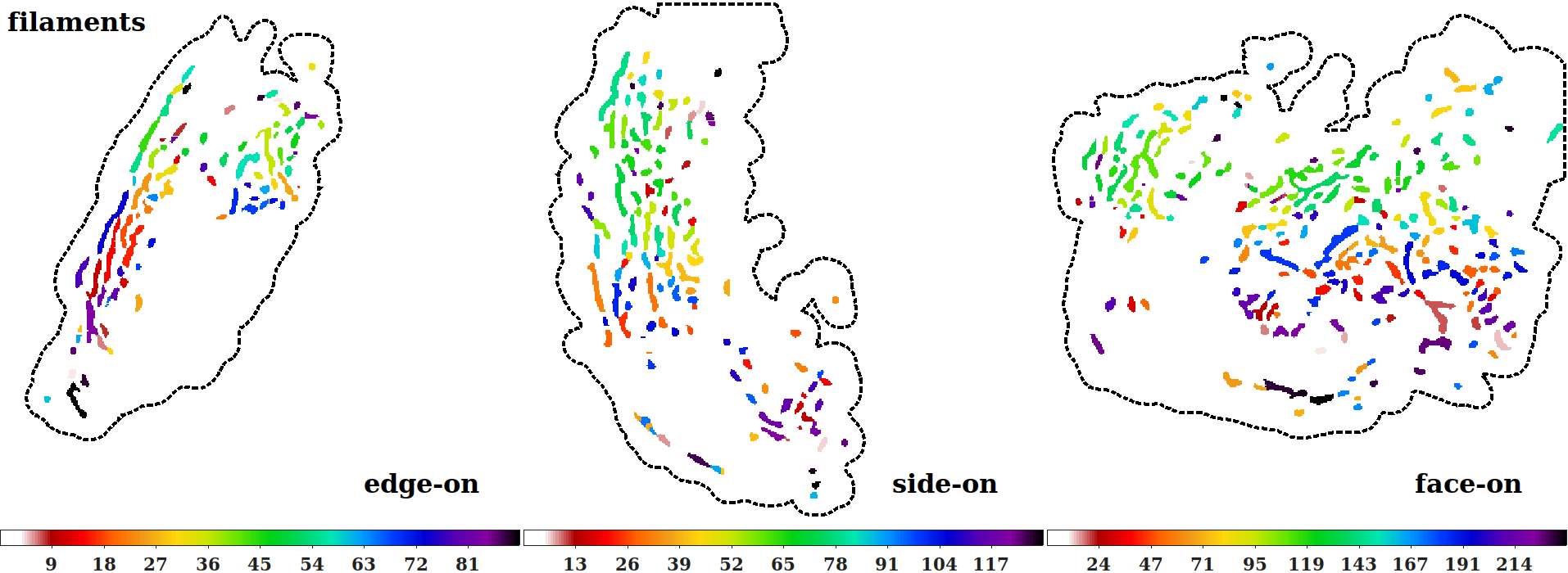} 
 \caption{2D analysis: The top row shows the radio power of the simulated relics seen along the three principle axis of the simulation box. As labelled, we call the different perspectives \textit{edge-on}, \textit{side-on} and \textit{face-on}. The white dashed contours show the regions, that we included in the 3D analysis. The bottom row shows the detected structures.}
 \label{fig::simulation_2D_maps}
\end{figure*}

First, we analysed the two dimensional projections (maps) of the simulated relic. To this end, we integrated the three dimensional radio power along the three principle axes of the simulation box. This procedure gives three maps of the relic seen from different perspectives. We show these maps in Fig.~\ref{fig::simulation_2D_maps}. As labelled, we call the three different projections: \textit{face-on}, \textit{edge-on} and \textit{side-on}. Strictly speaking, the side-on projection shows the relic seen edge-on. However, to make the distinction between the two edge-on projections easy, we call the second one side-on.

\begin{figure*}
 \includegraphics[width = 0.33\textwidth]{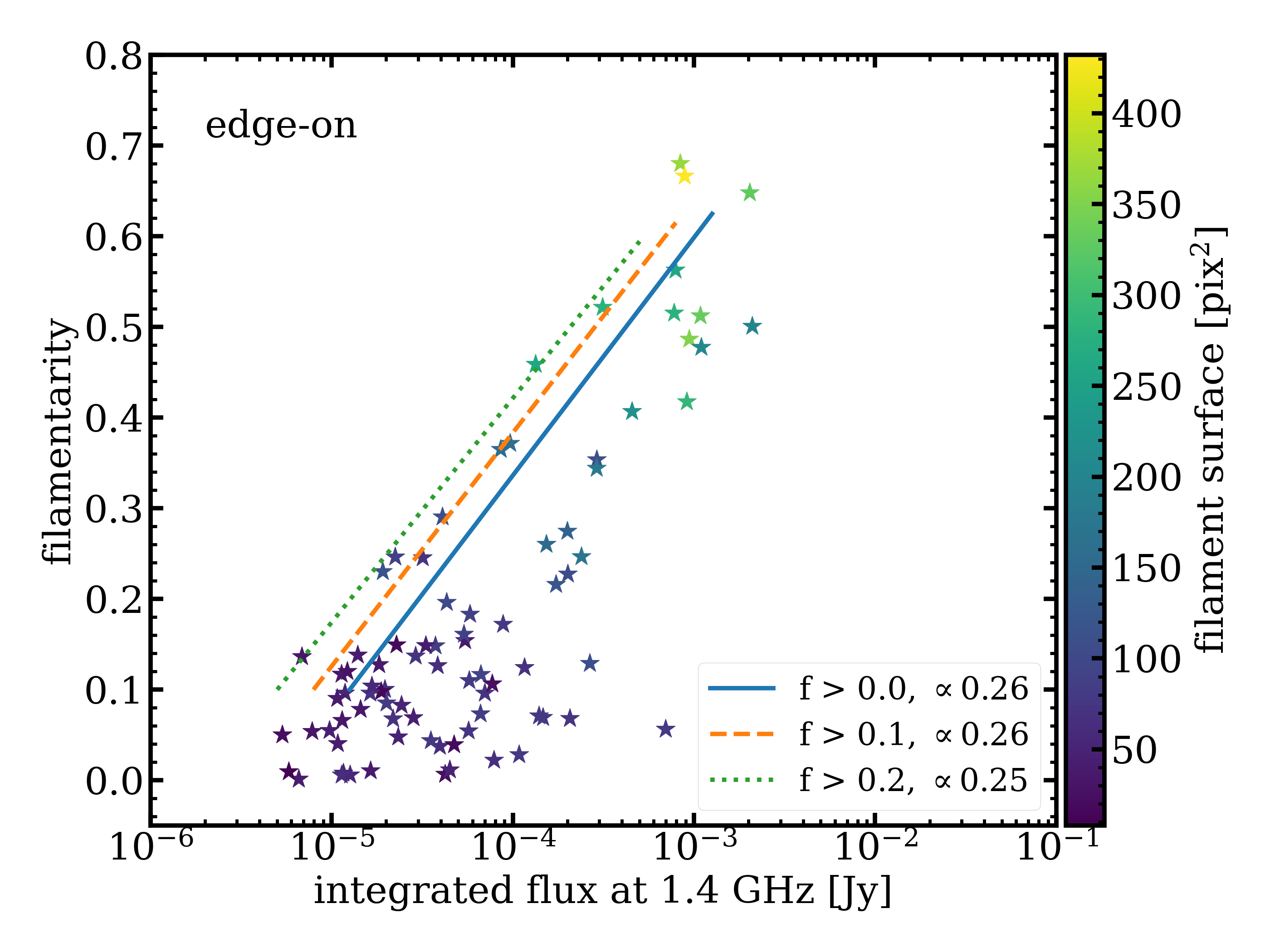}
 \includegraphics[width = 0.33\textwidth]{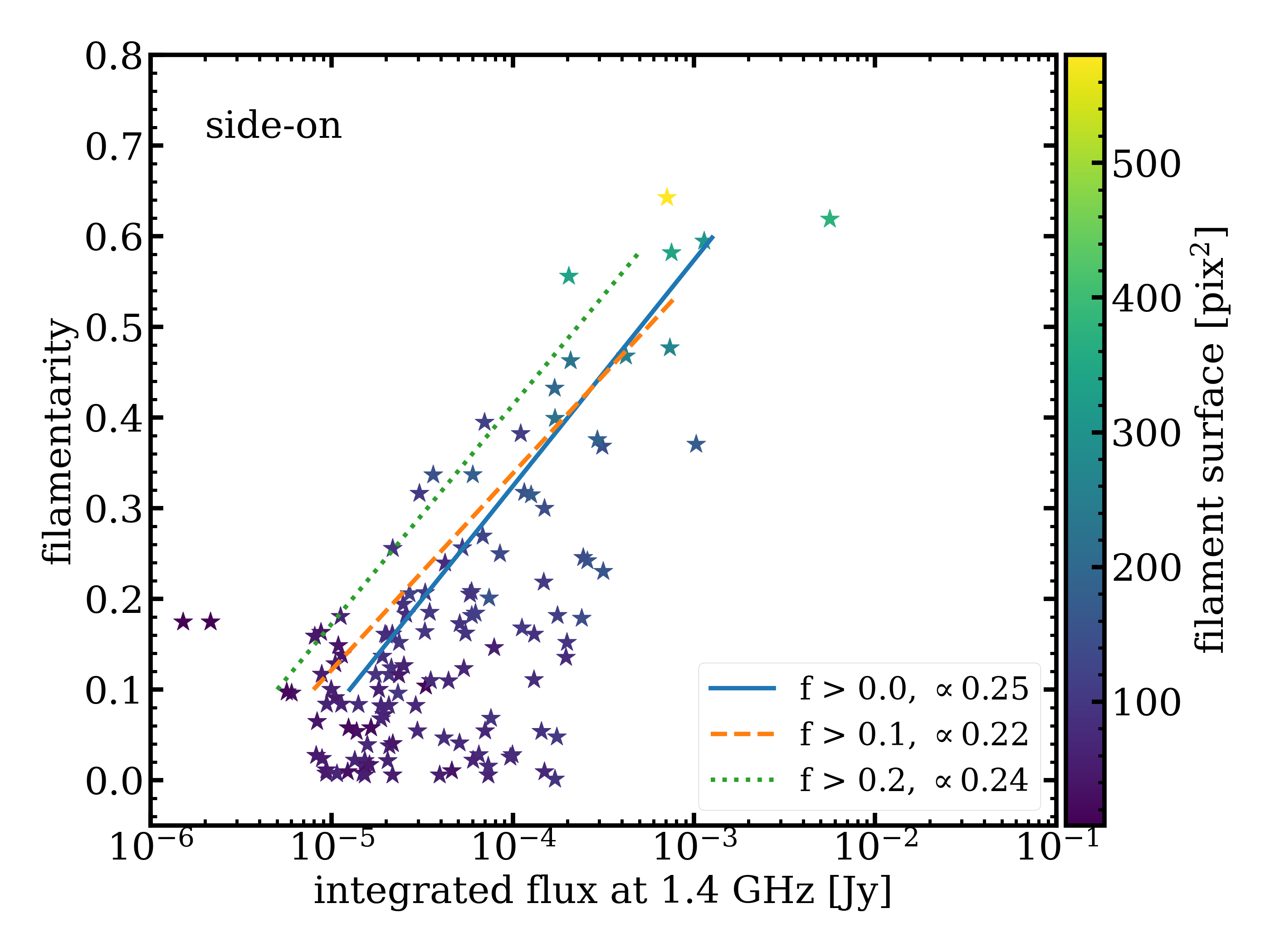}
 \includegraphics[width = 0.33\textwidth]{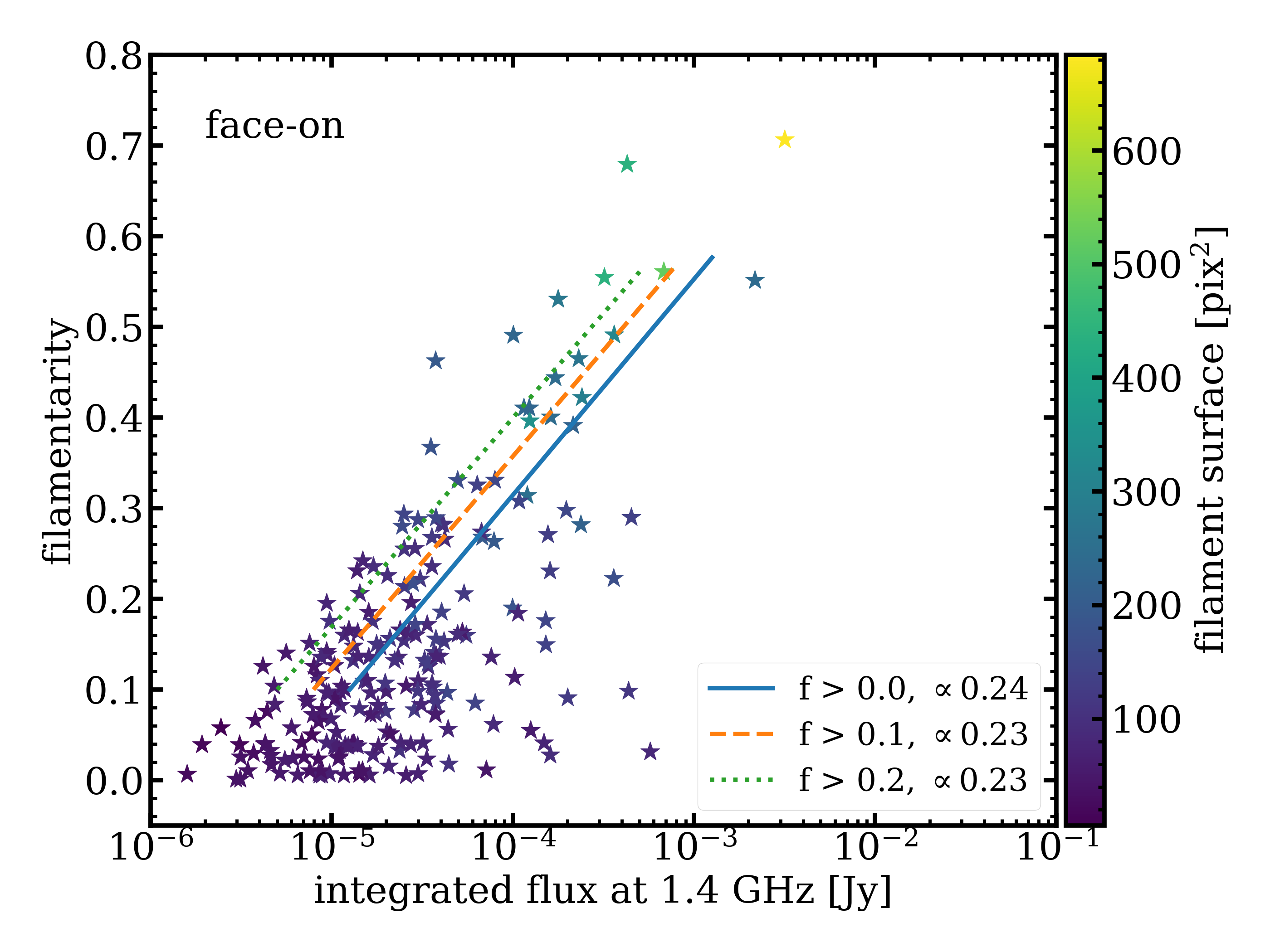}
 \caption{2D analysis: Filamentarity of the simulated radio relics. Each panel shows a different projection of the relic, as labelled (compare with Fig.~\ref{fig::simulation_2D_maps}). In each figure, we plot the 2D filamentarity against the integrated radio flux of each structure. The symbols are colour-coded by size. As in Fig.~\ref{fig::fila_vs_p14_2d_obs}, the different lines show the slopes of the $f_{\mathrm{2D}} \propto \kappa \log_{10}(S_{\mathrm{int}})$ fits.}
 \label{fig::fila_vs_p14_2d_sim}
\end{figure*}

Using \texttt{Sub-X}, we extract the different structures in the maps. The maps of the tagged structures are given in the bottom panels of Fig.~\ref{fig::simulation_2D_maps}. As for the observed relics, \texttt{Sub-X} detects the different structures very well. Using Eq.~\ref{eq::filamentarity_2D}, we compute the 2D filamentarity of the different structures. In Fig.~\ref{fig::fila_vs_p14_2d_sim}, we plot the 2D filamentarity against the integrated radio flux and size of each identified structure. We find the same trends as for the observed relics, Sec.~\ref{sec::observations}. The more luminous structures have a higher filamentarity and they tend to have a larger surface area. Using an ODR to fit $\fzd \propto {\kappa} \log_{10}(S_{\mathrm{int}})$, we find an average slope of $\langle \kappa \rangle \approx 0.25$. This slope agrees very well with the observational findings. Restricting the fit to sturctures with $\fzd > 0.1$ and $\fzd > 0.2$, we find average slopes of $\langle \kappa \rangle \approx 0.25$ and $\langle \kappa \rangle (\fzd > 0.1) \approx 0.24$ and $\langle \kappa (\fzd > 0.2) \rangle \approx 0.24$, respectively.

Comparing the different projections, we find that the fitted slope is similar among the three. The edge-on relic has $\kappa \approx 0.26$, the side-on has $\kappa \approx 0.25$, and the face-on relic has $\kappa \approx 0.24$. 

However, the plots show that there are more small and roundish structures detected in both the side-on case and the face-on case. However, these structures' contributions to the overall radio power and shape of the relic are negligible. We note that the side-on relic is tilted a bit with respect to the line-of-sight. Hence, it is not entirely seen edge-on but also not entirely face-on. This might indicate that actual filamentary structures are more prominent in edge-on relics.

In summary, we find that the structures that are classified as filaments tend to be both brighter and larger in size. Or in other words: brighter and larger structures tend to have a higher filamentarity. This findings are in good agreement of the morphological properties of the structures found in the radio maps of the observed relics.
  
\subsection{Intrinsic 3D morphology}\label{ssec::3Dmorph}

In the previous sections, we found that the morphological properties of the relics' structures are similar in the simulation and observations. Fortunately, the simulation provides the three dimensional shape of the simulated relics. Hence, we applied \texttt{Sub-X} to the three dimensional simulated radio data. Here, we only applied \texttt{Sub-X} in the radio emitting region, i.e. the region where we find the relic. We marked this region with the white contours in Fig.~\ref{fig::simulation_2D_maps}.

\begin{figure*}
    \centering
    \includegraphics[width = \textwidth]{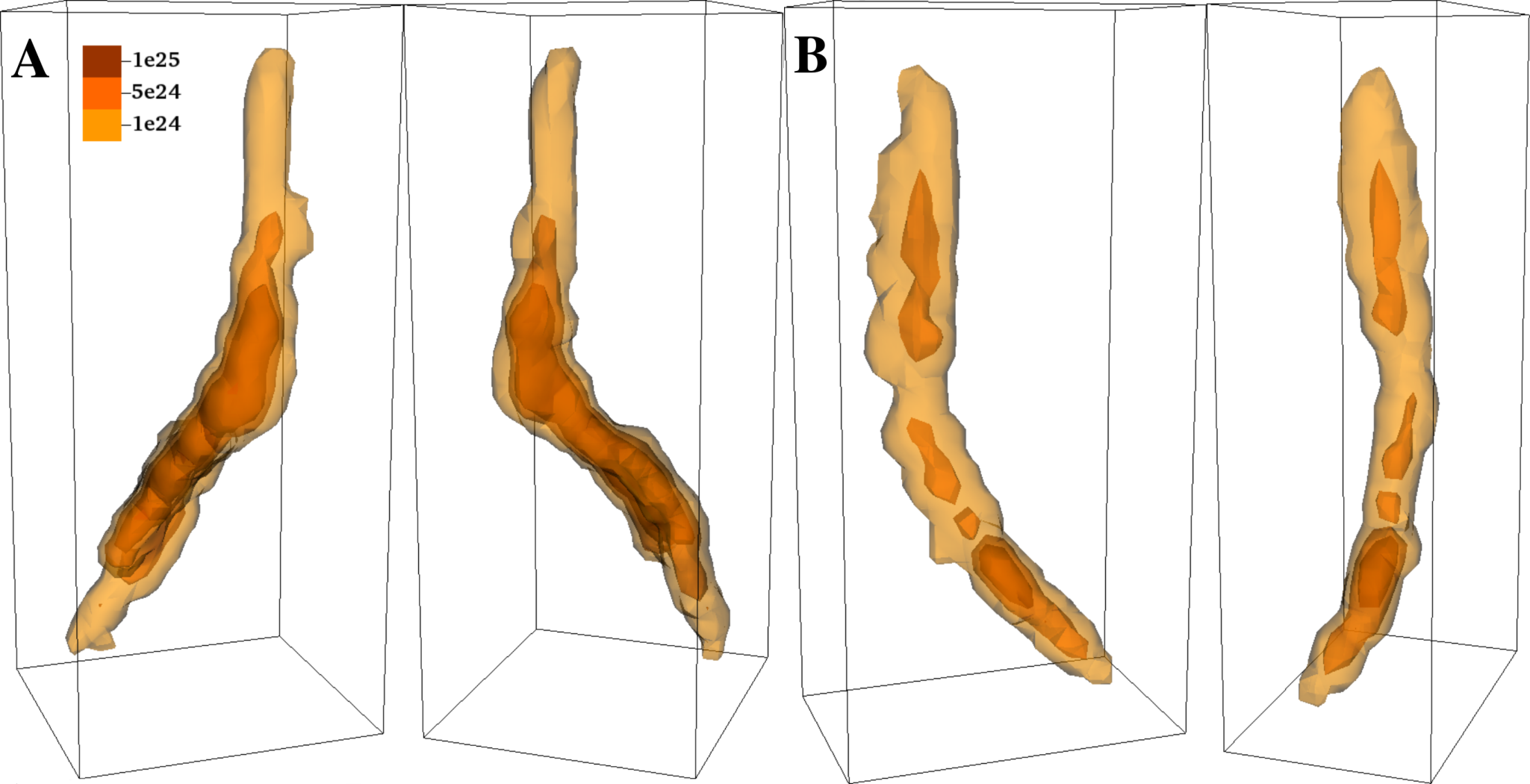}
    \caption{3D analysis: Examples of two filaments that were extracted from the simulated radio relic. The figures show 3D renderings of the $[1\cdot10^{24}, \ 5\cdot10^{24}, \ 1\cdot10^{25}] \ \erg/\sek/\Hz$ radio contours. The filaments' properties - i.e. filamentarity, planarity, ratio of aspect ratios and the three characteristic lengths scales - are summarized in Tab. \ref{tab::fila_examples}. The two examples are neither the longest nor the most filamentary substructures extracted but they were selected to display the typical (average) filamentary substructure.}
    \label{fig::fila_examples1}
\end{figure*}

\begin{figure*}
    \centering
    \includegraphics[width = \textwidth]{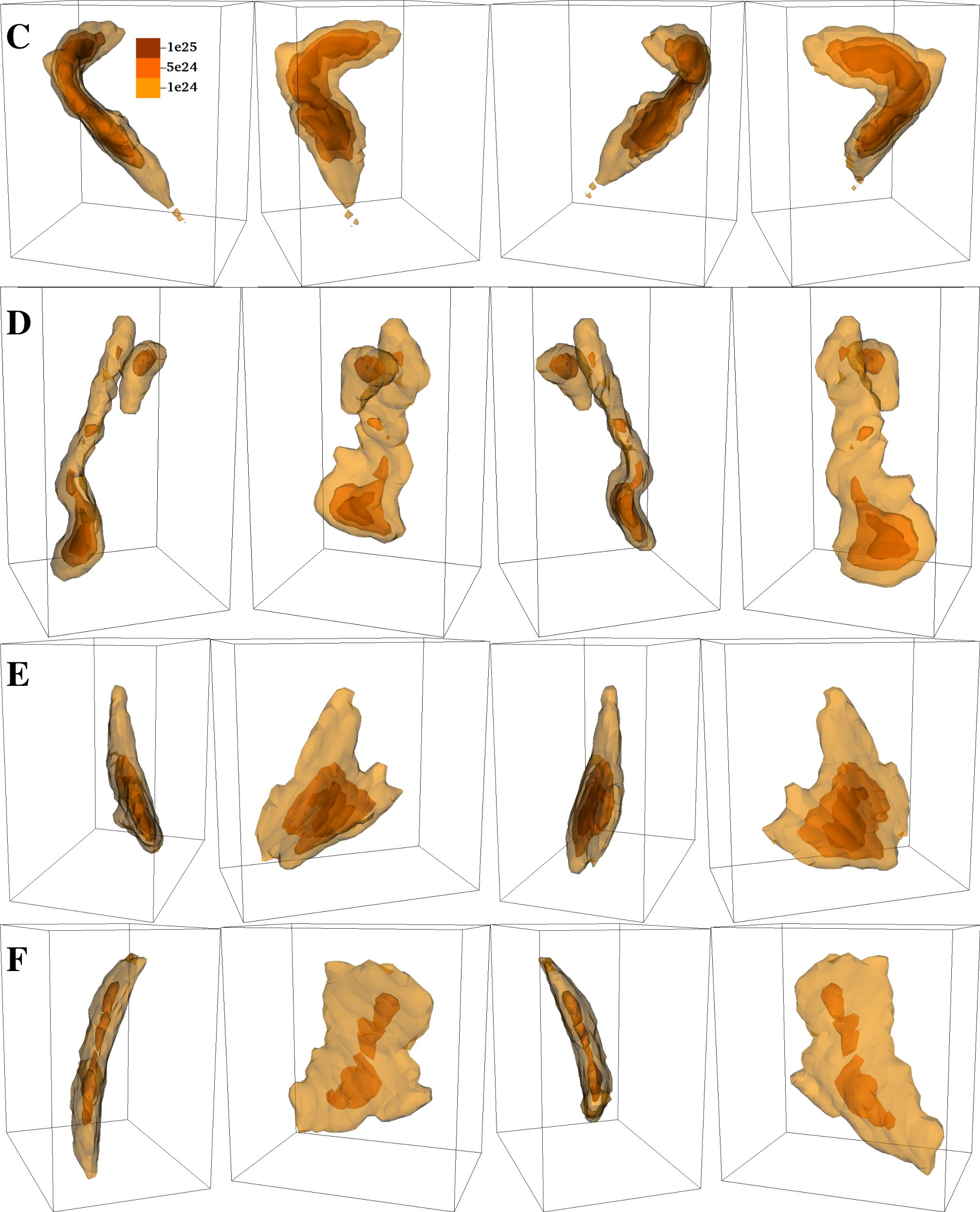}
    \caption{3D analysis: Renderings of substructures that were extracted from the simulated radio relic. Each row shows a different structure seen from four different sides. Albeit their local deformations, structure C and D are both classified as filaments. Structure E and F are both classified as ribbons. The figures show 3D renderings of the $[1\cdot10^{24}, \ 5\cdot10^{24}, \ 1\cdot10^{25}] \ \erg/\sek/\Hz$ radio contours. The properties - i.e. filamentarity, planarity, ratio of aspect ratios and the three characteristic lengths scales - are summarized in Tab. \ref{tab::fila_examples}.}
    \label{fig::fila_examples2}
\end{figure*}

In Fig.~\ref{fig::fila_examples1} and \ref{fig::fila_examples2}, we plot some examples of identified structures. Furthermore, we summarize the main geometrical properties of these examples in Tab. \ref{tab::fila_examples}. Neither of the displayed structures is an extreme case, e.g. in terms of length or any other property. Rather, they are all typical representations of the identified type of  structure. Structure A and structure B are both classified as filaments. Also the visual classification would identify both of them as filaments. Structures C and D are classified as filaments as well. However, the visual inspection shows that they deviate from perfect filaments, due to some irregular features. Finally, both structure E and F are classified as ribbons. The visual inspection also yields that they are ribbons, as there is no characteristic length which is significantly larger or smaller than the other two. Interestingly, depending on the orientation of structure E and F, they appear as filaments. However, it is expected that ribbons can resemble filaments, if they are seen projection.

\begin{table}
 \centering
 \begin{tabular}{c|c|c|c|c|c|c}
  ID & $\fdd$ & $\pdd$ & $a$  & $t$  & $w$  & $l$   \\ 
     &        &        &      & kpc$/$h  & kpc$/$h  & kpc$/$h   \\\hline \hline
  A  & 0.44   & 0.11   & 2.10 & 17.5 & 21.7 & 56.5 \\
  B  & 0.37   & 0.10   & 1.78 & 17.5 & 21.3 & 46.2 \\
  C  & 0.35   & 0.13   & 1.58 & 18.4 & 24.1 & 50.0 \\
  D  & 0.33   & 0.14   & 1.50 & 19.3 & 25.7 & 51.4 \\
  E  & 0.31   & 0.17   & 1.34 & 18.9 & 26.8 & 50.9 \\
  F  & 0.21   & 0.21   & 1.01 & 22.3 & 34.0 & 52.3 
 \end{tabular}
 \caption{3D analysis: Properties of the six structures shown in Fig.~\ref{fig::fila_examples1} and \ref{fig::fila_examples2}. The first column gives the object's ID. The second and third column give the 3D filamentarity, Eq.~\ref{eq::filamentarity_3D}, and planarity, Eq.~\ref{eq::planarity_3D}, respectively. The fourth column gives the resulting ratio of aspect ratios, Eq.~\ref{eq::a_factor}. The last three columns give the characteristic thickness, width and length.}
 \label{tab::fila_examples}
\end{table}

\begin{figure}
 \centering
 \includegraphics[width = 0.49 \textwidth]{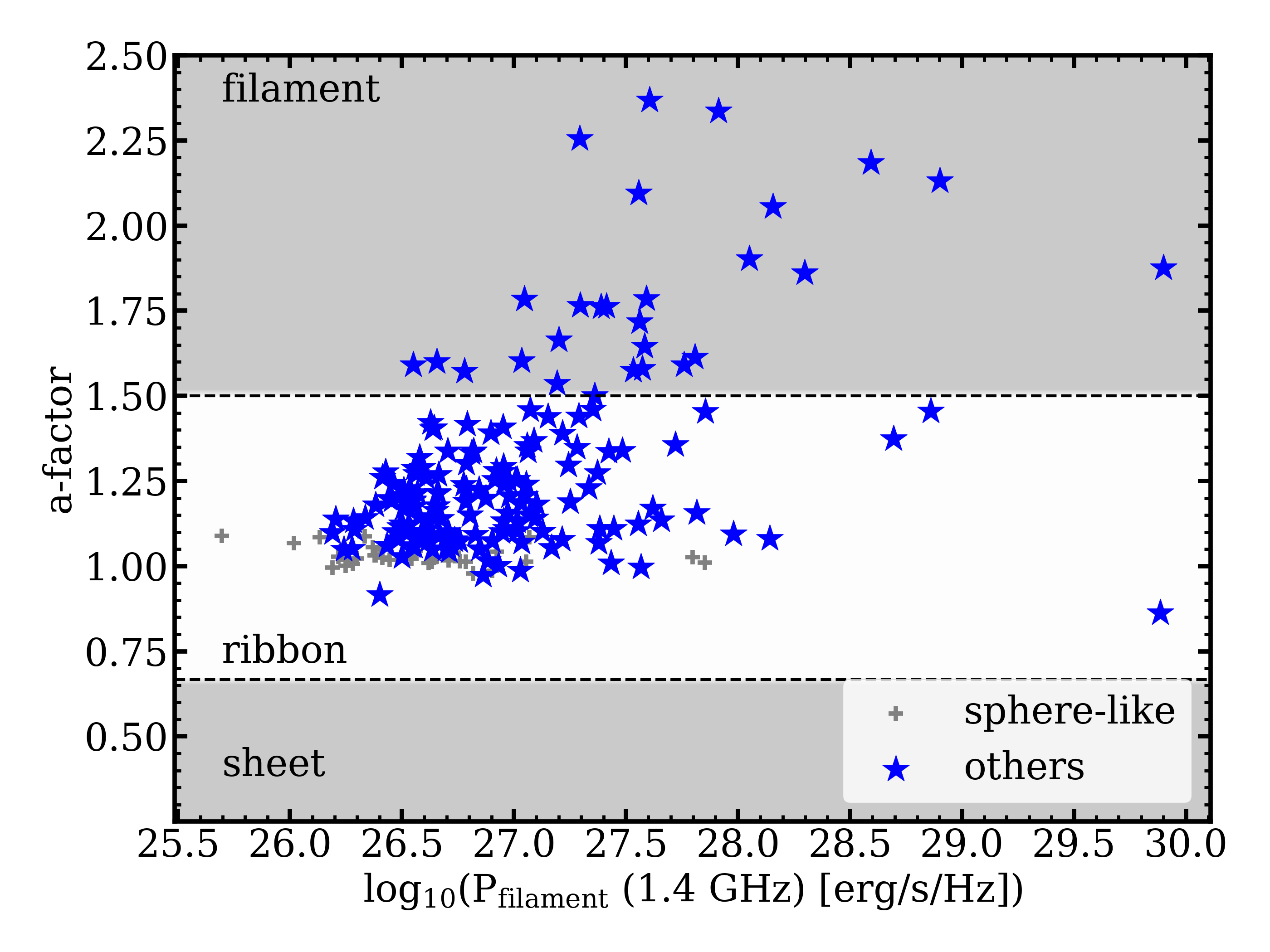}
 \caption{3D analysis: the plot shows the ratio of the aspect ratios, Eq.~\ref{eq::a_factor}, plotted against the integrated radio power of each detected structure. The shaded regions and contours show the classification of each object. The gray crosses are the objects that are classified as a sphere-like object.}
 \label{fig::a-factor}
\end{figure}

For each identified structure, we computed the 3D filamentarity Eq.~\ref{eq::filamentarity_3D} and 3D planarity Eq.~\ref{eq::planarity_3D}, as well as the ratio of the aspect ratios Eq.~\ref{eq::a_factor}. In Fig. \ref{fig::a-factor}, we plot the ratio of the aspect ratios, Eq.~\ref{eq::a_factor}, against the integrated radio power in Fig.~\ref{fig::a-factor}. First, we find that there are some structures that are classified as sphere-like, i.e. the grey crosses in Fig.~\ref{fig::a-factor}. On the other hand, all structures that are not sphere-like, are either identified as a ribbons or filaments. The relic does not consist of any sheet like radio structures. Moreover, the radio power increases for larger ratios of the aspect ratios. This yields that brighter structures are more likely to be a filament.

\begin{figure}
 \centering
 \includegraphics[width = 0.49\textwidth]{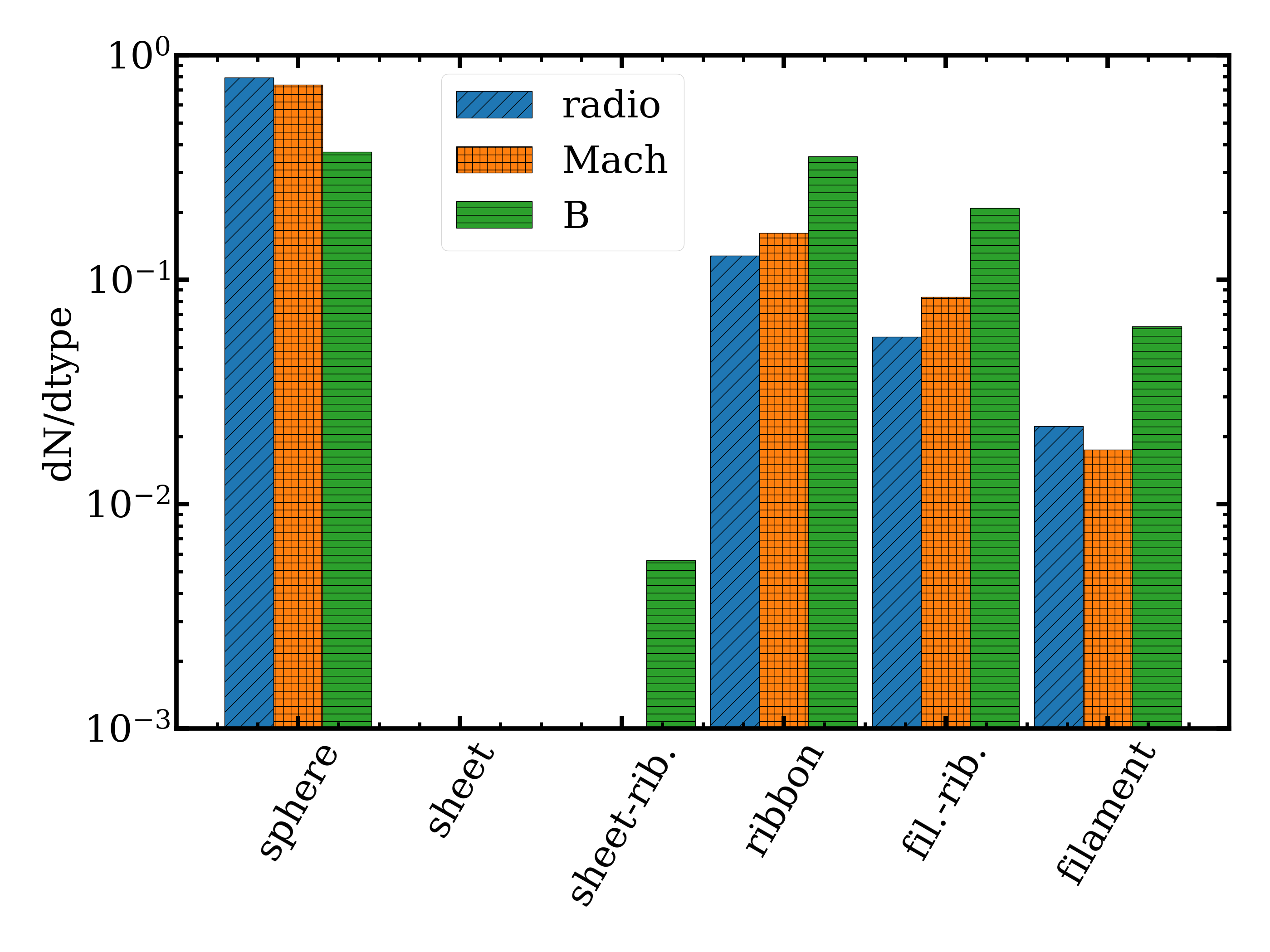}
 \caption{3D analysis: Occurrence of the different geometrical shapes. The shaded region given the normalized distributions for the radio emission (blue, diagonal lines), the Mach number (orange, squares) and the magnetic field (green, horizontal liens).}
 \label{fig::hist_number}
\end{figure}

As a next step, we determined the occurrence of the different types of structures in the simulation. Fig.~\ref{fig::hist_number} shows the distribution of the different type of structures. The majority of the structures is in fact ribbons, i.e. about $\sim 69 \ \%$. Filaments and spheres are represented by similar amounts, i.e. $\sim 14 \ \%$ and $\sim 16 \ \%$, respectively.

In Fig.~\ref{fig::hist_length}, we plot the distribution of the characteristic length scales for all structures detected in 3D. We find that the average length, width and thickness of the radio structures are $\langle l \rangle \approx 33.4 \ \kpc/h$, $\langle w \rangle \approx 21.4 \ \kpc/h$ and $\langle t \rangle \approx 17.5 \ \kpc/h$.

In summary, we find that the 3D structures of radio relics are either filaments, ribbons or sphere-like. In fact, the majority of the structures are ribbons, that might appear as 2D filaments, when seen in projection. However, we find that the brighter a radio structure, the more likely it is to be a filament. We note that the sphere-like radio structures are not to be mistaken with point sources. Rather they are small roundish substructures within the relic.

\subsection{3D morphology of the magnetic field and shock front}

In the literature, it is discussed whether the filamentary structures of radio relics are either tracing the magnetic field structure or the shape of the underlying shock \citep[e.g. see discussion][]{deGasperin_2022_A3667}.  The simulation provides both the 3D magnetic field and the 3D shock front. Hence, we applied \texttt{Sub-X} to both of them as well. Here, we used the same regions as in for the 3D radio emission, see white contours in Fig.~\ref{fig::simulation_2D_maps}. 

\begin{figure*}
  \centering
  \includegraphics[width = 0.49\textwidth]{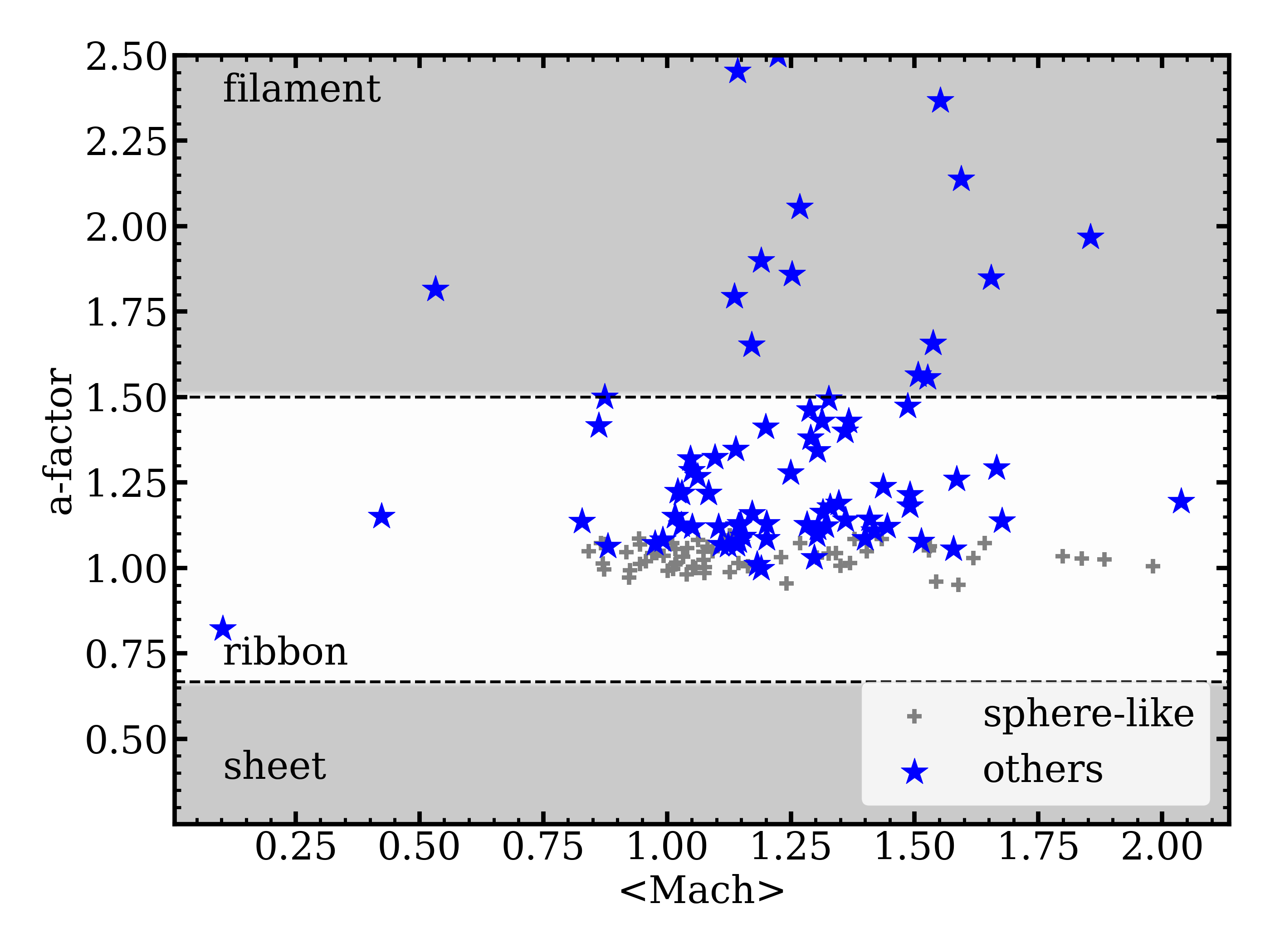}
  \includegraphics[width = 0.49\textwidth]{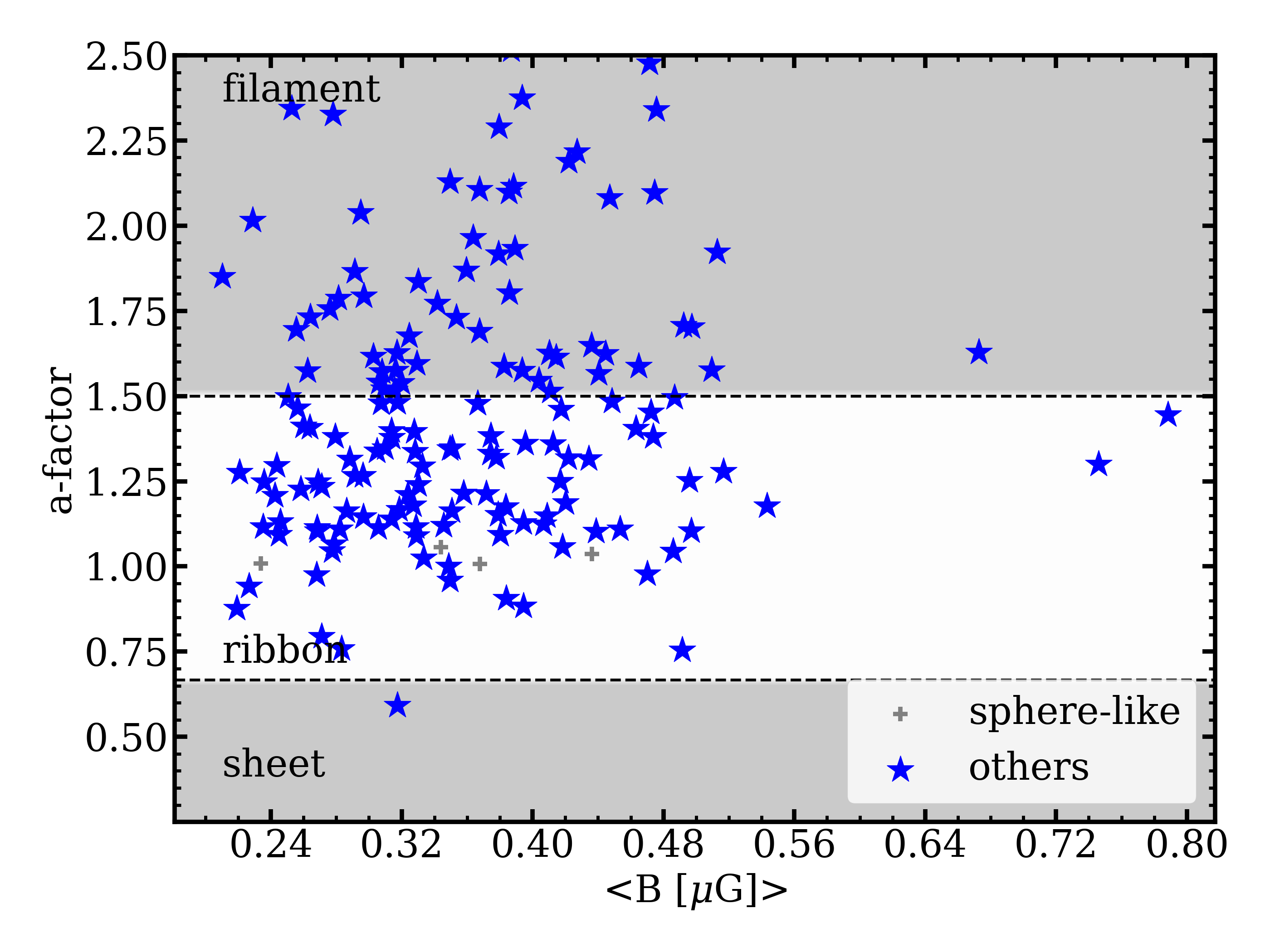}
  \caption{3D analysis: Classification of the different shock structures (left) and magnetic structures (right) in the simulation. Here, we plot the ratio of the aspect ratios of each object against its average Mach number (left) and magnetic field strength (right). Sphere-like objects are plotted as grey crosses and non-sphere like structures are plotted as grey stars.}
  \label{fig::simulation_3D_mach_mag}
 \end{figure*}

In Fig.~\ref{fig::simulation_3D_mach_mag}, we plot the ratio of the aspect ratios against the average Mach number and average magnetic field strength. We note that the average Mach number can be below our threshold of 1.3, see Sec. \ref{ssec::crater}. This is caused by \texttt{Sub-X}'s smoothing. The 3D grid data of the Mach number contains discontinuities, i.e. either grid cells have a value above 1.3 and belong to the shock front, or they are 0 and they do not contain the shock front. The smoothing softens these discontinuities. This softening introduces values below 1.3 in cells, that previously contained a 0 and shared a interface with a non-zero cell. However, this does not affect the results of our analysis.

We find that the number of individual structures is smaller for the Mach number than for the magnetic field. This is due to the fact, that the magnetic field occupies the entire analysed volume. On the other hand, the shock front only occupies a fraction of it. Hence, the number of identifiable structures is larger for the magnetic field.

The shock front mainly, i.e. the Mach number, consists of sphere-like structures, ribbons and filaments. We did not find any sheets in the shock front. The latter is also true for the magnetic field structure. However, the magnetic field consists of one sheet. Yet, we could not find anything special about this structure. Furthermore, we did not find a correlation between the morphology and neither the Mach number nor the magnetic field.

As for the radio emission, we computed the distribution of the magnetic field structures and Mach number structures, see Fig.~\ref{fig::hist_number}. The majority of the Mach number is in shape of ribbons, i.e. $\sim 45 \ \%$, or spheres, i.e. $\sim 42 \ \%$. Consequently, only a small fraction of the shock front consists of filaments, i.e. $\sim 13 \ \%$. On the other hand, the magnetic field structures are mostly shaped as ribbons, i.e. $\sim 53 \ \%$, or filaments, i.e. $\sim 43 \ \%$. Only small fractions of the magnetic field are shaped as sphere, i.e. $\sim 2 \ \%$, or as sheets, i.e. $\sim 0.5 \ \%$.

\begin{figure}
 \centering
 \includegraphics[width = 0.5\textwidth]{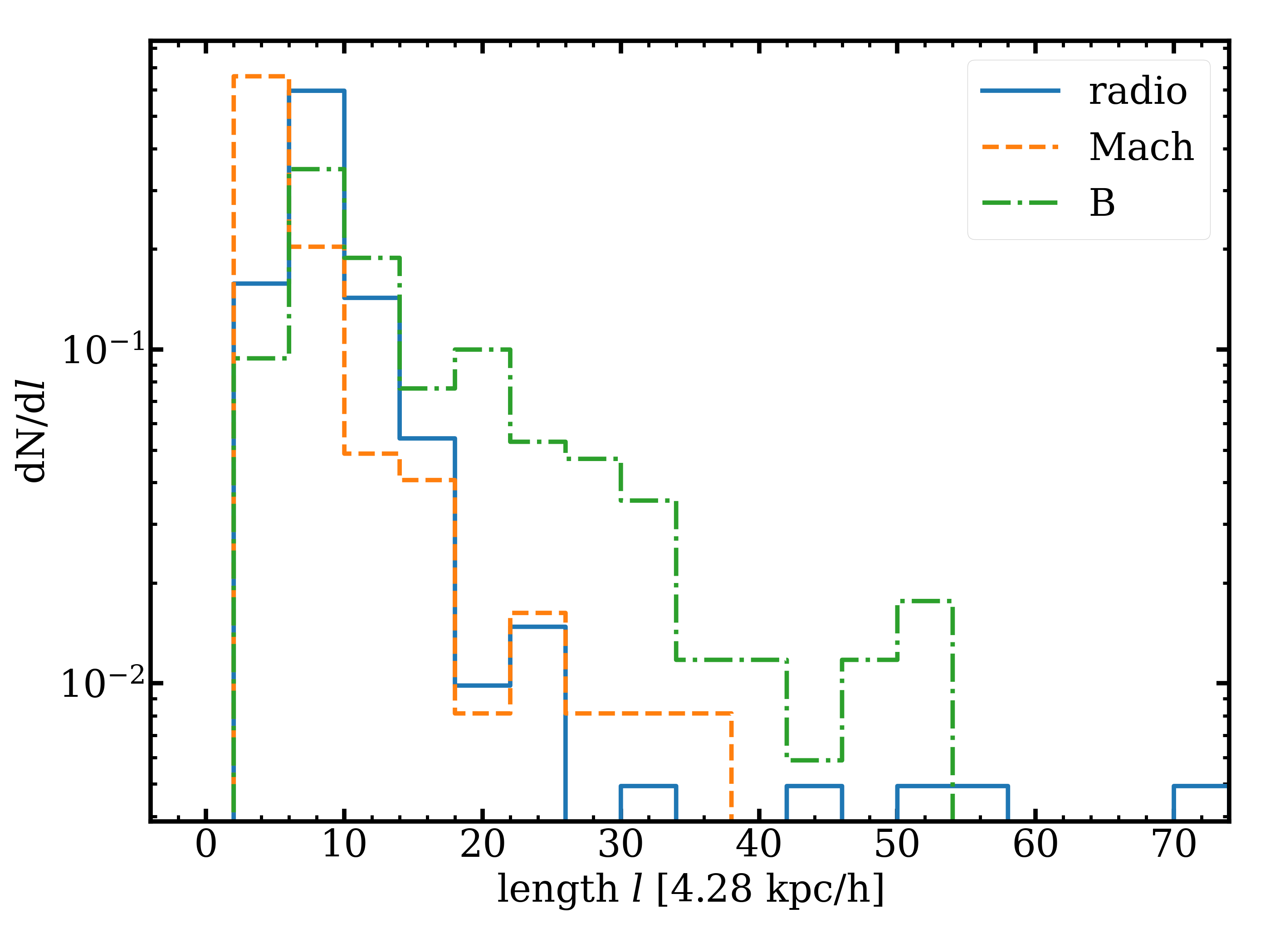} \\
 \includegraphics[width = 0.5\textwidth]{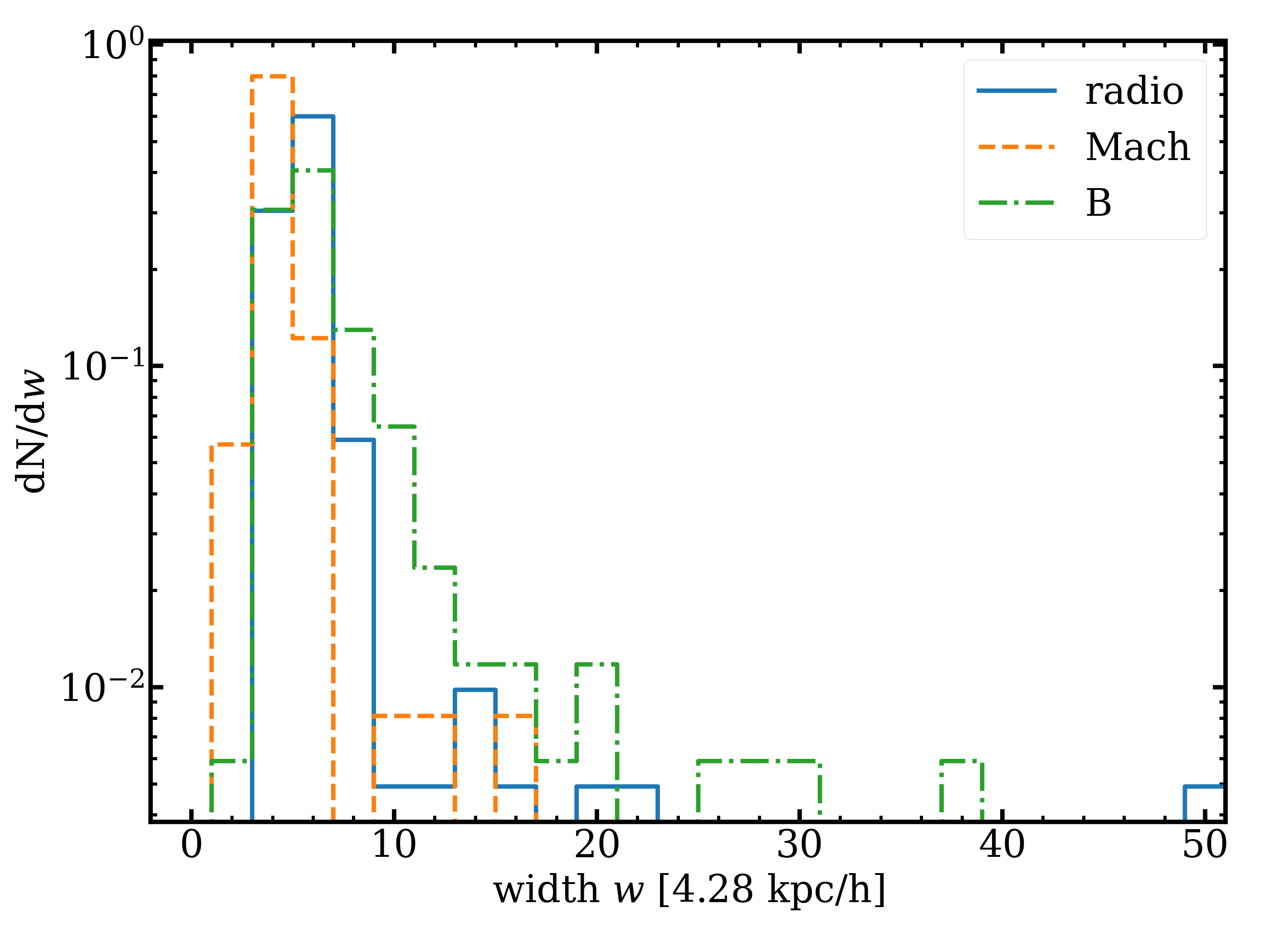} \\
 \includegraphics[width = 0.5\textwidth]{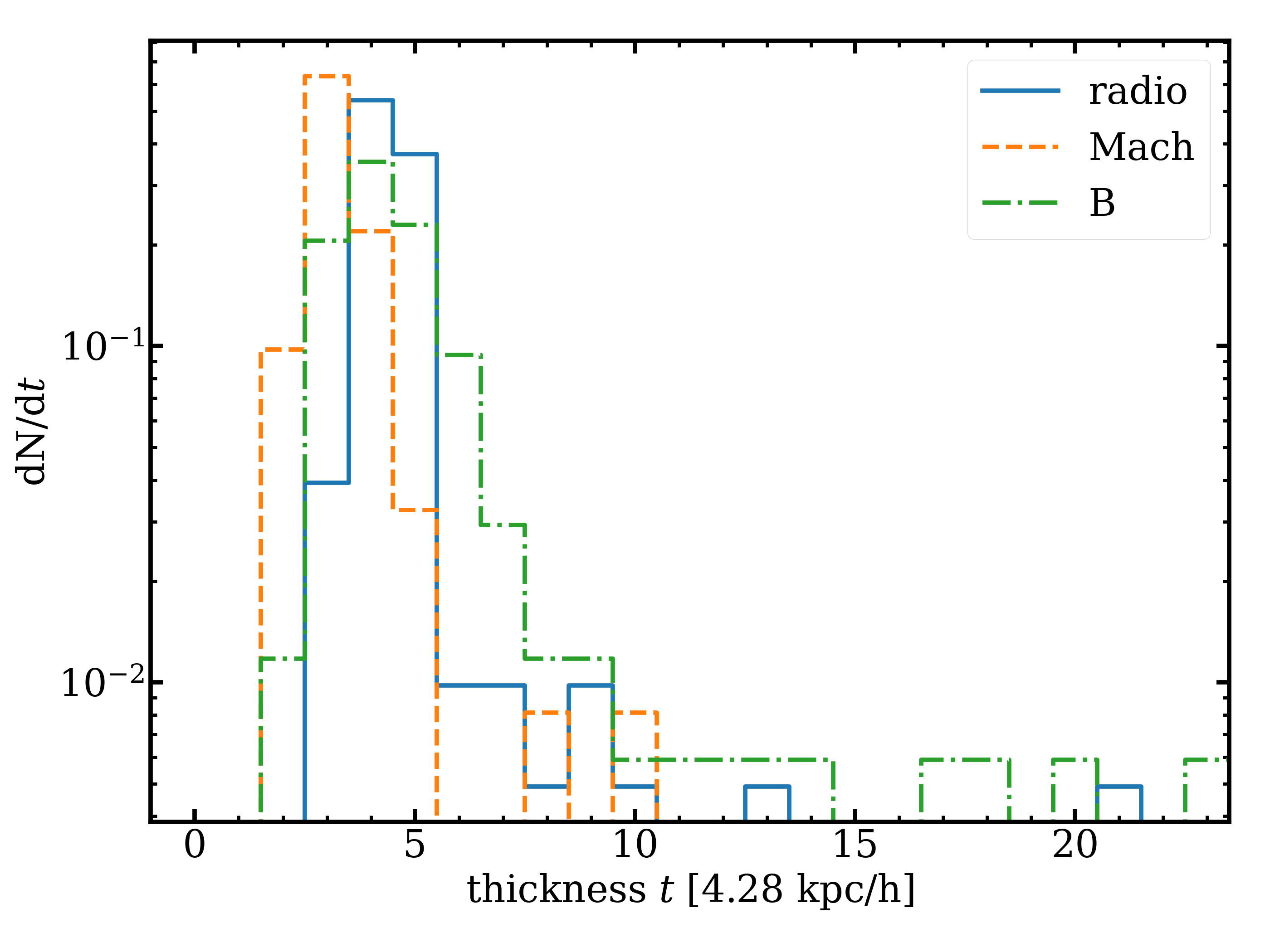} \\
 \caption{3D analysis: normalized distributions of the characteristic length (top), width (middle) and thickness (bottom) of the identified structures. All panels show the distributions for the radio emission (blue, diagonal lines), the Mach number (orange, squares) and the magnetic field (green, horizontal liens).}
 \label{fig::hist_length}
\end{figure}

Finally, we also measured the distribution if the characteristic length scales of the magnetic field structures and Mach number structures, see Fig.~\ref{fig::hist_length}. For the magnetic field, we find that $\langle l \rangle \approx 55.6 \ \kpc/h$, $\langle w \rangle \approx 25.9 \ \kpc/h$ and $\langle t \rangle \approx 19.2 \ \kpc/h$. For the Mach number, we find that $\langle l \rangle \approx 20.7 \ \kpc/h$, $\langle w \rangle \approx 14.3 \ \kpc/h$ and $\langle t \rangle \approx 11.9 \ \kpc/h$. Consequently, the magnetic structures are on average larger than the radio structures, and the structures of the Mach number are on average smaller than the radio structures. However, the reason for this could be of geometric nature. The magnetic field occupies the entire analysed volume. Hence, its structures can be large. On the other hand, the shock front only occupies a smaller volume, i.e. less than $2 \ \%$ of the analysed volume. Finally, the radio emission occupies an intermediate volume, i.e. about $10 \ \%$  of the analysed volume. Hence, the structures of the Mach number and of radio emission have less space then the magnetic field structures. Likewise, structures of the Mach number have less space than the structures of the radio emission.

In summary, we find that also the 3D magnetic field and the 3D shock front consist of filaments, ribbons and spheres. However, the magnetic field tends to have more filamentary structures than the shock front and the diffuse radio emission, that we analysed in Sec. \ref{ssec::3Dmorph}. Moreover, the magnetic structures are on average longer than the structure of both the shock front and the diffuse radio emission. On the other hand, the diffuse radio emission consists of significantly more ribbons than the other two quantities. Finally, the shock front consists of significantly more sphere-like objects than the other two. We did not find, that the shape of the magnetic field or shock front depend on the local magnetic field strength or Mach number, respectively. 

\section{Summary \& Discussion}\label{sec::conclusion}
 
In this work, we have made the first steps towards quantifying the filamentary structures that are observed in several radio relics \citep[e.g.][]{Rajpurohit_2022_A2256,deGasperin_2022_A3667}. To this end, we have developed a parameter-free structure extractor, \texttt{Sub-X}, that solely depends on the mathematical properties of the object, Sec.~\ref{sec::filament_finder}. This finder can be used in, both, two and three dimensions - in principle, it should work in arbitrary dimensions.

Here, we have studied the filamentary structures in six well-known radio relics, as well as in a simulated radio relic that is found in a cosmological simulations, Sec.~\ref{sec::observations} and Sec.~\ref{sec::results}. We have used \texttt{Sub-X} to identify the structures of the observed relics and of the simulated relic. For the latter, we applied \texttt{Sub-X} to both the projected 2D maps and the actual 3D data. 

Using Minkowski functionals, we have computed the geometrical shape parameters of the extracted structures. To characterize the shape of a 3D object, we have introduced a new variable, the ratio of aspect ratios, see Eq.~\ref{eq::a_factor}. The ratio of aspect ratios relates the two aspect ratios of a 3D object. Hence, it indicates if the length of an object is significantly larger than the corresponding width and thickness, or if the thickness is significantly shorter than the length and width. In the former case, the object is a filament, while the latter case indicates a sheet. If the aspect ratio between the length and the width is similar to the aspect ratio between the width and the thickness, the structures is classified as a ribbon. Finally, there is the extreme case of a sphere, where length, width and thickness are all close to equality. 

We found that the radio structures have similar properties in observations and simulations. These findings allow us to draw first conclusions about the origin of the filamentary structure observed in radio relics. Our main results are summarized as follows:
 
\begin{enumerate}
 \item In the maps both of the observed and simulated relics, the identified structures are classified as filaments. Moreover, the brighter the structure is, the more likely it is to be a filament. Fitting the 2D filamentarity against the logarithm of a structure's integrated radio flux yields that the 2D filamentarity increases by about $\sim 0.25$, if the integrated radio flux increases by one order of magnitude. Albeit, the scatter of this relation is large between the different objects, the overall trend is consistent between simulations and observations. 
 \item The analysis of the 3D simulated radio relic showed, that the relic consists of filaments, ribbons and sphere-like structures. Moreover, the relic does not consists of any sheets of radio emission. Hence, a sheet-like origin for the observed filamentary structures is discarded, and the observed filamentary structures are not produced by sheets seen in projection. Moreover, the brightest substructures of the radio emission are more likely to be actual 3D filaments. Finally, the visual inspection of some extracted structures showed that 3D ribbons can appear as 2D filaments or sheets, depending on their orientation. 
 \item We applied the same 3D analysis to the magnetic field and shock front. We found that, both, the Mach number and the magnetic field in the relic region consist of similar 3D structures as the radio emission, i.e. both of them consist of filaments, ribbons and sphere-like structures. However, we could not find a correlation between the Mach number/magnetic field strength of the identified structure and their shape.
 \item Finally, we measured the characteristic length scales of the identified radio structure, Mach number structures and magnetic field structures. On average, the different length scales are between $10 \ \kpc/h$ and a few $10 \ \kpc/h$ and the magnetic field structures are larger than the Mach number structures.
\end{enumerate}

We have shown that \texttt{Sub-X} is capable of robustly detecting structures both in 2D and 3D data. However, it remains unclear whether the filamentary structures in radio relics are attributed to the shape of the shock front or the shape of the magnetic field. Rather it seems to be a combination of both. This is supported by the fact that the two quantities are not independent of each other. The shock wave compresses the magnetic field, causing magnetic field amplification. Hence, regions of higher magnetic fields, i.e. magnetic filaments, should be located at the shock front. A detailed study that disentangles the contribution of the magnetic field from the shock front is beyond the scope of this work, and will be pursued in future studies. 

Finally, we emphasize that \texttt{Sub-X} is not only applicable to the study of radio relics. It can easily be applied to other sources of diffuse radio emission or diffuse emission in general. \texttt{Sub-X} is available on Github: https://github.com/dnswttr/Sub-X.

\section*{Acknowledgements}
We thank Matthias Hoeft for the constructive referee report and criticism. \\
The authors gratefully acknowledge the Gauss Centre for Supercomputing e.V. (www.gauss-centre.eu) for supporting this project by providing computing time through the John von Neumann Institute for Computing (NIC) on the GCS Supercomputer JUWELS at J\"ulich Supercomputing Centre (JSC), under project no. hhh44 and TuMiB. \\
D.W. is funded by the Deutsche Forschungsgemeinschaft (DFG, German Research Foundation) - 441694982. 

M.B. acknowledges funding by the DFG under Germany's Excellence Strategy -- EXC-2121 "Quantum Universe" -- 390833306.

This project has received funding from the European Union’s Horizon 2020 research and innovation programme under the Marie Sklodowska-Curie grant agreement No 101030214 (PI: P.G.).

K.R. acknowledgea ﬁnancial support from ERC starting grant “MAGCOW” No. 714196.

Finally, we wish to acknowledge the developers of the following python packages, which were used extensively during this project: \texttt{QuantImPy} \citep{QuantImPy}, \texttt{numpy} \citep{numpy}, \texttt{scipy} \citep{scipy}, \texttt{scikit-image} \citep{scikit-image}, \texttt{matplotlib} \citep{matplotlib}, \texttt{h5py} \citep{h5py} and \texttt{astropy} \citep{astropy}. \\
Finally, we made use of the Cosmological Calculator by E. Wright (http://www.astro.ucla.edu/~wright/CosmoCalc.html).
\section*{Data availability}
\texttt{Sub-X} is available on Github: https://github.com/dnswttr/Sub-X.
\bibliographystyle{mnras}
\bibliography{mybib}
 
\appendix
 
\section{Shape parameters} \label{app::fp_shapes}
 
The shape parameters defined in Sec.~\ref{sec::shape_parameters} appear as a natural definition of an object's geometry. However, it can be difficult to interprete their value correctly and objects have different shape parameters than expected. In the following, we want to put the values of filamentarity and planarity into context.
 
\subsection{2D Shape Parameters}
 
At first, we look at the 2D filamentarity, Eq.~\ref{eq::filamentarity_2D}, for different objects. Therefore, we define a rectangle with edge lengths of $x = 1$ and $a\cdot x$. Here, $a \ge 1$ is a scale factor that stretches the rectangle into one direction. Hence, for $a \rightarrow \infty$, the rectangle becomes a filament. For $a = 1$, the rectanlge is a square. The circumference and the surface area of such a rectangle are $C = 2 \cdot (1+a)$ and $S = a$, respectively. The 2D filamentarity of a rectangle depends solely on $a$:
 
\begin{align}
 \fzd &= \frac{ (1+a)^2- \pi a }{ (1+a)^2+ \pi a } .
\end{align}
 
\begin{figure}
 \includegraphics[width = 0.49\textwidth]{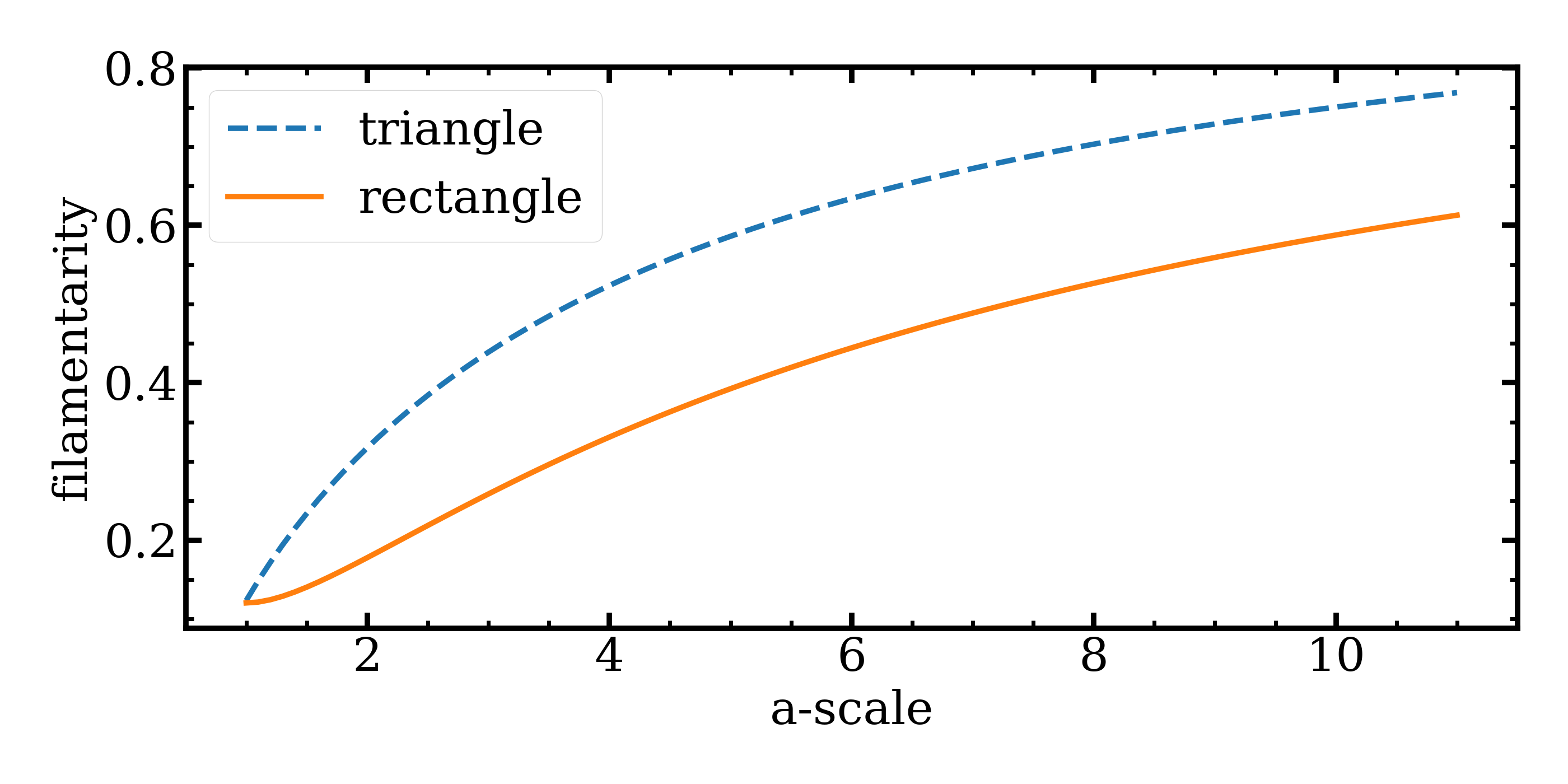}
 \caption{2D filamentarity for a rectangle (orange, solid) and a triangle (blue, dashed) both with varying length.}
 \label{fig::2dfil_anal}
\end{figure}

In Fig.~\ref{fig::2dfil_anal}, we plot the filamentarity for different values of $a$.  A square has a filamentarity of $\fzd (a = 1) \approx 0.12$ and the filamentarity approaches $1$ for larger values of $a$. However, the increase in filamentarity is smaller than expected. For example, a visual inspection of rectangles with $a = 5$ and $a = 10$ would classify both of them as filaments. However, the analytical values are $\fzd(a = 5) \approx 0.39$ and $\fzd(a = 10) \approx 0.59$. Hence, while the visual inspection would classify both as filaments, the analytical classification is not so clear. 
 
 We conducted the same ``experiment'' using a triangle with basis $x = 1$ and height $a$. The triangles circumference and surface area are $C = 1 + \sqrt{4 a^2 + 1}$ and $S = a/2$. Hence, the 2D filamentarity is: 
 
\begin{align}
 \fzd &=\frac{(1 +  \sqrt{4a^2 + 1})^2 - 2 \pi a }{(1 + \sqrt{4a^2 + 1})^2 + 2 \pi a }.
\end{align}
 
The filamentarity for different values of $a$ is plotted in Fig.~\ref{fig::2dfil_anal}. For $a = 1$, the filamentarity is the same as for the rectangle, i.e. $\fzd \approx 0.12$. However, the triangle's filamentarity increases much faster with $a$. Hence, we find that $\fzd(a = 5) \approx 0.59$ and $\fzd(a = 10) \approx 0.75$. 

Now, we try to compare the filamentarity of the rectangle and the triangle. Each of them can be characterised by two characteristic length scales. The rectangle is characterised by its length $a$ and its width $x = 1$. Similar, the triangle is characterised by its height $a$ and its basis $x = 1$. Hence, for the same values of $a$ the two characteristic length scales of the triangle and of the rectangle are the same. However except for $a = 1$, the filamentarity of the triangle is always larger, because the ratio of circumference to enclosed surface area is larger for the triangle than for the rectangle. Albeit, a visual inspection might yield that the two objects have a similar filamentarity - mainly because of the same sizes of the two characteristic length scales - the actual numerical value yields something different. 

These findings highlight that the filamentarity can be characterised by a single value. However, one has to pay attention when putting this single value into context, as a visual inspection might lead to a different conclusion.
 
\subsection{3D  Shape Parameters}\label{app::afactor}

Here, we look at the 3D shape parameters, i.e the the 3D filamentarity, Eq.~\ref{eq::filamentarity_3D}, 3D planarity, Eq.~\ref{eq::planarity_3D} and the ratio of the aspect ratios, i.e. Eq.~\ref{eq::a_factor}. To show the behaviour of these three measurements, we defined a ellipsoid that is characterised by its three major axes: $r_1$, $r_2$ and $r_3$. We set $r_1 = 1$, while $r_2$ and $r_3$ can take values between 0 and 1. In case, that $r_1 = r_2 = r_3 = 1$, the ellipsoid is a sphere. If, $r_1 = r_2 = 1$ and $r_3 \rightarrow 0$, the ellipsoid becomes a sheet. If $r_1 = 1$ and $r_2 = r_3 \rightarrow 0$, the ellipsoid becomes a filament. For the cases where, $r_2 \ne r_3$ and they are both below 1, it is difficult to make a direct statement about the objects shape and the ratio of the aspect ratios becomes useful. 

As examples, we plot some of the ellipsoids in Fig.~\ref{fig::triax}. Here, object a) is the sphere that has $r_1 = r_2 = r_3 = 1$. On the other hand, object g) has $r_1 = 1$ and $r_2 = r_3 = 0.03125$, and, hence, it is the most sheet-like object in the sample. The most filament-like object in the sample is object i), that has $r_1 = r_2 = 1$ and $r_3 = 0.03125$. 

In the first panel of Fig.~\ref{fig::triax}, we plot the planarity against the filamentarity. To guide the eye, we have connected the data points for which $r_3$ is constant. 

In Sec.~\ref{sec::shape_parameters}, we introduced the ratio of the aspect ratios, i.e. Eq.~\ref{eq::a_factor}, as a new measurement for the shape of a three dimensional object. To test this new characteristic, we constructed a simple test case. To this end, we

In Fig.~\ref{fig::triax}, we plot the ratio of the aspect ratios for different values of $r_2$ and $r_3$. As expected, if the two are equal and close to zero, the object is classified as a filament. Vice versa, if $r_3 = 1$ and $r_2 = 0.03125$, the object is classified as a sheet. For the intermediate cases, where both $r_2$ and $r_3$ are below 1 but are still different, the ratio of aspect ratios classifies the different objects as predicted. For example, $r_2 = 0.25$ and $r_3 = 0.5$ gives $a \approx 1$, and the ellipsoid is classified as a ribbon. As an other example, an ellipsoid with $r_2 = 0.03125$ and $r_3 = 0.5$ is classified as a sheet, i.e. $a \approx 0.22$. As a last example, an ellipsoid with $r_2 = 0.03125$ and $r_3 = 0.0625$ is classified as a filament, i.e. $a \approx 7.12$. These findings show that the ratio of aspect ratios is a robust parameter to classify the shape of a 3D object.
 
 \begin{table}
     \centering
     \begin{tabular}{c|c|c|c|c|c|c|c}
$r_1$ &  $r_2$   &  $r_3$    &  $a$   &  $\fdd$ &  $\pdd$  &  shape     & ID \\ \hline \hline
1.0   &  1.0 	 &  1.0      &  1.00  &  0.00   &  0.00    &  sphere    &  a    \\
1.0   &  1.0 	 &  0.5      &  0.94  &  0.03   &  0.06    &  ribbon    &  b    \\
1.0   &  1.0 	 &  0.25     &  0.70  &  0.06   &  0.23    &  ribbon    &  c    \\
1.0   &  1.0 	 &  0.125    &  0.43  &  0.08   &  0.47    &  sheet     &  d    \\
1.0   &  1.0 	 &  0.0625   &  0.23  &  0.09   &  0.67    &  sheet     &  e    \\
1.0   &  1.0 	 &  0.03125  &  0.12  &  0.10   &  0.82    &  sheet     &  f    \\
1.0   &  0.5 	 &  0.5      &  1.05  &  0.05   &  0.03    &  ribbon    &  g    \\
1.0   &  0.5 	 &  0.25     &  0.98  &  0.11   &  0.12    &  ribbon    &  h    \\
1.0   &  0.5 	 &  0.125    &  0.69  &  0.15   &  0.33    &  ribbon    &  i    \\
1.0   &  0.5 	 &  0.0625   &  0.40  &  0.17   &  0.56    &  sheet  &  j    \\
1.0   &  0.5 	 &  0.03125  &  0.22  &  0.18   &  0.74    &  sheet  &  k    \\
1.0   &  0.25 	 &  0.25     &  1.35  &  0.22   &  0.07    &  ribbon    &  l    \\
1.0   &  0.25 	 &  0.125    &  1.33  &  0.31   &  0.14    &  ribbon    &  m    \\
1.0   &  0.25 	 &  0.0625   &  0.94  &  0.34   &  0.38    &  ribbon    &  n    \\
1.0   &  0.25 	 &  0.03125  &  0.56  &  0.37   &  0.60    &  sheet  &  o    \\
1.0   &  0.125 	 &  0.125    &  2.13  &  0.45   &  0.10    &  filament  &  p    \\
1.0   &  0.125 	 &  0.0625   &  2.20  &  0.53   &  0.20    &  filament  &  q    \\
1.0   &  0.125 	 &  0.03125  &  1.57  &  0.57   &  0.40    &  filament  &  r    \\
1.0   &  0.0625  &  0.0625   &  3.76  &  0.66   &  0.12    &  filament  &  s    \\
1.0   &  0.0625  &  0.03125  &  4.01  &  0.72   &  0.21    &  filament  &  t    \\
1.0   &  0.03125 &  0.03125  &  7.12  &  0.80   &  0.13    &  filament  &  u    \\ \hline
     \end{tabular}
     \caption{Properties of the 21 ellipsoids plotted in Fig.~\ref{fig::triax0}. The first three columns give the length of the ellipsoids' three major axes. The fourth column gives the corresponding ratio of aspect ratios- The fifth and sixth column provide the 3D filamentarity and 3D planarity, respectively. The seventh column gives the geometrical classification of each ellipsoid. The last column gives the ellipsoids' ID, that is also used in Tab. \ref{tab::triax} and \ref{fig::triax}. }
     \label{tab::triax}
 \end{table}

\begin{figure}
    \centering
    \includegraphics[width = 0.5\textwidth]{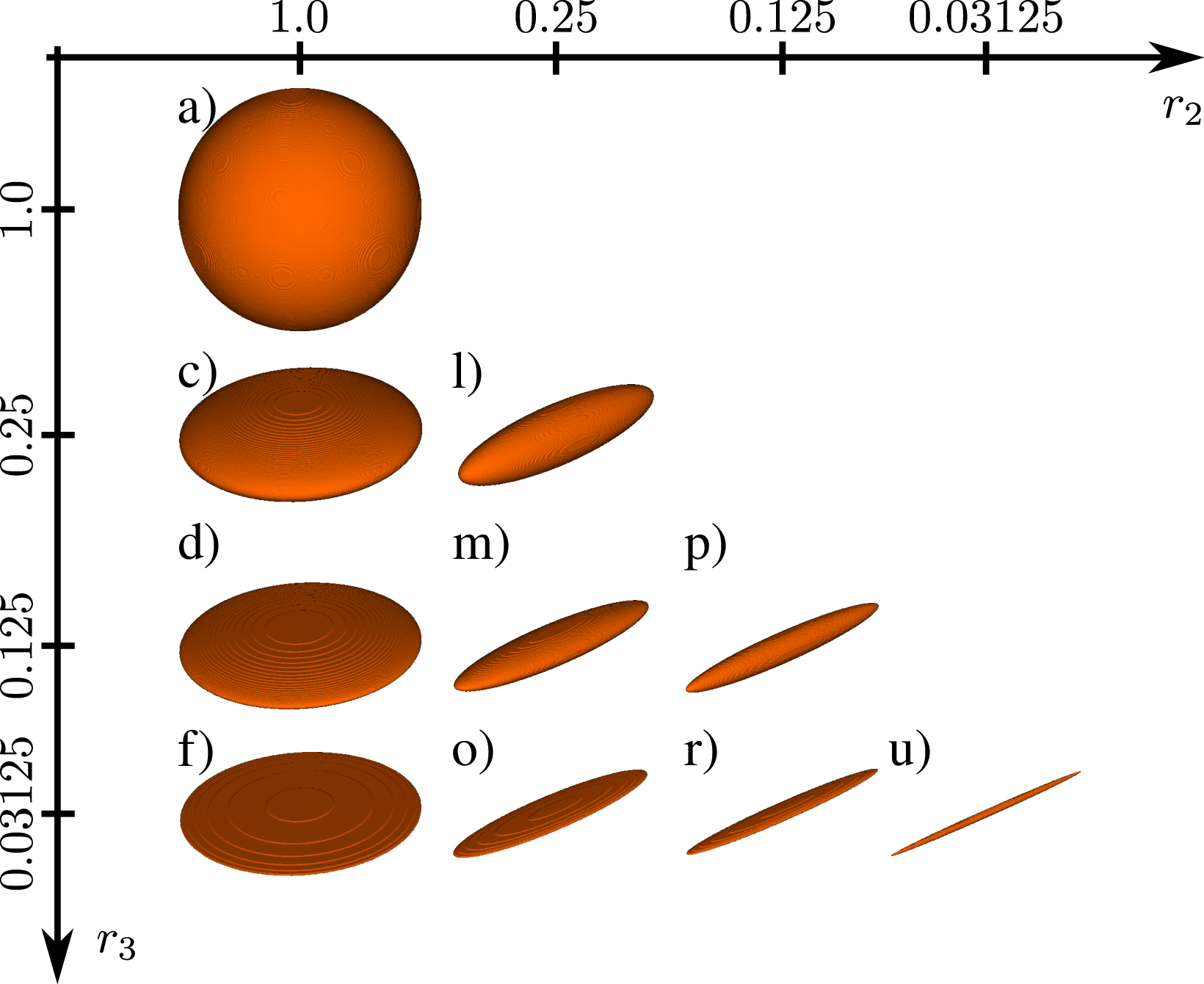}
    \caption{Examples of 11 different ellipsoids, that we analysed in App. \ref{app::afactor}. Ellipsoid \textit{a} is a sphere. The other ellipsoids show, how the sphere changes its shape when it is squeezed along its second, $r_2$, and third, $r_3$, major axes. If the sphere is only squeezed along $r_3$, vertical direction, it transforms into a sheet. If the sphere is squeezed along both axis, diagonal direction, it transforms into a filament. The properties of the different objects and their classification are summarized in Tab. \ref{tab::triax}.}
    \label{fig::triax0}
\end{figure}

\begin{figure}
 \centering
 \includegraphics[width = 0.49\textwidth]{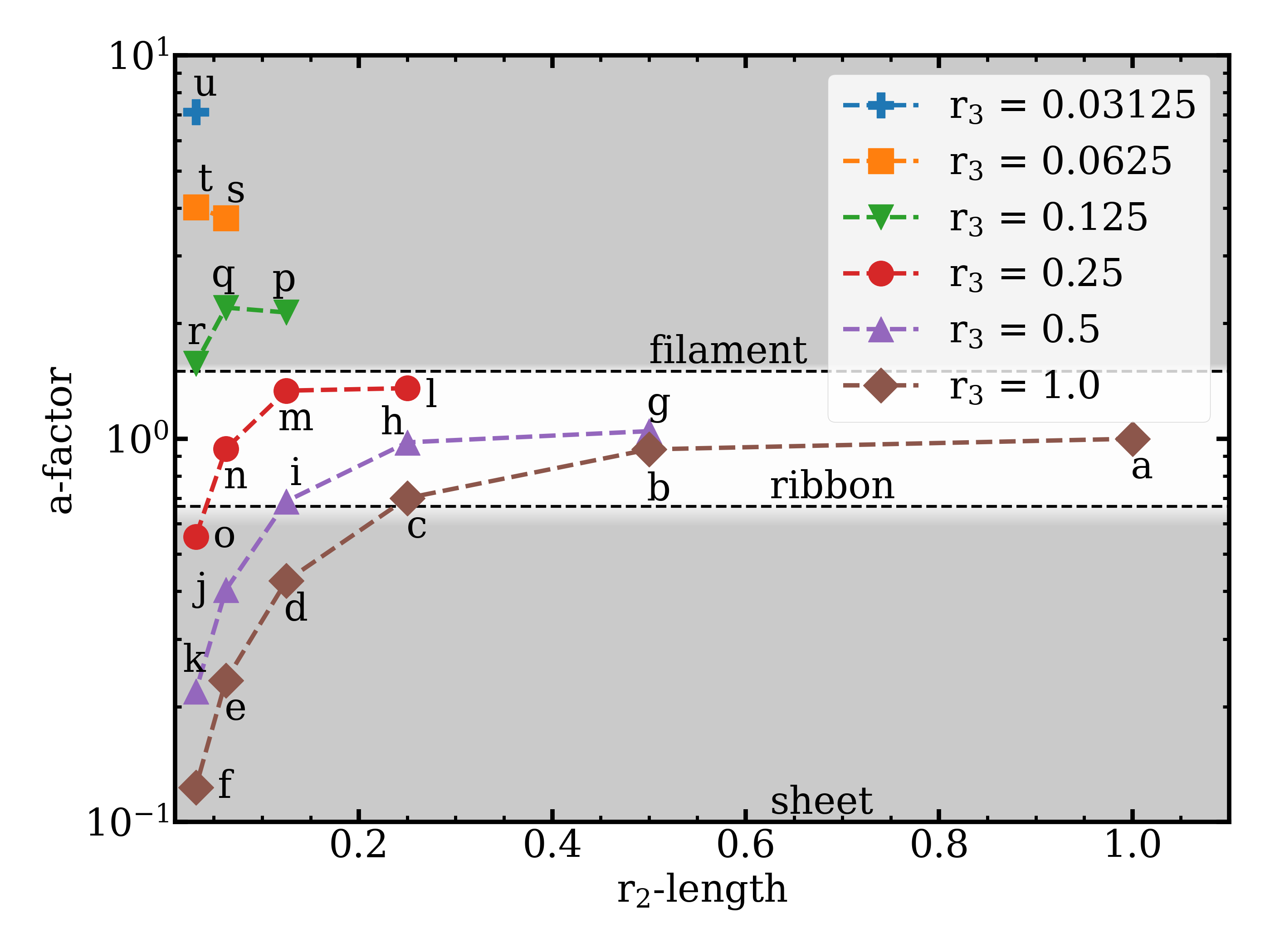}
 \caption{Classification of the 21 ellipsoids given in Fig. \ref{fig::triax0} and Tab. \ref{tab::triax}. The x-axis gives the length of the second major axes of each ellipsoid and the y-axis gives the ratio of aspect ratios. Ellipsoids with the third major axes of equal length are connected by the lines. The annotations match the IDs given in Fig. \ref{fig::triax0} and Tab. \ref{tab::triax}.}
 \label{fig::triax}
\end{figure}
 
\section{Usage of $\chi$} \label{app::chi}
 
As noted in Sec.~\ref{sec::shape_parameters}, there are multiple possibilities to define $l_3$, see Eq.~\ref{eq::r3}. Following \citet{Bag_2019}, it is defined as $l_{3,\chi = 1} = C/4\pi$, while \citet{Schmalzing_1999,Seta_2020} define it as $l_{3,\chi \ne 1} = C/4\pi\chi$. Here, $C$ is the curvature and $\chi$ is the Euler characteristic. The Euler characteristic is a measurement for the number of ``holes'' inside a body. Hence, a solid body has $\chi = 1$, while a torus has $\chi = 0$. Moreover, a hollow body, i.e. a body with a solid surface but empty interior, has $\chi = 2$. As discussed above, the choice of $l_3$ directly affects the computed shape parameters. Nevertheless, the radio emission of relics is continuous and, hence, it should not contain any holes. Therefore, we attribute all structures with $\chi \ne 1$ to numerical uncertainties. Hence, we have developed a filling algorithm to fill the holes in structures with $\chi \ne 1$.

\begin{figure} \centering
  \includegraphics[width = 0.3077\textwidth]{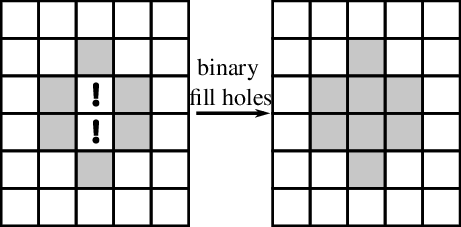} \\
  \includegraphics[width = 0.49\textwidth]{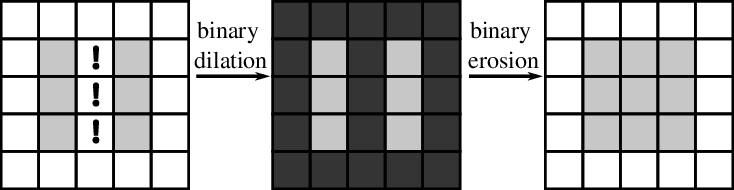}
  \caption{Sketch of the filling algorithm. For a better overview, we show only two deep maps. These can be seen as slices through a 3D structure. In the top panel, we display the \texttt{binary\_fill\_holes}. The cells, containing the exclamation mark, are empty regions within a structure. \texttt{binary\_fill\_holes} fills these cells. The bottom panel shows the \texttt{binary\_dilation} and \texttt{binary\_erosion} functions. Here the exclamation marks mark empty regions that that run through the entire structure, i.e. this plot can be thought of as a slice through a torus. As a first step, we apply \texttt{binary\_dilation} to fill all these holes. However, as a side effect all the dark shaded cells are filled. Hence in a second step, we apply \texttt{binary\_erosion} to empty the cells, that are not supposed to be filled. Fortunately, the cells containing the exclamation marks are not affect by the \texttt{binary\_erosion}.  }
  \label{fig::howtotfill}
 \end{figure}

Our filling algorithm is a two step procedure, that is sketched in Fig.~\ref{fig::howtotfill}. Our algorithm uses functions that are part of the \texttt{scipy ndimage} package \citep{scipy}. For details on the used functions, we point to the package documentation.

As a first step, we use the \texttt{binary\_fill\_holes} function, as shown in the first panel of Fig.~\ref{fig::howtotfill}. The \texttt{binary\_fill\_holes} function fills empty cells , that are surrounded by non-empty cells. However, the \texttt{binary\_fill\_holes} is not capable of filling any holes that run through the entire structure, e.g. it would not fill the central region of a torus. 

Hence, as a second step we use a combination of the \texttt{binary\_dilation} function and the \texttt{binary\_erosion} function. First, we apply the \texttt{binary\_dilation} function using the \texttt{generate\_binary\_structure(3,2)} as the used \texttt{structure}. However, the \texttt{binary\_dilation} function fills cells that are not allowed to be filled, i.e. see dark shaded regions in Fig \ref{fig::howtotfill}. Hence, we further apply the \texttt{binary\_erosion} function, using \texttt{generate\_binary\_structure(3,2)}, as well.

\begin{figure*}
 \includegraphics[width = \textwidth]{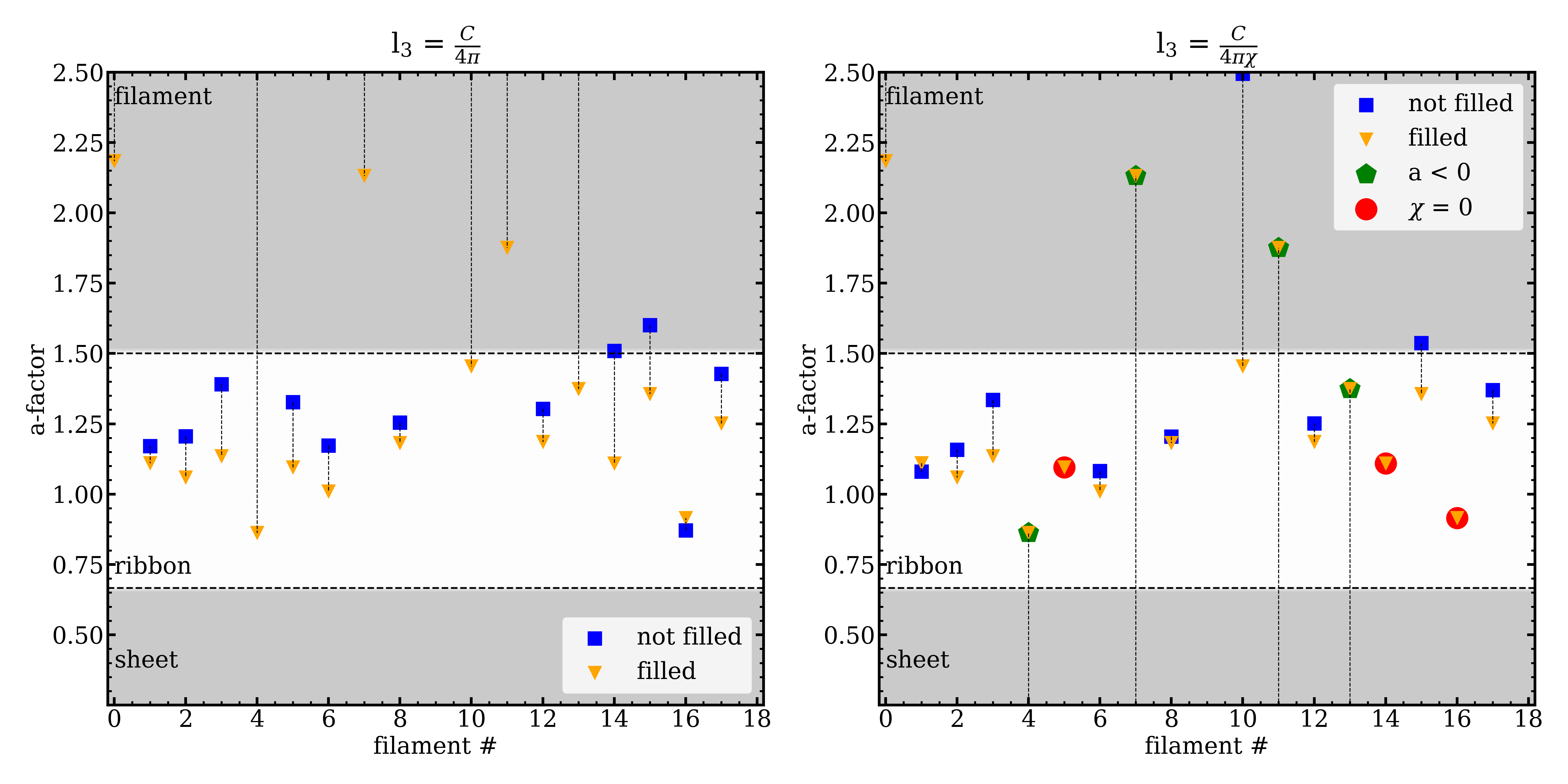}
 \caption{The plots show how the ratio of aspect ratios changes for filled and unfilled structures. The left plot uses the definition of $l_3$ as given in \citet{Bag_2019}, while the right plot uses the definition of $l_3$ as given in \citet{Schmalzing_1998}. In both plots, the x-axis gives the extracted structures that originally have $\chi \ne 1$. The y-axis gives the ratio of aspect ratios. The blue squares give the cases for the unfilled structures, while the yellow triangles give the results after filling the holes in the structures.}
 \label{fig::minkowski_fill_comparison}
\end{figure*}

In total, we found that 18 radio structures out of 205 identified structures have a Euler characteristic of $\chi \ne 1$. In Fig. \ref{fig::minkowski_fill_comparison}, we plot how this additional filling algorithm altered the estimated shapes of these 18 radio structures. 

First, we look the case of $l_3 = C/4\pi$. Here, the difference between the filled and unfilled structures is that the filled structures have a larger volume and a smaller surface. As seen in the first panel of Fig. \ref{fig::minkowski_fill_comparison}, the majority of structures do not change there class. There are four structures, namely number 4, 10, 14 and 15, that change from being filaments to ribbons. However, structures 14 and 15, are very close to boundary between ribbons and filaments. Hence, their actual shape is not drastically. Hence, the filling algorithm only changes the classification for two structures, i.e. number 4 and 10, which is less than $1 \ \%$ of the total number of identified structures. Hence, there classification does not alter the quantitative results of our analysis.

Second, we look the case of $l_3 = C/4\pi\chi$. Here, we have to differentiate three cases. The first case is $\chi \ne 1$, and $a > 0$. For this case, we find that solely structure 10 and 15 change their classification. However, also here structure 15 is very close to the boundary between being a filament or a ribbon. The second case is that $\chi \ne 1$, and $a < 0$. We identified four structures, namely 4, 7, 11 and 13, that have this property. However, a geometrical shape is not defined for negative ratio of aspect ratios. The ratio of aspect ratios can becomes negative, if the absolute of either the 3D filamentarity or the 3D planarity are larger than 1. This can only happen in the nonphysical case of  $l_3 < 0$ negative. The third case is that $\chi = 0$. In this case, $l_3$ becomes infinity and the ratio of aspect ratio is not defined. This is the case for 3 structures in our sample. Consequently, for the two cases that either $a < 0$ or $\chi < 0$, the filling algorithm is indeed needed to characterize the shape of a structures.

These examples have shown, that the filling algorithm is indeed helpful when characterizing the shape of structures with $\chi \ne 1$.
 
\section{Observations}\label{app::obs}

\begin{figure*}
 \includegraphics[width = \textwidth]{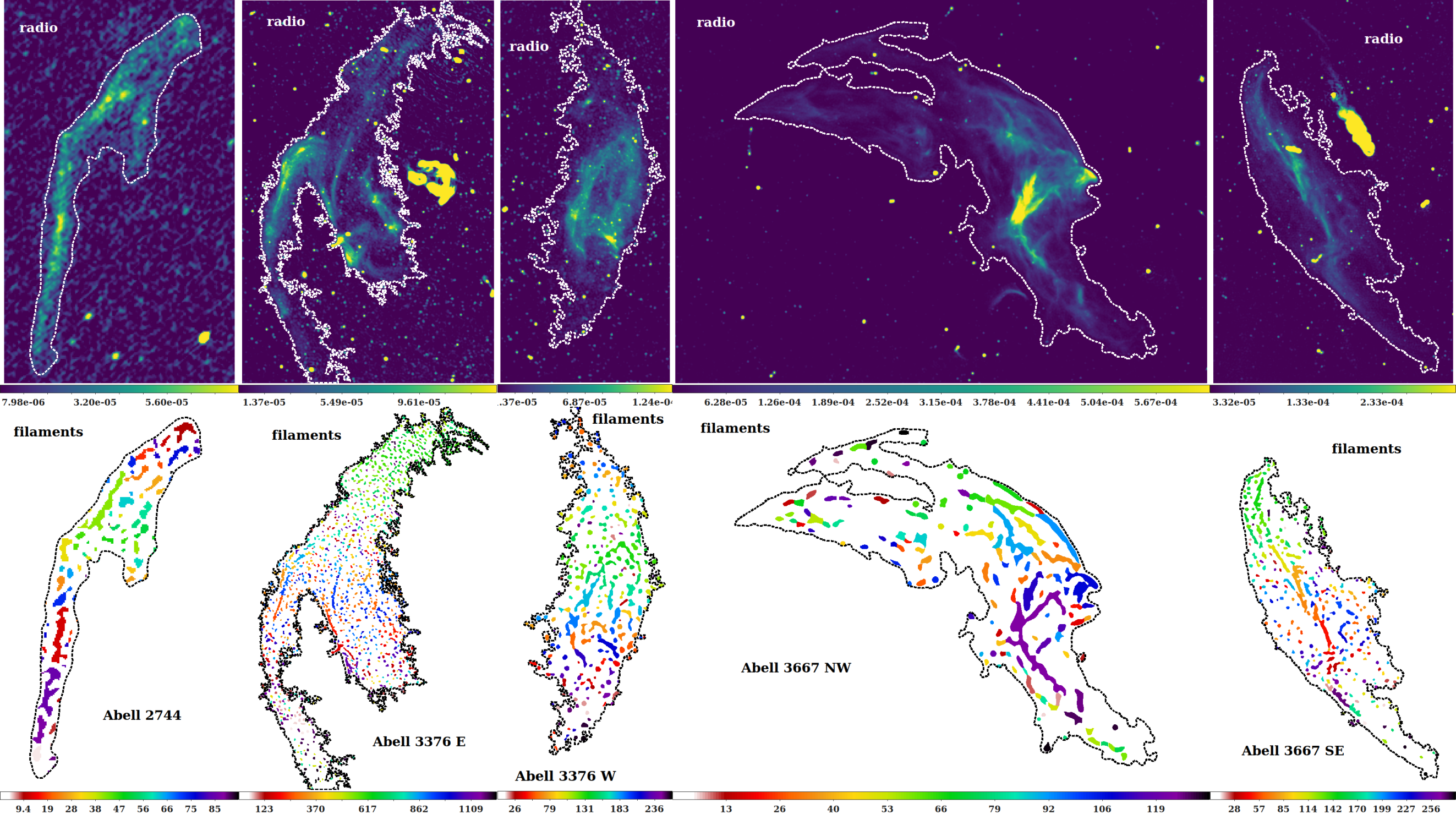}
 \caption{Observational analysis: Radio maps (top) and maps of the tagged structures (bottom) of the reaming relics not plotted in Sec. \ref{sec::observations}. The contours mark the region, where \texttt{Sub-X} searched for structures. The first column shows the relic in Abell 2744. The second and third column shows the relics in Abell 3376. The last two columns show the relics in Abell 3667.}
 \label{fig::Acollage}
\end{figure*}

In Sec. \ref{sec::observations}, we only provided maps of the radio emission and corresponding extracted structures of the relic in Abell 2256. For completeness, we plot the corresponding maps of the remaining five relics in Fig. \ref{fig::Acollage}. The maps highlight as well, how well \texttt{Sub-X} detects individual structures.

\section{Robustness of \texttt{Sub-X}} \label{app::robustness}

As presented in Sec. \ref{sec::filament_finder}, \texttt{Sub-X} uses different parameters, namely $S_{\mathrm{lim}}$, $c_{\max}$, $c_{\mathrm{mean}}$ and $c_{\min}$, to identify and group substructures. Consequently, the size and, hence, the shape of the identified substructures depends on the choice of parameters. To verify, that our results are not significantly affected by our choices, we have re-analysed the shape of the six observed relics, see Sec. \ref{sec::observations}, using different values for the parameters.

\begin{figure*}
 \includegraphics[width = \textwidth]{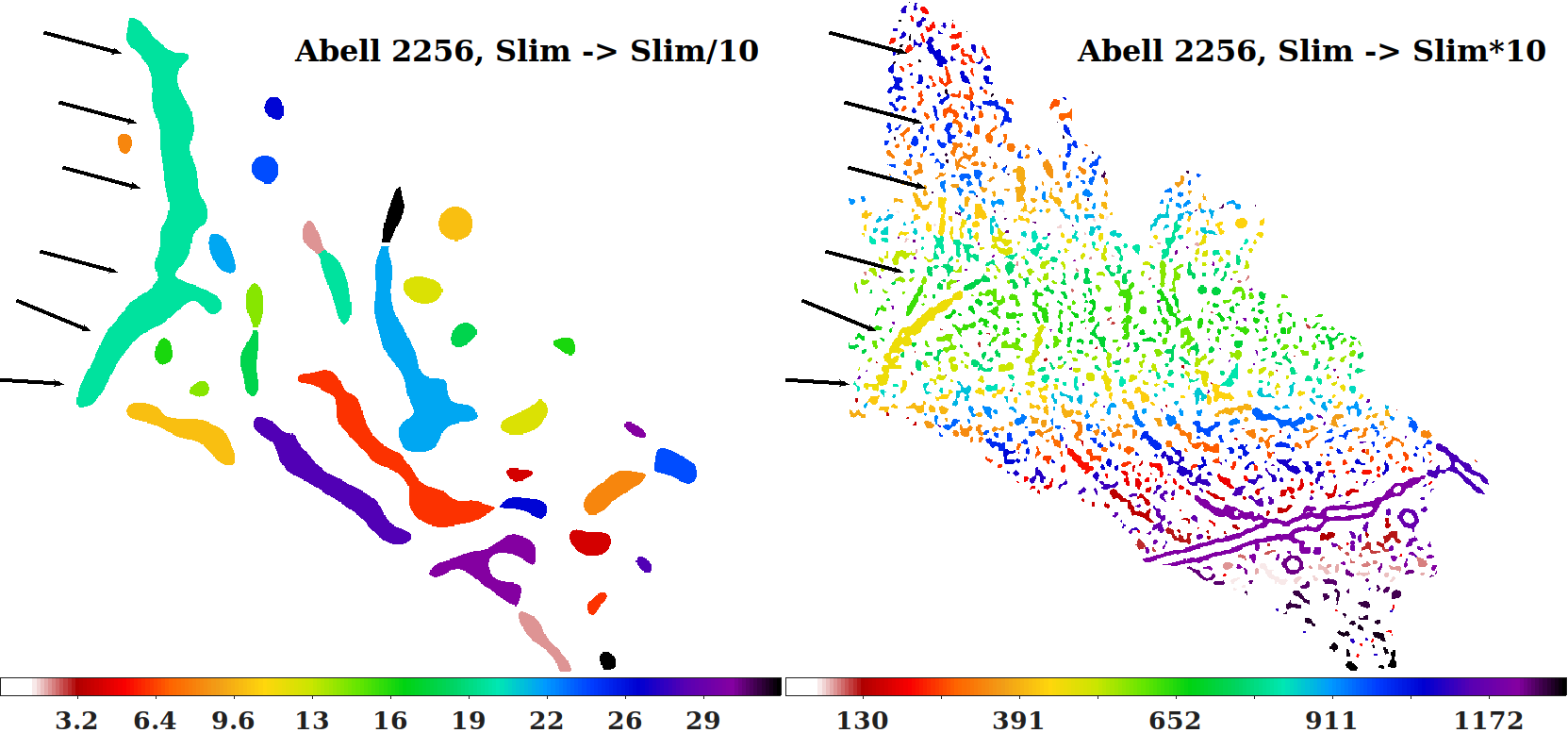}
 \caption{Extracted substructures in Abell 2256, when using different values for $S_{\mathrm{lim}}$. In the left plot, $S_{\mathrm{lim}}$ has been decreased by one order of magnitude. As a result, only large-scale substructures are identified. The arrows mark a region, that consist of multiple small-scale substructures, compare with Fig. \ref{fig::A2256_paper}. In the right plot, $S_{\mathrm{lim}}$ has been increased by one order of magnitude. As a result, numerous small-scale substructures are identified. As a consequence, true substructures can be broken up into several smaller substructures, see arrows and compare with Fig. \ref{fig::A2256_paper}.}
 \label{fig::slim_compare}
\end{figure*}

In the first test, we decreased $S_{\mathrm{lim}}$ by one order of magnitude. Specifically, in the first step of the the smoothing, we set all pixels with a value below $S_{\mathrm{lim}}/10$ to zero. Consequently, the median of the Euclidean distance transform, and the standard deviation of the Gaussian filter both increase. A visual comparison of the extracted filaments yields that lowering $S_{\mathrm{lim}}$ causes fewer small-scale substructures to be detected. Furthermore, clearly distinct substructures are grouped together. As an example, we plot the identified substructures for Abell 2256 in Fig. \ref{fig::slim_compare}. For each relic, we computed the 2D filamentarity of each substructure, and performed an ODR, i.e. $f_{\mathrm{2D}} \propto \kappa \log_{10  as well as}(S_{\mathrm{int}})$. As in Sec. \ref{sec::observations}, we performed three ODR including only substructures with $\fzd > 0.0$,  $\fzd > 0.1$ and $\fzd > 0.2$. On average, we obtain a slope of $\kappa \approx 0.27 \pm 0.04$, that is similar to the one found in Sec. \ref{sec::observations}.

In the second test, we have increased $S_{\mathrm{lim}}$ by one order of magnitude. Specifically, in the first step of the the smoothing, we set all pixels with a value below $S_{\mathrm{lim}}\cdot10$ to zero. Consequently, the median of the Euclidean distance transform and the standard deviation of the Gaussian filter decrease. As expected, significantly more small-scale substructures are identified, see Fig. \ref{fig::slim_compare}. In some cases, a single substructure is even identified as several smaller ones. The corresponding ODRs yield an average slope of $\kappa \approx 0.20 \pm 0.01$. This slope is a bit flatter than the one found in Sec. \ref{sec::observations}. However, the ODR is highly contaminated by all the small-scale substructures, that are not all individual substructures.

\begin{figure}
 \includegraphics[width = 0.5\textwidth]{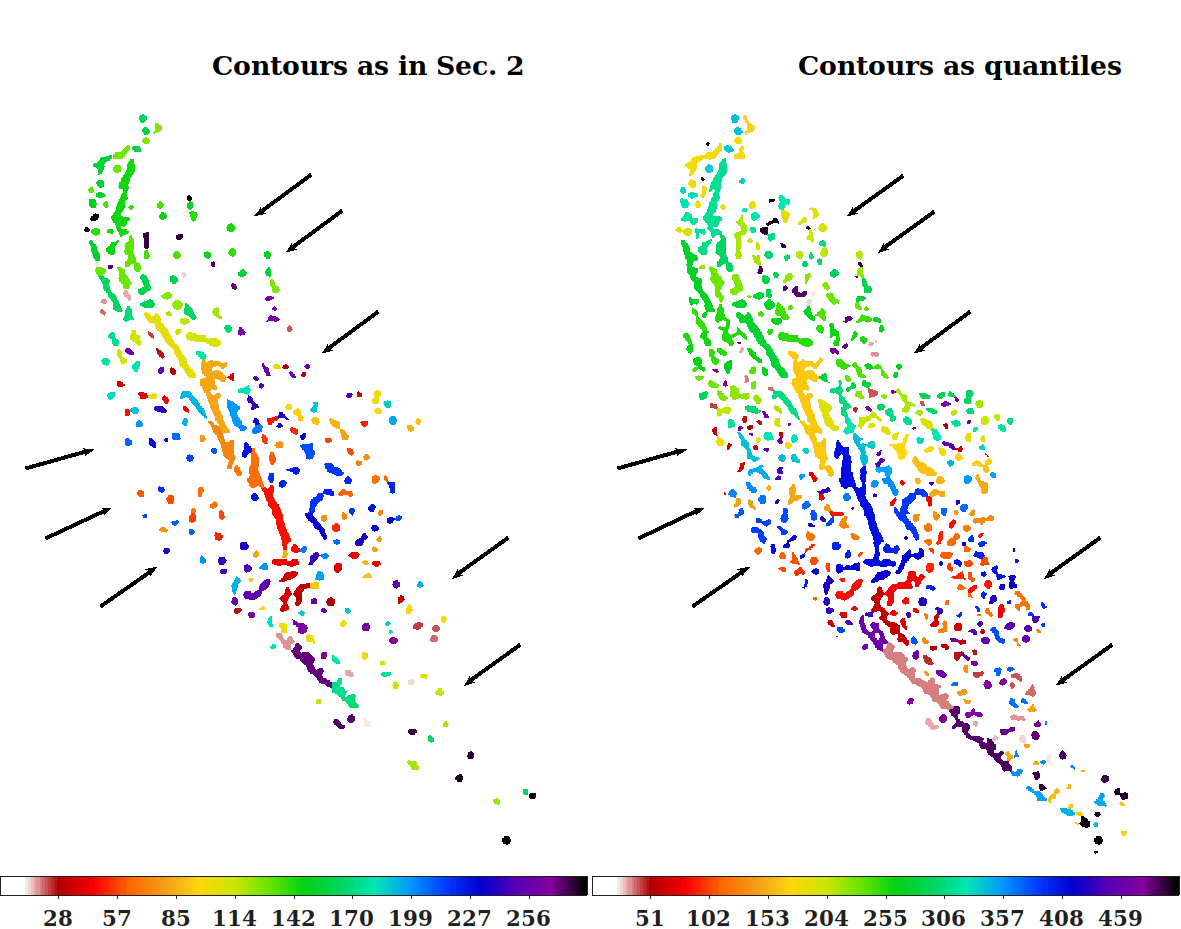}
 \caption{Visual comparison of extracted substructures in Abell 3667 SE, when using different values of $c_{\max}$, $c_{\mathrm{mean}}$ and $c_{\min}$. The left panel shows the substructures, when $c_{\max}$, $c_{\mathrm{mean}}$ and $c_{\min}$ are chosen as in Sec. \ref{sec::filament_finder}. The right panel shows the substructures, when the quantiles are used. For the later case, significantly more small scale filaments are extracted in front and behind the relic, as marked by the arrows.}
 \label{fig::contour_compare2}
\end{figure}

In the third test, we used different choices for the contour values. For the comparison, we chose that $c_{\max}$, $c_{\mathrm{mean}}$ and $c_{\min}$ are the $70 \ \%$ quantile, $50 \ \%$ quantile, and $30 \ \%$ quantile of the pixels with $- \mathcal{L}\left( \mathcal{G}_{\sigma} \right) > 0$, respectively. Here, we kept $S_{\mathrm{lim}}$ as defined in Sec. \ref{sec::filament_finder}. A visual comparison of the extracted filaments showed that a larger number of small-scale substructures is detected. As an example, we plot the detected substructures for Abell 3667 SE in Fig. \ref{fig::contour_compare2}. Again, we computed the 2D filamentarity of each substructure and the ODR, i.e. $f_{\mathrm{2D}} \propto \kappa \log_{10}(S_{\mathrm{int}})$. The average slope of the ODRs is $\kappa \approx 0.26 \pm 0.02$, that is in agreement with the results from Sec. \ref{sec::observations}.

Albeit some small differences, our checks showed that our results of the main paper are robust against the specific choices of $S_{\mathrm{lim}}$, $c_{\max}$, $c_{\mathrm{mean}}$ and $c_{\min}$.

\section{Shock Finder}\label{app::shock_finder}

In the scope of this work, we have implemented a new shock finder into \CRaTer. In the following, we present this new shock finder. The new algorithm works as follows.

A tracer particle is aware of being inside a shock or not. After each advection, a tracer particle measures the local divergence, $\nabla \cdot \mathbf{v}_0$, as well as the local gradient of temperature, $\nabla T_0$ , and entropy, $\nabla S_0$. If the velocity divergence is negative, $\nabla \cdot \mathbf{v}_0 < 0$, and the two gradients point into the same direction, $\nabla T_0 \cdot \nabla S_0 > 0$, the tracer has entered a shock front. As long as the two criteria are fulfilled, the tracer is still considered to be inside the shock front. Once, either of the criteria is not met anymore, i.e. $\nabla \cdot \mathbf{v}_1 > 0$ or $\nabla T_1 \cdot \nabla S_1 < 0$, the tracer has exited the shock. The corresponding Mach number is computed from the entropy jump:

\begin{align}
 M = \sqrt{\frac{4}{5} \frac{T_{\mathrm{1}}}{T_{\mathrm{0}}} \frac{\rho_{\mathrm{1}}}{\rho_{\mathrm{0}}} + 0.2}.
\end{align}

\end{document}

%% file: commands.tex
\newcommand{\enzo}{\texttt {ENZO}} 
\newcommand{\CRaTer}{\texttt{CRaTer}}

\newcommand{\music}{\texttt{MUSIC}}
\newcommand{\dd}{\mathrm{d}}
\newcommand{\Mpc}{\mathrm{Mpc}}
\newcommand{\Msun}{\mathrm{M}_{\odot}}

\newcommand{\kpc}{\mathrm{kpc}}

\newcommand{\km}{\mathrm{km}}
\newcommand{\sek}{\mathrm{s}}

\newcommand{\MeV}{\mathrm{MeV}}
\newcommand{\GeV}{\mathrm{GeV}}

\newcommand{\erg}{\mathrm{erg}}
\newcommand{\Hz}{\mathrm{Hz}}

\newcommand{\GHz}{\mathrm{GHz}}

\newcommand{\fzd}{f_{\mathrm{2D}}}
\newcommand{\fdd}{f_{\mathrm{3D}}}
\newcommand{\pdd}{p_{\mathrm{3D}}}

\definecolor{myred}{rgb}{1,0,0} 
\definecolor{myblue}{rgb}{0,0,1}
\definecolor{mygreen}{rgb}{0,1,0}

%% file: main.bbl
\providecommand{\SortNoop}[1]{}
\begin{thebibliography}{52}
\expandafter\ifx\csname natexlab\endcsname\relax\def\natexlab#1{#1}\fi

\bibitem[{{Astropy Collaboration} {et~al}\mbox{.}(2022){Astropy Collaboration},
  {Price-Whelan}, {Lim}, {Earl}, {Starkman}, {Bradley}, {Shupe}, {Patil},
  {Corrales}, {Brasseur}, {N{\"o}the}, {Donath}, {Tollerud}, {Morris},
  {Ginsburg}, {Vaher}, {Weaver}, {Tocknell}, {Jamieson}, {van Kerkwijk},
  {Robitaille}, {Merry}, {Bachetti}, {G{\"u}nther}, {Aldcroft},
  {Alvarado-Montes}, {Archibald}, {B{\'o}di}, {Bapat}, {Barentsen},
  {Baz{\'a}n}, {Biswas}, {Boquien}, {Burke}, {Cara}, {Cara}, {Conroy},
  {Conseil}, {Craig}, {Cross}, {Cruz}, {D'Eugenio}, {Dencheva}, {Devillepoix},
  {Dietrich}, {Eigenbrot}, {Erben}, {Ferreira}, {Foreman-Mackey}, {Fox},
  {Freij}, {Garg}, {Geda}, {Glattly}, {Gondhalekar}, {Gordon}, {Grant},
  {Greenfield}, {Groener}, {Guest}, {Gurovich}, {Handberg}, {Hart},
  {Hatfield-Dodds}, {Homeier}, {Hosseinzadeh}, {Jenness}, {Jones}, {Joseph},
  {Kalmbach}, {Karamehmetoglu}, {Ka{\l}uszy{\'n}ski}, {Kelley}, {Kern},
  {Kerzendorf}, {Koch}, {Kulumani}, {Lee}, {Ly}, {Ma}, {MacBride}, {Maljaars},
  {Muna}, {Murphy}, {Norman}, {O'Steen}, {Oman}, {Pacifici}, {Pascual},
  {Pascual-Granado}, {Patil}, {Perren}, {Pickering}, {Rastogi}, {Roulston},
  {Ryan}, {Rykoff}, {Sabater}, {Sakurikar}, {Salgado}, {Sanghi}, {Saunders},
  {Savchenko}, {Schwardt}, {Seifert-Eckert}, {Shih}, {Jain}, {Shukla}, {Sick},
  {Simpson}, {Singanamalla}, {Singer}, {Singhal}, {Sinha}, {Sip{\H{o}}cz},
  {Spitler}, {Stansby}, {Streicher}, {{\v{S}}umak}, {Swinbank}, {Taranu},
  {Tewary}, {Tremblay}, {Val-Borro}, {Van Kooten}, {Vasovi{\'c}}, {Verma}, {de
  Miranda Cardoso}, {Williams}, {Wilson}, {Winkel}, {Wood-Vasey}, {Xue},
  {Yoachim}, {Zhang}, {Zonca}, \& {Astropy Project Contributors}}]{astropy}
{Astropy Collaboration} {et~al.}, 2022, \apj, 935, 167

\bibitem[{{Bag} {et~al}\mbox{.}(2019){Bag}, {Mondal}, {Sarkar}, {Bharadwaj},
  {Choudhury}, \& {Sahni}}]{Bag_2019}
{Bag} S., {Mondal} R., {Sarkar} P., {Bharadwaj} S., {Choudhury} T.~R., {Sahni}
  V., 2019, \mnras, 485, 2235

\bibitem[{{Banfi} {et~al}\mbox{.}(2020){Banfi}, {Vazza}, \&
  {Wittor}}]{Banfi2020}
{Banfi} S., {Vazza} F., {Wittor} D., 2020, \mnras, 496, 3648

\bibitem[{{Bharadwaj} {et~al}\mbox{.}(2000){Bharadwaj}, {Sahni},
  {Sathyaprakash}, {Shandarin}, \& {Yess}}]{Bharadwaj_2000_2D_fila}
{Bharadwaj} S., {Sahni} V., {Sathyaprakash} B.~S., {Shandarin} S.~F., {Yess}
  C., 2000, \apj, 528, 21

\bibitem[{{Boelens} \& {Tchelepi}(2021)}]{QuantImPy}
{Boelens} A. M.~P., {Tchelepi} H.~A., 2021, SoftwareX, 16, 100823

\bibitem[{Boggs \& Rogers(1990)}]{boggs1990orthogonal}
Boggs P.~T., Rogers J.~E., 1990, Contemporary Mathematics, 112, 183

\bibitem[{{Botteon} {et~al}\mbox{.}(2020){Botteon}, {Brunetti}, {Ryu}, \&
  {Roh}}]{2020A&A...634A..64B}
{Botteon} A., {Brunetti} G., {Ryu} D., {Roh} S., 2020, \aap, 634, A64

\bibitem[{{Brienza} {et~al}\mbox{.}(2021){Brienza}, {Shimwell}, {de Gasperin},
  {Bikmaev}, {Bonafede}, {Botteon}, {Br{\"u}ggen}, {Brunetti}, {Burenin},
  {Capetti}, {Churazov}, {Hardcastle}, {Khabibullin}, {Lyskova},
  {R{\"o}ttgering}, {Sunyaev}, {van Weeren}, {Gastaldello}, {Mandal}, {Purser},
  {Simionescu}, \& {Tasse}}]{Brienza_2021_Nest}
{Brienza} M. {et~al.}, 2021, Nature Astronomy, 5, 1261

\bibitem[{{Br{\"u}ggen} \& {Vazza}(2020)}]{Bruggen_2020}
{Br{\"u}ggen} M., {Vazza} F., 2020, \mnras, 493, 2306

\bibitem[{{Brummel-Smith} {et~al}\mbox{.}(2019){Brummel-Smith}, {Bryan},
  {Butsky}, {Corlies}, {Emerick}, {Forbes}, {Fujimoto}, {Goldbaum}, {Grete},
  {Hummels}, {Kim}, {Koh}, {Li}, {Li}, {Li}, {OShea}, {Peeples}, {Regan},
  {Salem}, {Schmidt}, {Simpson}, {Smith}, {Tumlinson}, {Turk}, {Wise}, {Abel},
  {Bordner}, {Cen}, {Collins}, {Crosby}, {Edelmann}, {Hahn}, {Harkness},
  {Harper-Clark}, {Kong}, {Kritsuk}, {Kuhlen}, {Larrue}, {Lee}, {Meece},
  {Norman}, {Oishi}, {Paschos}, {Peruta}, {Razoumov}, {Reynolds}, {Silvia},
  {Skillman}, {Skory}, {So}, {Tasker}, {Wagner}, {Wang}, {Xu}, \&
  {Zhao}}]{2019JOSS....4.1636B}
{Brummel-Smith} C. {et~al.}, 2019, The Journal of Open Source Software, 4, 1636

\bibitem[{{Bryan} {et~al}\mbox{.}(2014){Bryan}, {Norman}, {O'Shea}, {Abel},
  {Wise}, {Turk}, {Reynolds}, {Collins}, {Wang}, {Skillman}, {Smith},
  {Harkness}, {Bordner}, {Kim}, {Kuhlen}, {Xu}, {Goldbaum}, {Hummels},
  {Kritsuk}, {Tasker}, {Skory}, {Simpson}, {Hahn}, {Oishi}, {So}, {Zhao},
  {Cen}, {Li}, \& {Enzo Collaboration}}]{ENZO_2014}
{Bryan} G.~L. {et~al.}, 2014, \apjs, 211, 19

\bibitem[{Collette(2013)}]{h5py}
Collette A., 2013, Python and HDF5. O'Reilly

\bibitem[{Danielsson(1980)}]{DANIELSSON1980227}
Danielsson P.-E., 1980, Computer Graphics and Image Processing, 14, 227

\bibitem[{{de Gasperin} {et~al}\mbox{.}(2022){de Gasperin}, {Rudnick},
  {Finoguenov}, {Wittor}, {Akamatsu}, {Br{\"u}ggen}, {Chibueze}, {Clarke},
  {Cotton}, {Cuciti}, {Dom{\'\i}nguez-Fern{\'a}ndez}, {Knowles}, {O'Sullivan},
  \& {Sebokolodi}}]{deGasperin_2022_A3667}
{de Gasperin} F. {et~al.}, 2022, \aap, 659, A146

\bibitem[{{Dedner} {et~al}\mbox{.}(2002){Dedner}, {Kemm}, {Kr{\"o}ner}, {Munz},
  {Schnitzer}, \& {Wesenberg}}]{2002JCoPh.175..645D}
{Dedner} A., {Kemm} F., {Kr{\"o}ner} D., {Munz} C.-D., {Schnitzer} T.,
  {Wesenberg} M., 2002, Journal of Computational Physics, 175, 645

\bibitem[{{Di Gennaro} {et~al}\mbox{.}(2018){Di Gennaro}, {van Weeren},
  {Hoeft}, {Kang}, {Ryu}, {Rudnick}, {Forman}, {R{\"o}ttgering}, {Br{\"u}ggen},
  {Dawson}, {Golovich}, {Hoang}, {Intema}, {Jones}, {Kraft}, {Shimwell}, \&
  {Stroe}}]{digennaro2018saus}
{Di Gennaro} G. {et~al.}, 2018, \apj, 865, 24

\bibitem[{{Eisenstein} \& {Hut}(1998)}]{Eisenstein_1998}
{Eisenstein} D.~J., {Hut} P., 1998, \apj, 498, 137

\bibitem[{{Ensslin} {et~al}\mbox{.}(1998){Ensslin}, {Biermann}, {Klein}, \&
  {Kohle}}]{ensslin1998}
{Ensslin} T.~A., {Biermann} P.~L., {Klein} U., {Kohle} S., 1998, \aap, 332, 395

\bibitem[{{Giacintucci} {et~al}\mbox{.}(2022){Giacintucci}, {Venturi},
  {Markevitch}, {Bourdin}, {Mazzotta}, {Merluzzi}, {Dallacasa}, {Bardelli},
  {Sikhosana}, {Smirnov}, \& {Bernardi}}]{2022ApJ...934...49G}
{Giacintucci} S. {et~al.}, 2022, \apj, 934, 49

\bibitem[{{Hahn} \& {Abel}(2011)}]{music}
{Hahn} O., {Abel} T., 2011, \mnras, 415, 2101

\bibitem[{Harris {et~al}\mbox{.}(2020)Harris, Millman, van~der Walt, Gommers,
  Virtanen, Cournapeau, Wieser, Taylor, Berg, Smith, Kern, Picus, Hoyer, van
  Kerkwijk, Brett, Haldane, del R{\'{i}}o, Wiebe, Peterson,
  G{\'{e}}rard-Marchant, Sheppard, Reddy, Weckesser, Abbasi, Gohlke, \&
  Oliphant}]{numpy}
Harris C.~R. {et~al.}, 2020, Nature, 585, 357

\bibitem[{{Hoeft} \& {Br{\"u}ggen}(2007)}]{2007MNRAS.375...77H}
{Hoeft} M., {Br{\"u}ggen} M., 2007, \mnras, 375, 77

\bibitem[{Hunter(2007)}]{matplotlib}
Hunter J.~D., 2007, Computing in Science \& Engineering, 9, 90

\bibitem[{{Knowles} {et~al}\mbox{.}(2022){Knowles}, {Cotton}, {Rudnick},
  {Camilo}, {Goedhart}, {Deane}, {Ramatsoku}, {Bietenholz}, {Br{\"u}ggen},
  {Button}, {Chen}, {Chibueze}, {Clarke}, {de Gasperin}, {Ianjamasimanana},
  {J{\'o}zsa}, {Hilton}, {Kesebonye}, {Kolokythas}, {Kraan-Korteweg}, {Lawrie},
  {Lochner}, {Loubser}, {Marchegiani}, {Mhlahlo}, {Moodley}, {Murphy},
  {Namumba}, {Oozeer}, {Parekh}, {Pillay}, {Passmoor}, {Ramaila}, {Ranchod},
  {Retana-Montenegro}, {Sebokolodi}, {Sikhosana}, {Smirnov}, {Thorat},
  {Venturi}, {Abbott}, {Adam}, {Adams}, {Aldera}, {Bauermeister}, {Bennett},
  {Bode}, {Botha}, {Botha}, {Brederode}, {Buchner}, {Burger}, {Cheetham}, {de
  Villiers}, {Dikgale-Mahlakoana}, {du Toit}, {Esterhuyse}, {Fadana},
  {Fanaroff}, {Fataar}, {Foley}, {Fourie}, {Frank}, {Gamatham}, {Gatsi},
  {Geyer}, {Gouws}, {Gumede}, {Heywood}, {Hlakola}, {Hokwana}, {Hoosen},
  {Horn}, {Horrell}, {Hugo}, {Isaacson}, {Jonas}, {Jordaan}, {Joubert},
  {Julie}, {Kapp}, {Kasper}, {Kenyon}, {Kotz{\'e}}, {Kotze}, {Kriek}, {Kriel},
  {Krishnan}, {Kusel}, {Legodi}, {Lehmensiek}, {Liebenberg}, {Lord}, {Lunsky},
  {Madisa}, {Magnus}, {Main}, {Makhaba}, {Makhathini}, {Malan}, {Manley},
  {Marais}, {Maree}, {Martens}, {Mauch}, {McAlpine}, {Merry}, {Millenaar},
  {Mokone}, {Monama}, {Mphego}, {New}, {Ngcebetsha}, {Ngoasheng}, {Ockards},
  {Otto}, {Patel}, {Peens-Hough}, {Perkins}, {Ramanujam}, {Ramudzuli},
  {Ratcliffe}, {Renil}, {Robyntjies}, {Rust}, {Salie}, {Sambu}, {Schollar},
  {Schwardt}, {Schwartz}, {Serylak}, {Siebrits}, {Sirothia}, {Slabber},
  {Sofeya}, {Taljaard}, {Tasse}, {Tiplady}, {Toruvanda}, {Twum}, {van Balla},
  {van der Byl}, {van der Merwe}, {van Dyk}, {Van Tonder}, {Van Wyk}, {Venter},
  {Venter}, {Welz}, {Williams}, \& {Xaia}}]{Knowles_2022_MeerKAT}
{Knowles} K. {et~al.}, 2022, \aap, 657, A56

\bibitem[{{Locatelli} {et~al}\mbox{.}(2020){Locatelli}, {Rajpurohit}, {Vazza},
  {Gastaldello}, {Dallacasa}, {Bonafede}, {Rossetti}, {Stuardi}, {Bonassieux},
  {Brunetti}, {Br{\"u}ggen}, \& {Shimwell}}]{locatelli2020dsa}
{Locatelli} N.~T. {et~al.}, 2020, \mnras, 496, L48

\bibitem[{{Mecke}(2000)}]{Mecke_2000_Mink_Book}
{Mecke} K.~R., 2000, in Statistical Physics and Spatial Statistics, {Mecke}
  K.~R., {Stoyan} D., eds., Springer Berlin Heidelberg, Berlin, Heidelberg, pp.
  111--184

\bibitem[{{Minkowksi}(1903)}]{Minkowski_1903}
{Minkowksi} H., 1903, Math. Ann, 57, 447

\bibitem[{{Nuza} {et~al}\mbox{.}(2017){Nuza}, {Gelszinnis}, {Hoeft}, \&
  {Yepes}}]{2017MNRAS.470..240N}
{Nuza} S.~E., {Gelszinnis} J., {Hoeft} M., {Yepes} G., 2017, \mnras, 470, 240

\bibitem[{{Nuza} {et~al}\mbox{.}(2012){Nuza}, {Hoeft}, {van Weeren},
  {Gottl{\"o}ber}, \& {Yepes}}]{2012MNRAS.420.2006N}
{Nuza} S.~E., {Hoeft} M., {van Weeren} R.~J., {Gottl{\"o}ber} S., {Yepes} G.,
  2012, \mnras, 420, 2006

\bibitem[{{Owen} {et~al}\mbox{.}(2014){Owen}, {Rudnick}, {Eilek}, {Rau},
  {Bhatnagar}, \& {Kogan}}]{2014ApJ...794...24O}
{Owen} F.~N., {Rudnick} L., {Eilek} J., {Rau} U., {Bhatnagar} S., {Kogan} L.,
  2014, \apj, 794, 24

\bibitem[{{Parekh} {et~al}\mbox{.}(2020){Parekh}, {Thorat}, {Kale}, {Hugo},
  {Oozeer}, {Makhathini}, {Kleiner}, {White}, {J{\'o}zsa}, {Smirnov}, {van der
  Heyden}, {Perkins}, {Andati}, {Ramaila}, \&
  {Ramatsoku}}]{2020MNRAS.499..404P}
{Parekh} V. {et~al.}, 2020, \mnras, 499, 404

\bibitem[{{Pearce} {et~al}\mbox{.}(2017){Pearce}, {van Weeren},
  {Andrade-Santos}, {Jones}, {Forman}, {Br{\"u}ggen}, {Bulbul}, {Clarke},
  {Kraft}, {Medezinski}, {Mroczkowski}, {Nonino}, {Nulsen}, {Randall}, \&
  {Umetsu}}]{2017ApJ...845...81P}
{Pearce} C.~J.~J. {et~al.}, 2017, \apj, 845, 81

\bibitem[{{Planck Collaboration} {et~al}\mbox{.}(2018){Planck Collaboration},
  {Aghanim}, {Akrami}, {Ashdown}, {Aumont}, {Baccigalupi}, {Ballardini},
  {Banday}, {Barreiro}, {Bartolo}, {Basak}, {Battye}, {Benabed}, {Bernard},
  {Bersanelli}, {Bielewicz}, {Bock}, {Bond}, {Borrill}, {Bouchet}, {Boulanger},
  {Bucher}, {Burigana}, {Butler}, {Calabrese}, {Cardoso}, {Carron},
  {Challinor}, {Chiang}, {Chluba}, {Colombo}, {Combet}, {Contreras}, {Crill},
  {Cuttaia}, {de Bernardis}, {de Zotti}, {Delabrouille}, {Delouis}, {Di
  Valentino}, {Diego}, {Dor{\'e}}, {Douspis}, {Ducout}, {Dupac}, {Dusini},
  {Efstathiou}, {Elsner}, {En{\ss}lin}, {Eriksen}, {Fantaye}, {Farhang},
  {Fergusson}, {Fernandez-Cobos}, {Finelli}, {Forastieri}, {Frailis},
  {Fraisse}, {Franceschi}, {Frolov}, {Galeotta}, {Galli}, {Ganga},
  {G{\'e}nova-Santos}, {Gerbino}, {Ghosh}, {Gonz{\'a}lez-Nuevo}, {G{\'o}rski},
  {Gratton}, {Gruppuso}, {Gudmundsson}, {Hamann}, {Handley}, {Hansen},
  {Herranz}, {Hildebrandt}, {Hivon}, {Huang}, {Jaffe}, {Jones}, {Karakci},
  {Keih{\"a}nen}, {Keskitalo}, {Kiiveri}, {Kim}, {Kisner}, {Knox},
  {Krachmalnicoff}, {Kunz}, {Kurki-Suonio}, {Lagache}, {Lamarre}, {Lasenby},
  {Lattanzi}, {Lawrence}, {Le Jeune}, {Lemos}, {Lesgourgues}, {Levrier},
  {Lewis}, {Liguori}, {Lilje}, {Lilley}, {Lindholm}, {L{\'o}pez-Caniego},
  {Lubin}, {Ma}, {Mac{\'\i}as-P{\'e}rez}, {Maggio}, {Maino}, {Mandolesi},
  {Mangilli}, {Marcos-Caballero}, {Maris}, {Martin}, {Martinelli},
  {Mart{\'\i}nez-Gonz{\'a}lez}, {Matarrese}, {Mauri}, {McEwen}, {Meinhold},
  {Melchiorri}, {Mennella}, {Migliaccio}, {Millea}, {Mitra},
  {Miville-Desch{\^e}nes}, {Molinari}, {Montier}, {Morgante}, {Moss}, {Natoli},
  {N{\o}rgaard-Nielsen}, {Pagano}, {Paoletti}, {Partridge}, {Patanchon},
  {Peiris}, {Perrotta}, {Pettorino}, {Piacentini}, {Polastri}, {Polenta},
  {Puget}, {Rachen}, {Reinecke}, {Remazeilles}, {Renzi}, {Rocha}, {Rosset},
  {Roudier}, {Rubi{\~n}o-Mart{\'\i}n}, {Ruiz-Granados}, {Salvati}, {Sandri},
  {Savelainen}, {Scott}, {Shellard}, {Sirignano}, {Sirri}, {Spencer},
  {Sunyaev}, {Suur-Uski}, {Tauber}, {Tavagnacco}, {Tenti}, {Toffolatti},
  {Tomasi}, {Trombetti}, {Valenziano}, {Valiviita}, {Van Tent}, {Vibert},
  {Vielva}, {Villa}, {Vittorio}, {Wand elt}, {Wehus}, {White}, {White},
  {Zacchei}, \& {Zonca}}]{PlanckVI2018}
{Planck Collaboration} {et~al.}, 2018, arXiv e-prints, arXiv:1807.06209

\bibitem[{{Rajpurohit} {et~al}\mbox{.}(2020){Rajpurohit}, {Hoeft}, {Vazza},
  {Rudnick}, {van Weeren}, {Wittor}, {Drabent}, {Brienza}, {Bonnassieux},
  {Locatelli}, {Kale}, \& {Dumba}}]{rajpurohit2020toothbrush}
{Rajpurohit} K. {et~al.}, 2020, \aap, 636, A30

\bibitem[{{Rajpurohit} {et~al}\mbox{.}(2022){Rajpurohit}, {van Weeren},
  {Hoeft}, {Vazza}, {Brienza}, {Forman}, {Wittor},
  {Dom{\'\i}nguez-Fern{\'a}ndez}, {Rajpurohit}, {Riseley}, {Botteon}, {Osinga},
  {Brunetti}, {Bonnassieux}, {Bonafede}, {Rajpurohit}, {Stuardi}, {Drabent},
  {Br{\"u}ggen}, {Dallacasa}, {Shimwell}, {R{\"o}ttgering}, {Gasperin},
  {Miley}, \& {Rossetti}}]{Rajpurohit_2022_A2256}
{Rajpurohit} K. {et~al.}, 2022, \apj, 927, 80

\bibitem[{{Rajpurohit} {et~al}\mbox{.}(2021){Rajpurohit}, {Vazza}, {van
  Weeren}, {Hoeft}, {Brienza}, {Bonnassieux}, {Riseley}, {Brunetti},
  {Bonafede}, {Br{\"u}ggen}, {Formann}, {Rajpurohit}, {R{\"o}ttgering},
  {Drabent}, {Dom{\'\i}nguez-Fern{\'a}ndez}, {Wittor}, \&
  {Andrade-Santos}}]{rajpurohit2021A2744}
{Rajpurohit} K. {et~al.}, 2021, \aap, 654, A41

\bibitem[{{Sahni} {et~al}\mbox{.}(1998){Sahni}, {Sathyaprakash}, \&
  {Shandarin}}]{Sahni_1998}
{Sahni} V., {Sathyaprakash} B.~S., {Shandarin} S.~F., 1998, \apjl, 495, L5

\bibitem[{{Schmalzing} {et~al}\mbox{.}(1999){Schmalzing}, {Buchert}, {Melott},
  {Sahni}, {Sathyaprakash}, \& {Shandarin}}]{Schmalzing_1999}
{Schmalzing} J., {Buchert} T., {Melott} A.~L., {Sahni} V., {Sathyaprakash}
  B.~S., {Shandarin} S.~F., 1999, \apj, 526, 568

\bibitem[{{Schmalzing} \& {Gorski}(1998)}]{Schmalzing_1998}
{Schmalzing} J., {Gorski} K.~M., 1998, \mnras, 297, 355

\bibitem[{{Seta} {et~al}\mbox{.}(2020){Seta}, {Bushby}, {Shukurov}, \&
  {Wood}}]{Seta_2020}
{Seta} A., {Bushby} P.~J., {Shukurov} A., {Wood} T.~S., 2020, Physical Review
  Fluids, 5, 043702

\bibitem[{{Sheth} {et~al}\mbox{.}(2003){Sheth}, {Sahni}, {Shandarin}, \&
  {Sathyaprakash}}]{Sheth_2003}
{Sheth} J.~V., {Sahni} V., {Shandarin} S.~F., {Sathyaprakash} B.~S., 2003,
  \mnras, 343, 22

\bibitem[{{Skillman} {et~al}\mbox{.}(2013){Skillman}, {Xu}, {Hallman},
  {O'Shea}, {Burns}, {Li}, {Collins}, \& {Norman}}]{2013ApJ...765...21S}
{Skillman} S.~W., {Xu} H., {Hallman} E.~J., {O'Shea} B.~W., {Burns} J.~O., {Li}
  H., {Collins} D.~C., {Norman} M.~L., 2013, \apj, 765, 21

\bibitem[{{Stuardi} {et~al}\mbox{.}(2019){Stuardi}, {Bonafede}, {Wittor},
  {Vazza}, {Botteon}, {Locatelli}, {Dallacasa}, {Golovich}, {Hoeft}, {van
  Weeren}, {Br{\"u}ggen}, \& {de Gasperin}}]{Stuardi2019}
{Stuardi} C. {et~al.}, 2019, \mnras, 489, 3905

\bibitem[{van~der Walt {et~al}\mbox{.}(2014)van~der Walt, {S}ch\"onberger,
  {Nunez-Iglesias}, {B}oulogne, {W}arner, {Y}ager, {G}ouillart, {Y}u, \& the
  scikit-image contributors}]{scikit-image}
van~der Walt S. {et~al.}, 2014, PeerJ, 2, e453

\bibitem[{{van Weeren} {et~al}\mbox{.}(2019){van Weeren}, {de Gasperin},
  {Akamatsu}, {Br{\"u}ggen}, {Feretti}, {Kang}, {Stroe}, \&
  {Zandanel}}]{vanweeren2019review}
{van Weeren} R.~J., {de Gasperin} F., {Akamatsu} H., {Br{\"u}ggen} M.,
  {Feretti} L., {Kang} H., {Stroe} A., {Zandanel} F., 2019, \ssr, 215, 16

\bibitem[{{Vazza} {et~al}\mbox{.}(2018){Vazza}, {Brunetti}, {Br{\"u}ggen}, \&
  {Bonafede}}]{2018MNRAS.474.1672V}
{Vazza} F., {Brunetti} G., {Br{\"u}ggen} M., {Bonafede} A., 2018, \mnras, 474,
  1672

\bibitem[{Virtanen {et~al}\mbox{.}(2020)Virtanen, Gommers, Oliphant, Haberland,
  Reddy, Cournapeau, Burovski, Peterson, Weckesser, Bright, {van der Walt},
  Brett, Wilson, Millman, Mayorov, Nelson, Jones, Kern, Larson, Carey, Polat,
  Feng, Moore, {VanderPlas}, Laxalde, Perktold, Cimrman, Henriksen, Quintero,
  Harris, Archibald, Ribeiro, Pedregosa, {van Mulbregt}, \& {SciPy 1.0
  Contributors}}]{scipy}
Virtanen P. {et~al.}, 2020, Nature Methods, 17, 261

\bibitem[{{Wittor} {et~al}\mbox{.}(2021{\natexlab{a}}){Wittor}, {Ettori},
  {Vazza}, {Rajpurohit}, {Hoeft}, \&
  {Dom{\'\i}nguez-Fern{\'a}ndez}}]{Wittor_2021_Mach}
{Wittor} D., {Ettori} S., {Vazza} F., {Rajpurohit} K., {Hoeft} M.,
  {Dom{\'\i}nguez-Fern{\'a}ndez} P., 2021{\natexlab{a}}, \mnras, 506, 396

\bibitem[{{Wittor} {et~al}\mbox{.}(2021{\natexlab{b}}){Wittor}, {Hoeft}, \&
  {Br{\"u}ggen}}]{Wittor_2021_Spec}
{Wittor} D., {Hoeft} M., {Br{\"u}ggen} M., 2021{\natexlab{b}}, Galaxies, 9, 111

\bibitem[{{Wittor} {et~al}\mbox{.}(2019){Wittor}, {Hoeft}, {Vazza},
  {Br{\"u}ggen}, \& {Dom{\'\i}nguez-Fern{\'a}ndez}}]{wittor2019pol}
{Wittor} D., {Hoeft} M., {Vazza} F., {Br{\"u}ggen} M.,
  {Dom{\'\i}nguez-Fern{\'a}ndez} P., 2019, \mnras, 490, 3987

\bibitem[{{Wittor} {et~al}\mbox{.}(2017){Wittor}, {Vazza}, \&
  {Br{\"u}ggen}}]{2017MNRAS.464.4448W}
{Wittor} D., {Vazza} F., {Br{\"u}ggen} M., 2017, \mnras, 464, 4448

\bibitem[{{Wittor} {et~al}\mbox{.}(2020){Wittor}, {Vazza}, {Ryu}, \&
  {Kang}}]{wittor2020gammas}
{Wittor} D., {Vazza} F., {Ryu} D., {Kang} H., 2020, \mnras, 495, L112

\end{thebibliography}
